%% file: QEM_analysis.tex
\documentclass[journal,romanappendices]{IEEEtran}
\input{SupportDocuments/YFPreamble.tex}

\usepackage{winsnotation}
\usepackage{url}
\newcommand{\Sop}[1]{{\mathcal{#1}}}
\newcommand{\vectorize}[1]{{\rm vec}\{#1\}}
\newcommand{\ivectorize}[1]{{\rm vec}^{-1}\{#1\}}
\newcommand{\mdiag}[1]{{\rm mdiag}\left\{#1\right\}}
\newcommand{\uconj}[2]{#1#2#1^\dagger}
\renewenvironment{IEEEbiography}[1]
  {\IEEEbiographynophoto{#1}}
  {\endIEEEbiographynophoto}
\begin{document}
\title{Sampling Overhead Analysis of Quantum Error Mitigation: Uncoded vs. Coded Systems}

\author{Yifeng Xiong, Daryus Chandra, Soon Xin Ng, \IEEEmembership{Senior Member, IEEE}, and Lajos Hanzo, \IEEEmembership{Fellow, IEEE}
\thanks{Authors are with School of Electronics and Computer Science, University of Southampton, SO17 1BJ, Southampton (UK). Corresponding author: Lajos Hanzo (e-mail: lh@ecs.soton.ac.uk).}
\thanks{L. Hanzo would like to acknowledge the financial support of the Engineering and Physical Sciences Research Council projects EP/N004558/1, EP/P034284/1, EP/P034284/1, EP/P003990/1 (COALESCE), of the Royal Society's Global Challenges Research Fund Grant as well as of the European Research Council's Advanced Fellow Grant QuantCom.}
}

\maketitle

\begin{abstract}
Quantum error mitigation (QEM) is a promising technique of protecting hybrid quantum-classical computation from decoherence, but it suffers from sampling overhead which erodes the computational speed. In this treatise, we provide a comprehensive analysis of the sampling overhead imposed by QEM. In particular, we show that Pauli errors incur the lowest sampling overhead among a large class of realistic quantum channels having the same average fidelity. Furthermore, we show that depolarizing errors incur the lowest sampling overhead among all kinds of Pauli errors. Additionally, we conceive a scheme amalgamating QEM with quantum channel coding, and analyse its sampling overhead reduction compared to pure QEM. Especially, we observe that there exist a critical number of gates contained in quantum circuits, beyond which their amalgamation is preferable to pure QEM.
\end{abstract}

\begin{IEEEkeywords}
Quantum error mitigation, sampling overhead, quantum error correction codes, quantum error detection codes, hybrid quantum-classical computation.
\end{IEEEkeywords}

\maketitle

\section*{Acronyms}
\begin{tabular}{ll}
CPTnI & Completely Positive Trace-nonIncreasing \\
CPTP & Completely Positive Trace-Preserving \\
GEP & Gate Error Probability \\
GGEP & Generalized Gate Error Probability \\
QEM & Quantum Error Mitigation \\
QECC & Quantum Error Correction Code \\
QEDC & Quantum Error Detection Code \\
SOF & Sampling Overhead Factor \\
\end{tabular}

\section*{Notations}
\begin{itemize}
\item Scalars, vectors and matrices are represented by $x$, $\V{x}$, and $\M{X}$, respectively. Sets and operators are
denoted as $\Set{X}$ and $\Sop{X}$, respectively.
\item The notations $\M{I}_{k}$, $\V{1}_n$, and $\V{0}_{n}$ denote the $k\times k$ identity matrix, an $n$-dimensional
all-one vector, and the $n$-dimensional all-zero vector, respectively. The subscripts are omitted when there is no confusion.
\item The notation $\|\V{x}\|_p$ represents the $\ell_p$-norm of vector $\V{x}$. Then notation $1/\V{x}$ denotes the element-wise reciprocal of vector $\V{x}$.
\item The notation $[\M{A}]_{i,j}$ denotes the $(i,j)$-th entry of matrix $\V{A}$. For a vector $\V{x}$, $[\V{x}]_i$ denotes
its $i$-th component. The submatrix obtained by extracting the $i_1$-th to $i_2$-th rows and the $j_1$-th to $j_2$-th columns from $\M{A}$ is denoted as $[\M{A}]_{i_1:i_2,j_1:j_2}$. The notation $[\M{A}]_{:,i}$ denotes the $i$-th column of $\M{A}$, and $[\M{A}]_{i,:}$ denotes the $i$-th row, respectively.
\item The notation ${\mathrm{diag}}(\cdot)$ denotes a diagonal matrix obtained by placing its argument on the main diagonal, and ${\mathrm{mdiag}}(\M{A})$ denotes the matrix obtained by setting all entries in matrix $\M{A}$ to zero apart from the main diagonal.
\item The notation ${\mathrm{vec}}(\M{A})$ denotes the vector obtained by vectorizing matrix $\M{A}$, and ${\mathrm{vec}}^{-1}(\cdot)$ denotes the inverse operation.
\item The trace of matrix $\M{A}$ is denoted as ${\mathrm{Tr}}\{\M{A}\}$, and the complex conjugate of $\M{A}$ is denoted as $\M{A}^\dagger$. Similarly, the complex adjoint of an operator $\Sop{X}$ is also denoted as $\Sop{X}^\dagger$.
\item The notation $\M{A}\otimes \M{B}$ represents the Kronecker product between matrices $\M{A}$ and $\M{B}$. The tensor product of operators $\Sop{A}$ and $\Sop{B}$ is also denoted as $\Sop{A}\otimes\Sop{B}$. Furthermore, the Cartesian product between sets $\Set{A}$ and $\Set{B}$ is denoted as $\Set{A}\otimes \Set{B}$ as well.
\item The sign function ${\mathrm{sgn}}(x)$ is defined as
$$
{\mathrm{sgn}}(x) = \left\{
                 \begin{array}{ll}
                   1, & \hbox{$x>0$;} \\
                   0, & \hbox{$x=0$;} \\
                   -1, & \hbox{$x<0$.}
                 \end{array}
               \right.
$$
\item The Bachmann-Landau notations \cite{intro_algo} used in this treatise are given as follows:

\vspace{2mm}
\noindent
\begin{tabular}{ll}
    $a(n)= O(b(n))$ & $\mathop{\lim\sup}_{n\rightarrow \infty} a(n)(b(n))^{-1} < \infty$ \\
    $a(n)=o(b(n))$ & $\mathop{\lim\inf}_{n\rightarrow \infty} a(n)(b(n))^{-1} = 0$
\end{tabular}
\end{itemize}

\section{Introduction}
\IEEEPARstart{R}{ecent} years have witnessed an astonishing development in the area of quantum computing. State-of-the-art quantum computers are typically equipped with qubits scaling from fifty to a few hundreds, and have been shown difficult to be simulated on classical supercomputers \cite{quantum_supremacy}. This marks the beginning of the era of noisy intermediate-scale quantum computing \cite{nisq}. The word ``noisy'' indicates that noisy intermediate-scale quantum computers would suffer from the notorious quantum decoherence effects, which impose a perturbation on each and every quantum operation carried out by a ``quantum gate''.

In principle, noisy quantum gates do not necessarily prevent quantum computation from being sufficiently accurate. A classical result, namely the threshold theorem \cite{threshold_thm}, states that quantum computation may be carried out in the presence of decoherence with the help of \acp{qecc} \cite{qecc,qit,qecc_survey,qtecc}, given that the error rate of each quantum gate is below a certain threshold. Generally speaking, \acp{qecc} protect a logical quantum bit (qubit) by mapping it to a larger set of physical qubits. The redundancy of the physical qubits ensures that errors perturbing a small fraction of the qubits can be detected and corrected with the help of some ancillary qubits (ancillas). Moreover, the error-correction capability can be further enhanced by concatenating several \acp{qecc}, albeit naturally, at the expense of higher qubit overhead \cite{concatenation,qtc1,qtc2}.

\begin{figure}[t]
    \centering
   \begin{overpic}[width=.45\textwidth]{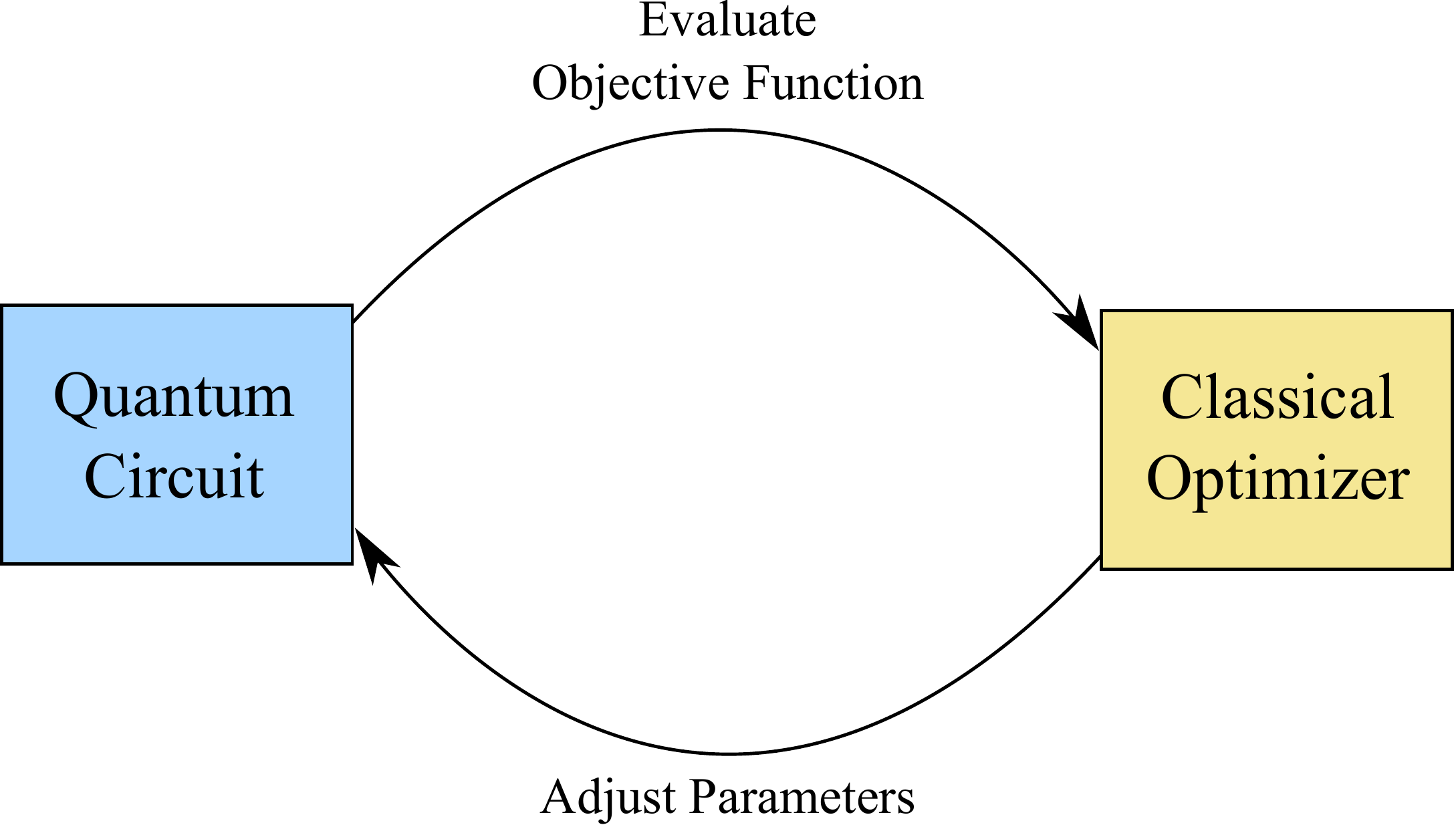}
    \put(40,25){\large $\min_{\V{\theta}} f(\V{\theta})$}
    \put(49,6){$\V{\theta}$}
    \put(46,43){$f(\V{\theta})$}
    \end{overpic}
    \caption{Illustration of the iterations in hybrid quantum-classical algorithms.}
    \label{fig:hqc}
\end{figure}

However, compared to the idealized quantum computing models considered in the threshold theorem, practical noisy intermediate-scale quantum computers may not be capable of supporting fully fault-tolerant operations, due to their limited number of qubits. Consequently, they may not be able to execute algorithms that require relatively long processing time, such as Shor's factorization algorithm \cite{shor} and Grover's quantum search algorithm \cite{grover}. Fortunately, there is evidence that algorithms tailored for noisy intermediate-scale quantum computers may yield superior performance compared to those of classical computers \cite{nisq,qaoa}. Most of these algorithms belong to the category of hybrid quantum-classical algorithms, which exploit the power of classical computation to compensate for the short coherence time of quantum processors. As portrayed in Fig. \ref{fig:hqc}, a typical hybrid quantum-classical algorithm would be performed in an iterative fashion. The quantum circuit, which will be referred to as the function-evaluation circuit in this treatise, is designed to evaluate an objective function, given a set of input parameters \cite{ansatz}. The value of the objective function is then utilized in a classical optimizer, which computes an adjusted set of parameters for the next iteration. In general, the design of the function-evaluation circuit determines the application of the algorithm. Popular designs include the alternating operator circuit used in the quantum approximate optimization algorithm \cite{qaoa}, the ``unitary coupled-cluster ansatz'' circuit applied in the computation of molecular energy based on variational eigensolver \cite{vqe,vqe002,vqe003}, as well as other heuristic designs aiming for quantum machine learning \cite{qml1,qml2}. Remarkably, the quantum approximate optimization algorithm has been applied to communication-related problems as well, such as channel decoding \cite{qaoa_decoding}.

Although the quantum circuits in hybrid quantum-classical algorithms have short depth, their decoherence may still inflict non-negligible computational errors \cite{nisq_error1,nisq_error2}. This necessitates the design of low-qubit-overhead techniques for protecting quantum gates. Recently, \ac{qem} has been proposed, which may correct the computational result without using any ancilla \cite{qem,practical_qem,qem_exp}. Without loss of generality, we may decompose a realistic imperfect quantum gate into a perfect gate followed by a quantum channel. \ac{qem} mitigates the deleterious effect of the channel by applying an ``inverse channel'' right after the imperfect gate, which is implemented using a probabilistic mixture of gates \cite{qem}.

In contrast to the qubit overhead of \acp{qecc}, \ac{qem} introduces another type of computational overhead, namely the sampling overhead \cite{qem}. This overhead originates from the fact that the ``inverse channel'' is typically not \ac{cptp} (unless the original channel is of unitary nature, and hence it does not impose decoherence) \cite{qem,practical_qem}. Consequently, \ac{qem} leads to an increased variance in the final computational result, hence additional measurements are required at the output quantum state for achieving a satisfactory accuracy. In effect, increasing the number of measurements will slow down the computation process. As the depth of the quantum circuit grows, the sampling overhead may accumulate dramatically. Ultimately, the benefit of quantum speedup will be neutralized for computation tasks that require extremely long coherence time.\footnote{Another issue limiting the performance of \ac{qem} is that there may be residual error channels, when the gates used by \ac{qem} are themselves imperfect and/or the channel estimation is imperfect. This issue may be partially alleviated via quantum gate set tomography, which enables us to implement \ac{qem} using imperfect gates \cite{practical_qem}. Yet in this treatise, we will focus on the issue of sampling overhead.}

In general, the sampling overhead required depends on the channel characteristics. Naturally, a fundamental question concerning the practicality of \ac{qem} is: ``Can we predict and control the sampling overhead given a limited number of channel parameters?'' In this treatise, we investigate this deep-rooted research question from both theoretical and practical perspectives. We first introduce the notion of the so-called \ac{sof} for characterizing the sampling overhead incurred by a quantum channel, and then provide a comprehensive analysis of the \ac{sof} of general \ac{cptp} channels. Finally, we discuss potential techniques of reducing the \ac{sof} of quantum channels.

Our main contributions are summarized as follows.
\begin{itemize}
\item We present the general design philosophy of \ac{qem} from a communication theoretical perspective, emphasizing its application in hybrid quantum-classical computation, and introduce the notion of \ac{sof} to quantify its sampling overhead. Specifically, we highlight that by invoking a ``quantum channel precoder'' in the quantum circuit, the \ac{sof} required by \ac{qem} may be reduced.
\item We formulate a coherent--triangular error decomposition of memoryless quantum channels, which are tensor products of single-qubit channels, with the aid of their Pauli transfer matrix \cite{ptm} representation. Specifically, the coherent component of a quantum channel has a zero \ac{sof}, meaning that it can be mitigated without any sampling overhead when compensated using \ac{qem} in the ideal case, while the triangular component always has non-zero \ac{sof}.
\item We prove that Pauli channels have the lowest \ac{sof} among all triangular channels, and show that depolarizing channels have the lowest \ac{sof} among Pauli channels. Furthermore, we provide upper and lower bounds on the \ac{sof} of Pauli channels.
\item We conceive an amalgam of \ac{qem} and quantum codes, including \acp{qecc} and \acp{qedc}, aiming for reducing the overall \ac{sof}. In particular, we show that there exist critical sizes of quantum circuits, beyond which the amalgamation is preferable to pure \ac{qem}.
\end{itemize}

\begin{figure}[t]
    \centering
   \includegraphics[width=.47\textwidth]{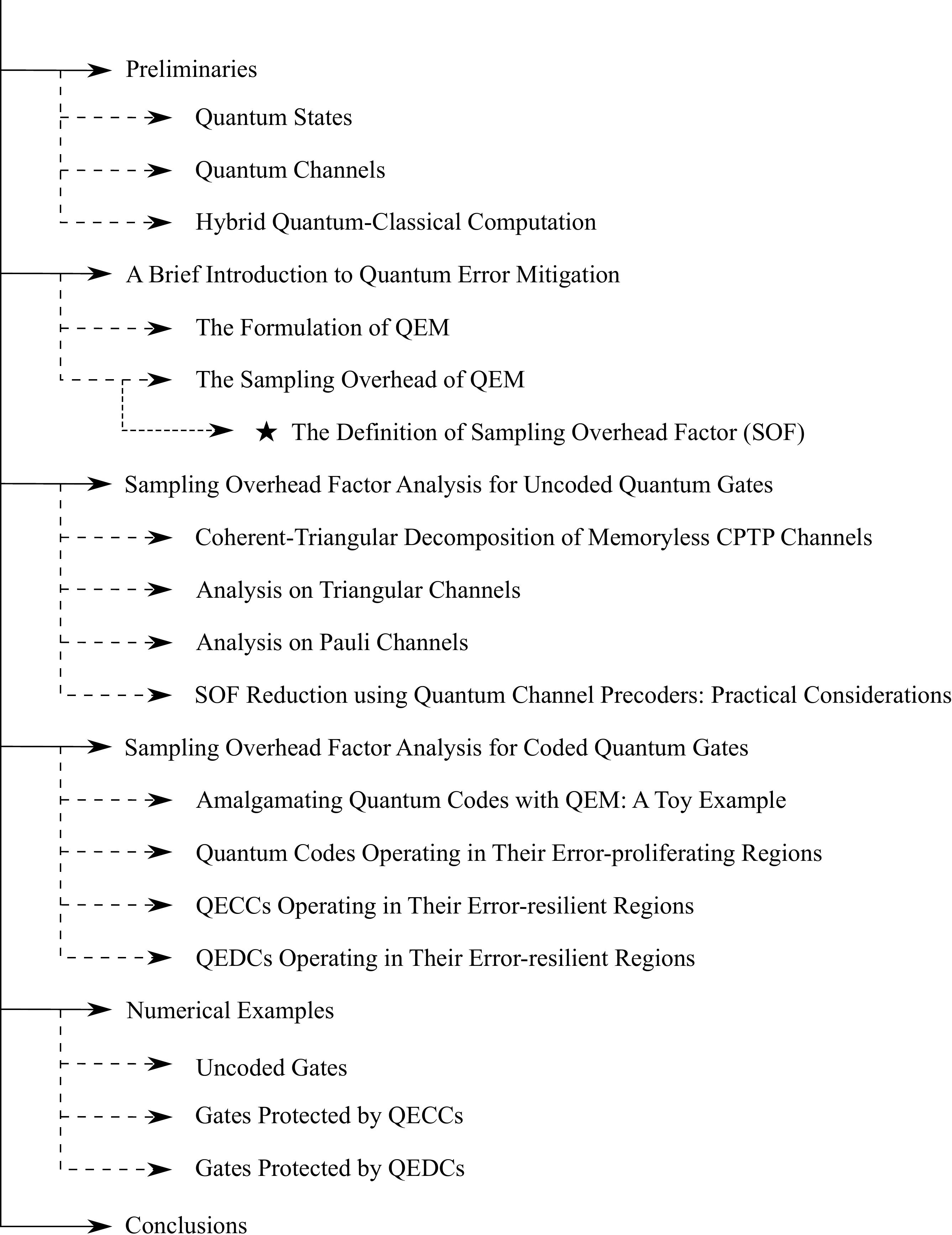}
    \caption{The structure of this treatise.}
    \label{fig:structure}
\end{figure}

The structure of the treatise is demonstrated in Fig. \ref{fig:structure}, and the rest of this treatise is organized as follows. In Section \ref{sec:preliminaries}, we present preliminary concepts that will be used extensively thereafter, such as quantum states, channels and hybrid quantum-classical computation. Then, in Section \ref{sec:qem} we present the general design and formulation of \ac{qem}. Based on this formulation, we analyse on the \ac{sof} of uncoded quantum gates in Section \ref{sec:uncoded}. In Section \ref{sec:coded}, we conceive and analyse the amalgam of \ac{qem} and \acp{qecc} as well as \acp{qedc}, which will be referred to as the \ac{qecc}-\ac{qem} and \ac{qedc}-\ac{qem} schemes, respectively. The analytical results are then demonstrated via numerical simulations in Section \ref{sec:numerical}. Finally, we conclude in Section \ref{sec:conclusions}.

\section{Preliminaries}\label{sec:preliminaries}
\subsection{Quantum states}
The basic carrier of quantum information is a qubit, namely a two-level quantum system. Ideally, the state of a qubit can be characterized by a vector as
\begin{equation}
\ket{\psi} = \alpha\ket{0} + \beta\ket{1},
\end{equation}
satisfying the normalization property of $|\alpha|^2+|\beta|^2=1$. Under the conventional computational basis, the basis vectors $\ket{0}$ and $\ket{1}$ can be expressed as
$$
\ket{0} = [1~0]^{\mathrm{T}},~\ket{1} = [0~1]^{\mathrm{T}}.
$$

In practice, the interaction between the qubits and the environment will cause decoherence, namely turning a deterministic quantum state into a probabilistic mixture of states. Especially, the deterministic states are termed as pure states, while the probabilistic mixtures are termed as mixed states. A mixed state of a qubit can be fully characterized by a $2\times 2$ matrix termed as the density matrix $\rho$ given by
\begin{equation}
\rho = \sum_i p_i \ket{\psi_i}\bra{\psi_i},
\end{equation}
satisfying $p_i\ge 0$ for all $i$ and $\sum_i p_i = 1$. Hence the density matrix is positive semi-definite and has unit trace. The pure states $\ket{\psi_i}$ are the components of the probabilistic mixture. Additionally, under the computational basis, it can be expressed as the linear combination of the following matrices:
$$
\begin{aligned}
\M{S}_{\Sop{I}} &= \left[
          \begin{array}{cc}
            1 & 0 \\
            0 & 1 \\
          \end{array}
        \right],~\M{S}_{\Sop{X}} = \left[
          \begin{array}{cc}
            0 & 1 \\
            1 & 0 \\
          \end{array}
        \right], \\
\M{S}_{\Sop{Y}} &= \left[
          \begin{array}{cc}
            0 & -i \\
            i & 0 \\
          \end{array}
        \right],~\M{S}_{\Sop{Z}} = \left[
          \begin{array}{cc}
            1 & 0 \\
            0 & -1 \\
          \end{array}
        \right].
\end{aligned}
$$

In general, a mixed state of an $n$-qubit system can be characterized by a $2^n\times 2^n$ density matrix. Similar to the single qubit case, the $n$-fold tensor products of $\M{S}_{\Sop{I}}$, $\M{S}_{\Sop{X}}$, $\M{S}_{\Sop{Y}}$ and $\M{S}_{\Sop{Z}}$ form a basis for the space of $2^n\times 2^n$ density matrices, termed as the $n$-qubit Pauli group. To facilitate further discussion, we denote $\M{S}_i^{(n)}$ as the $i$-th operator in the $n$-qubit Pauli group. The superscript $(n)$ is omitted when it is unambiguous from the context.

The difference between two quantum states $\rho_1$ and $\rho_2$ is typically quantified using fidelity, defined as

\begin{equation}
F(\rho_1,\rho_2)=\mathrm{Tr}\left\{\left(\rho_1^{\frac{1}{2}}\rho_2\rho_1^{\frac{1}{2}}\right)^{\frac{1}{2}}\right\}^2,
\end{equation}
and can be simplified as
$$
F(\rho_1,\rho_2) = {\mathrm{Tr}}\{\rho_1^\dagger\rho_2\}
$$
if either $\rho_1$ or $\rho_2$ represents a pure state.

\subsection{Quantum channels}
Formally speaking, quantum channels are typically modelled by linear operators acting on quantum states. Naturally, the output of a quantum channel has to be a legitimate quantum state, which is positive semi-definite, and has trace one. Therefore, quantum channels are required to be completely positive, trace preserving (\ac{cptp}) operators \cite[Sec. 8.2.4]{ncbook}. Any \ac{cptp} operator $\Sop{C}$ admits an operator-sum representation formulated as follows:
\begin{equation}\label{kraus_channel}
\Sop{C}(\rho) = \sum_{i} \M{K}_i \rho \M{K}_i^\dagger,
\end{equation}
where the operation elements $\M{K}_i$ satisfy the completeness condition given by
\begin{equation}\label{kraus_completeness}
\sum_{i}\M{K}_i^\dagger \M{K}_i = \M{I}.
\end{equation}

Alternatively, quantum channels can also be expressed in a matrix form. To elaborate, if we represent a quantum state having a density matrix of $\rho$ as a vector $\V{x}$, a quantum channel $\Sop{C}$ acting on $\V{x}$ can be written as a matrix $\M{C}$ satisfying

\begin{equation}\label{channel_vecform}
\M{C}\V{x} = \M{T}^\dagger\left[\sum_i (\M{K}_i^* \otimes \M{K}_i)\right]\M{T} \V{x},
\end{equation}
where $\M{T}$ is a basis transition matrix which determines the specific matrix form of the quantum channel. In general, the specific matrix form of a quantum channel depends on the set of bases we choose. In this treatise, we use the Pauli transfer matrix representation \cite{ptm} of quantum channels, given by
\begin{equation}\label{ptm_rep}
[\M{C}]_{i,j} = \frac{1}{2^n}{\mathrm{Tr}}\{\M{S}_i\Sop{C}(\M{S}_j)\},
\end{equation}
for which the basis transition matrix $\M{T}$ is given by
$$
\M{T} = \frac{1}{\sqrt{2^n}} \left[\vectorize{\M{S}_1}~\vectorize{\M{S}_2}~\dotsc~\vectorize{\M{S}_{4^n}}\right].
$$

In this sense, the vector representation $\V{x}$ of a density matrix $\rho$ can be expressed as
\begin{equation}\label{ptm_vec}
[\V{x}]_i = \frac{1}{\sqrt{2^n}}{\mathrm{Tr}}\{\M{S}_i\rho\}.
\end{equation}

A key quality indicator of a quantum channel is its average fidelity. As the terminology ``average fidelity'' suggests, it is the fidelity between the input and the output states, integrated over the space of all legitimate input states. Formally, the average fidelity of a quantum channel $\Sop{C}$ is defined as \cite{average_fidelity}
\begin{equation}
\bar{F}(\Sop{C}) = \int F(\ket{\psi}\!\bra{\psi},\Sop{C}(\ket{\psi}\!\bra{\psi}))~{\mathrm{d}}\!\ket{\psi}.
\end{equation}
Using the Pauli transfer matrix representation, the average fidelity of $\Sop{C}$ can be written in closed form as
\begin{equation}
\bar{F}(\Sop{C}) = \frac{{\mathrm{Tr}}\{\M{C}\}+2^n}{4^n+2^n},
\end{equation}
where $n$ is the number of qubits that $\Sop{C}$ acts upon. In general, $\bar{F}(\Sop{C})$ satisfies $0\le\bar{F}(\Sop{C})\le 1$, and $1-\bar{F}(\Sop{C})$ is often referred to as the ``average infidelity'' of $\Sop{C}$ \cite{average_fidelity}.

For Pauli channels, another important quality metric is the \ac{gep}, namely the probability that the output state does not coincide with the input state. For example, for the following channel
\begin{equation}\label{gep}
\Sop{C}(\rho) = (1-p)\rho + \frac{p}{3} \M{S}_{\Sop{X}}\rho\M{S}_{\Sop{X}}+ \frac{p}{3} \M{S}_{\Sop{Y}}\rho\M{S}_{\Sop{Y}}+ \frac{p}{3} \M{S}_{\Sop{Z}}\rho\M{S}_{\Sop{Z}},
\end{equation}
the \ac{gep} is $p$. Many important results on quantum coding, including the threshold theorem, are based on \ac{gep}.

A somewhat perplexing issue is that the \ac{gep} is inconsistent with the average fidelity. More precisely, for a Pauli channel $\Sop{C}$, we have $\bar{F}(\Sop{C}) \neq 1-{\mathrm{GEP}}$. To avoid the difficulty of using two different metrics for Pauli and non-Pauli channels, in this treatise we introduce a generalization of the \ac{gep}, which will be referred to as the \ac{ggep} hereafter. Specifically, we define the \ac{ggep} of channel $\Sop{C}$ as
\begin{equation}
\epsilon(\Sop{C}) = 1-\frac{1}{4^n} {\mathrm{Tr}}\{\M{C}\}.
\end{equation}
As a channel quality metric, \ac{ggep} has the following advantages.
\begin{enumerate}
  \item When $\Sop{C}$ is a Pauli channel, the \ac{ggep} degenerates to the conventional \ac{gep}, which is $p$ in \eqref{gep}.
  \item For a general channel $\Sop{C}$, which is not necessarily a Pauli channel, the \ac{ggep} is proportional to the average infidelity of $\Sop{C}$, in the sense that
  \begin{equation}
  \epsilon(\Sop{C}) = \left(1+\frac{2^n}{4^n}\right)[1-\bar{F}(\Sop{C})].
  \end{equation}
  Thus an operation preserving the average fidelity would also preserve the \ac{ggep}.
\end{enumerate}

\subsection{Hybrid quantum-classical computation}
The evaluation of eigenvalues and eigenvectors is a fundamental subroutine in many existing quantum algorithms, such as Shor's algorithm, the Harrow-Hassidim-Lloyd (HHL) algorithm, and in Hamiltonian simulation \cite{shor,hhl,hamiltonian_sim,hhl2,hhl3}. In early contributions, quantum phase estimation \cite[Sec. 5.2]{ncbook} was the default algorithm for eigenvalue evaluation, which requires a long coherence time. To enable eigenvalue evaluation on noisy intermediate-scale quantum computers, hybrid quantum-classical algorithms based on variational optimization have been proposed in \cite{vqe_theory,vqe_excited,vqe_spectra}.

Mathematically, for a Hermitian matrix $\M{H}$, the eigenvector $\V{\psi}_0$ corresponding to the smallest eigenvalue can be calculated as follows
\begin{subequations}\label{eigenvalue_problem}
\begin{align}
\V{\psi}_{\mathrm{g}} = \mathop{\arg\min}_{\V{\psi}} ~ &\V{\psi}^\dagger \M{H}\V{\psi}, \\
{\mathrm{subject~to~}} &~\|\V{\psi}\|_2 = 1. \label{normalization_cons}
\end{align}
\end{subequations}
This problem is referred to as the variational formulation of eigenvalue evaluation in the literature \cite{vqe,vqe_theory,vqe2}. If the eigenvectors are reparametrized using a vector $\V{\theta}$, yielding $\V{\psi}=\V{\psi}(\V{\theta})$, the task of finding the smallest eigenvalue or the corresponding eigenvector may be achieved by searching for the minimum of the objective function
\begin{equation}\label{obj_function}
J(\V{\theta}) = \V{\psi}^\dagger(\V{\theta})\M{H}\V{\psi}(\V{\theta})
\end{equation}
in the space of $\V{\theta}$, while satisfying the normalization constraint \eqref{normalization_cons}. The formulation \eqref{obj_function} has been applied to the electronic structure computation of the hydrogen molecule\footnote{The problem of hydrogen molecular electronic structure computation aims to find the bond length corresponding to the ground state of the hydrogen molecule. In this case, the objective function $J(\theta)$ is the ground state energy, while the parameter $\theta$ (which is a scalar) corresponds to the inter-atom distance.} in \cite{qcc}.  In general, this would be a non-convex problem with respect to $\V{\theta}$, which may be solved using iterative non-convex optimization solvers, such as the classic gradient descent and the Nelder-Mead simplex method \cite{nelder_mead}. In each iteration, the objective function or another function (e.g. the gradient) is first evaluated at a specific point in the parameter space, and then the parameters are updated according to the function values.

Under the framework of quantum computation, the problem \eqref{eigenvalue_problem} can be recast as
\begin{equation}\label{eigenvalue_problem_pure}
\ket{\psi_{\mathrm{g}}(\V{\theta})} = \mathop{\arg\min}_{\ket{\psi(\V{\theta})}}~\bra{\psi(\V{\theta})}\Sop{H}\ket{\psi(\V{\theta})},
\end{equation}
where $\Sop{H}$ is a quantum observable \cite[Sec. 2.2.5]{ncbook} representing the matrix $\M{H}$, and the state $\ket{\psi_{\mathrm{g}}(\V{\theta})}$ here corresponds to the ground state of $\Sop{H}$. The quadratic form $J(\V{\theta})=\bra{\psi(\V{\theta})}\Sop{H}\ket{\psi(\V{\theta})}$ can be viewed as the expectation value of the observable $\Sop{H}$. The constraint \eqref{normalization_cons} is automatically satisfied due to the normalization property of quantum states. Taking decoherence into account, \eqref{eigenvalue_problem_pure} can be reformulated equivalently for mixed states as
\begin{equation}\label{eigenvalue_problem_mixed}
\rho_{\mathrm{g}}(\V{\theta}) = \mathop{\mathrm{ argmin}}_{\rho(\V{\theta})}~{\mathrm{Tr}}\{\Sop{H}\rho(\V{\theta})\}.
\end{equation}
For conciseness of our discussion, we shall use the pure-state formulation \eqref{eigenvalue_problem_pure} hereafter, whenever there is no confusion.

\begin{figure}[t]
\centering
\includegraphics[width=.48\textwidth]{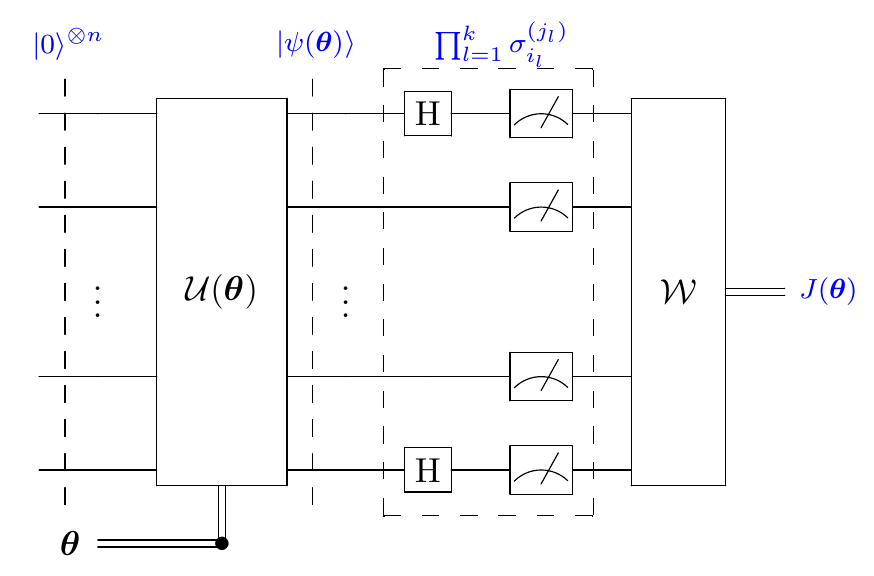}
\caption{Portrayal of a quantum circuit evaluating the objective function $J(\V{\theta})$ in a hybrid quantum-classical algorithm. The operator $\Sop{W}$ represents the weighted averaging operation. The gates in the dashed box are chosen according to the observable decomposition \eqref{decompose_observable}.}
\label{fig:qee}
\end{figure}

The essence of hybrid quantum-classical computation is to evaluate the functions using a quantum circuit, whereas the parameter values are updated using a classical computer, as illustrated in Fig. \ref{fig:hqc}. To be more specific, a schematic of the quantum circuit evaluating the objective function $J(\V{\theta})$ is portrayed in Fig. \ref{fig:qee}. The input state of the circuit is typically the all-zero state $\ket{0}^{\otimes n}$. The function-evaluation circuit $\Sop{U}(\V{\theta})$ encodes the parameter vector $\V{\theta}$, and transforms the input state to the state $\ket{\psi(\V{\theta})}$. The expectation value $J(\V{\theta})$ is then computed based on the result of multiple measurements. This is achieved by decomposing the observable $\Sop{H}$ (involving at most $K$-qubit interactions) as
\begin{equation}\label{decompose_observable}
\Sop{H}=\sum_{k=1}^K \bigg\{\sum_{i_1,j_1,\dotsc,i_k,j_k} h_{i_1,i_2,\dotsc,i_k}^{(j_1,j_2,\dotsc,j_k)} \prod_{l=1}^k \Sop{\sigma}_{i_l}^{(j_l)}\bigg\},
\end{equation}
where $\Sop{\sigma}_{i_l}^{(j_l)}$ denotes a  Pauli-$j_l$ operator (i.e., $j_l$ may be $\Sop{X}$, $\Sop{Y}$ or $\Sop{Z}$) acting on the $i_l$-th qubit. In light of this, the term $\prod_{l=1}^k \Sop{\sigma}_{i_l}^{(j_l)}$ can be implemented using a simple quantum circuit consisting of $k$ single-qubit gates followed by measurements, as shown in the dashed box of Fig. \ref{fig:qee}. For example, a Pauli-$\Sop{Z}$ operator in \eqref{decompose_observable} corresponds to a direct measurement, whereas a Pauli-$\Sop{X}$ operator corresponds to a Hadamard gate followed by measurement. Thus, the expectation value can be obtained by measuring the outputs of these simple circuits, and evaluating a weighted sum over them using the weights $h_{i_1,i_2,\dotsc,i_k}$.

In contrast to ``fully quantum'' algorithms (e.g. Shor's algorithm and Grover's search algorithm \cite{grover_cst}) aiming to arrive at one of the computational basis states at the very end of computation, hybrid quantum-classical algorithms aim to compute the expectation values. Therefore, the measurement results have to be averaged over a number of independent circuit executions. In this treatise, we will refer to this process as ``circuit sampling''.

To portray the potential advantage of hybrid quantum-classical computation, we provide a sketchy complexity comparison between classical computation and hybrid quantum-classical computation. Using classical computation, evaluating $J(\V{\theta})$ for $\V{\psi}\in\mathbb{C}^{2^n}$ requires on the order of $O(2^{2n})$ operations. By contrast, the complexity of the hybrid scheme depends both on the complexity of the function-evaluation circuit as well as on the structure of $\Sop{H}$. More precisely, denoting the complexity of the function-evaluation circuit in terms of quantum gates as $T$, the total complexity of evaluating $J(\V{\theta})$ would be $O(T\prod_{k=1}^K N_k)$, where $N_k$ denotes the number of weight-$k$ Pauli strings in $\Sop{H}$. Therefore, for application scenarios where the observable $\Sop{H}$ is ``sparse'' in the sense that the number of terms $\prod_{k=1}^K N_k$ is small (e.g. polynomial in $n$), a substantial speedup over classical computation may be achieved, when using the hybrid approach.

\section{A Brief Introduction to Quantum Error Mitigation}\label{sec:qem}
When contaminated by decoherence, the quantum circuits of hybrid quantum-classical computation would produce erroneous expectation values. Fortunately, the weighted-averaging nature of hybrid quantum-classical computation facilitates the conception of a qubit-overhead-free method that mitigates the deviation from the true expectation value, namely the \ac{qem} \cite{qem}. In this section, we introduce the formulation of \ac{qem} and the computational overhead it incurs -- namely the sampling overhead.

\subsection{The Basic Formulation of QEM}
\begin{figure}[t]
    \centering
    \includegraphics[width=.45\textwidth]{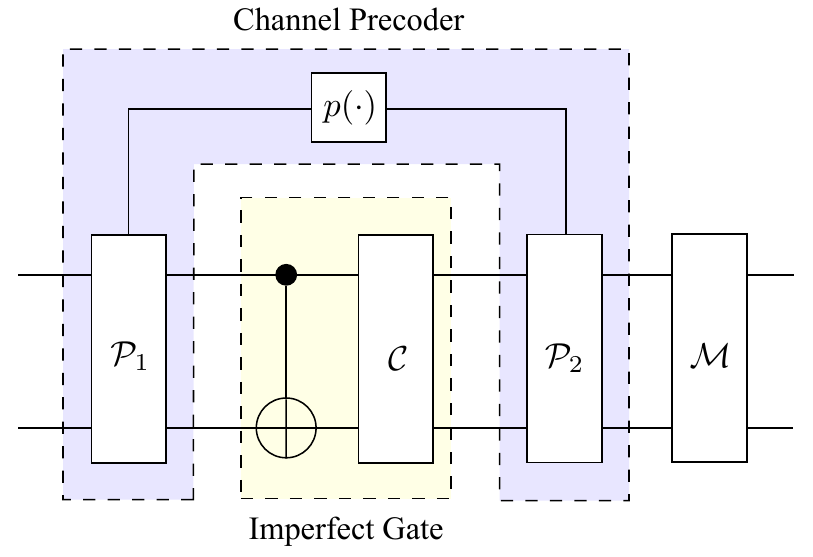}
    \caption{Schematic of a \ac{qem}-protected imperfect CNOT gate equipped with a quantum channel precoder.}
    \label{fig:precoder}
\end{figure}

The philosophy of \ac{qem} is to insert a probabilistic quantum circuit right after every quantum gate, which reverts the effect of the quantum channel modelling the imperfection inflicted by the gate. Conceptually, a \ac{qem}-protected gate can be portrayed as in Fig. \ref{fig:precoder}. The imperfect gate (in this case an imperfect CNOT gate) can be decomposed into a perfect gate and a quantum channel $\Sop{C}$. Given an input state having the density matrix $\rho_{\mathrm{in}}$, according to \eqref{kraus_channel}, the output state after the imperfect gate is given by
\begin{equation}
\rho_{\mathrm{out}} = \sum_i \M{K}_i\M{U}_{\mathrm{g}}\rho_{\mathrm{in}}\M{U}_{\mathrm{g}}^\dagger\M{K}_i^\dagger,
\end{equation}
where the matrix $\M{U}_{\mathrm{g}}$ corresponds to the effect of the perfect gate. In a vectorized form, the output state can be expressed as
\begin{equation}
\V{x}_{\mathrm{out}} = \M{C}\M{G}\V{x}_{\mathrm{in}},
\end{equation}
where we have $\M{G} = \M{U}_{\mathrm{g}}^*\otimes \M{U}_{\mathrm{g}}$, and the Pauli transfer matrix $\M{C}$ is given in \eqref{channel_vecform}. If we have an estimate $\hat{\M{C}}$ of $\M{C}$, potentially obtained using methods such as quantum process tomography \cite{qpt}, an estimate of the output of the decoherence-free gate $\M{G}$ can be obtained as
\begin{equation}
\begin{aligned}
\hat{\V{x}} = \M{M}\V{x}_{\mathrm{out}}= \hat{\M{C}}^{-1}\M{C}\M{G}\V{x}_{\mathrm{in}},
\end{aligned}
\end{equation}
where $\M{M}=\hat{\M{C}}^{-1}$ is the Pauli transfer matrix representation of the probabilistic quantum circuit $\Sop{M}$ constructed for inverting the channel.

To elaborate further, if a gate is followed directly by measurement, $\Sop{M}$ is implemented by applying different circuits according to a probability distribution in different circuit executions and performing a weighted averaging over the measurement outcomes. This can be formulated as
\begin{equation}
\sum_{k=1}^K \V{m}_k^{\mathrm{T}}\M{M}\V{x}_{\mathrm{out}} = \sum_{k=1}^K w_k p_k \V{m}_k^{\mathrm{T}}\M{M}_k\V{x}_{\mathrm{out}},
\end{equation}
where $\M{M}_k$ is the $k$-th candidate circuit applied at a probability of $p_k$, $w_k$ is the weight of the measurement outcome of the $k$-th candidate circuit, and $\V{m}_k$ is the Pauli transfer matrix representation of the measurement operator corresponding to the outcome. For a circuit constructed by multiple gates, the weights and probability distributions follow directly by linearity. For example, for a simple circuit containing two consecutive imperfect gates $\tilde{\M{G}}^{(1)}$ and $\tilde{\M{G}}^{(2)}$, we have
\begin{equation}
\begin{aligned}
&\hspace{-3mm}\sum_{j=1}^J \!\sum_{k=1}^K\V{m}_{jk}^{\mathrm{T}} \M{M}^{(2)}\tilde{\M{G}}^{(2)}\M{M}^{(1)}\tilde{\M{G}}^{(1)}\V{x}_{\mathrm{in}} \\
&\!=\! \sum_{j=1}^J \!\sum_{k=1}^K \!w_j^{(1)} \!w_k^{(2)} \!p_j^{(1)} \!p_k^{(2)} \!\V{m}_{jk}^{\mathrm{T}} \!\M{M}_k^{(2)}\!\tilde{\M{G}}^{(2)}\!\M{M}_j^{(1)}\!\tilde{\M{G}}^{(1)}\!\V{x}_{\mathrm{in}},
\end{aligned}
\end{equation}
where the superscripts ``$(1)$'' and ``$(2)$'' indicate the first and the second gate, respectively.

Since the expectation evaluation in hybrid quantum-classical computation is implemented by applying a linear transformation (weighted averaging) to the measurement outcomes, it fits nicely with \ac{qem}. By contrast, ``fully quantum'' algorithms such as Shor's algorithm and Grover's algorithm operate in different ways, hence they might not be protectable by \ac{qem}.

Optionally, one may apply a quantum channel precoder to the imperfect gate, yielding
\begin{equation}
\begin{aligned}
\hat{\V{x}}_{\mathrm{P}} &= \M{M}_{\mathrm{P}}\M{P}_2\M{C}\M{G}\M{P}_1\V{x}_{\mathrm{in}} \\
&= (\M{P}_2\hat{\M{C}}\M{G}\M{P}_1\M{G}^{-1})^{-1}\M{P}_2\M{C}\M{G}\M{P}_1\V{x}_{\mathrm{in}},
\end{aligned}
\end{equation}
as shown in Fig. \ref{fig:precoder}. The quantum channel precoder turns the channel $\M{C}$ into another (possibly more preferable) channel $\M{P}_2\M{C}\M{G}\M{P}_1\M{G}^{-1}$. For example, the so-called Pauli twirling of \cite{ptwirl,practical_qem,qem_exp} may be viewed as a quantum channel precoder turning an arbitrary channel into a Pauli channel. Similarly, Clifford twirling \cite{ctwirl1} turns an arbitrary channel into a depolarizing channel. According to the operator-sum representation \cite[Sec. 8.2.4]{ncbook}, a quantum channel precoder can be implemented by a probabilistic mixture of gates applied both before and after the imperfect gate to be protected\footnote{For the moment, we assume that the gates used to implement the quantum channel precoder are decoherence-free. The practical case of erroneous gates will be considered in Section \ref{ssec:qcp_practical}.}.

In order to obtain $\M{M}$, we may first choose a basis matrix $\M{B}$, in which each column is the vectorized Pauli transfer matrix of a candidate circuit \cite{practical_qem}. For example, if the first column of $\M{B}$ represents an operator $\Sop{O}$, we have $[\M{B}]_{:,1}={\mathrm{vec}}(\M{O})$, where $\M{O}$ is the Pauli transfer matrix of $\Sop{O}$. Next, we determine the coefficients as follows:
\begin{equation}\label{sampling_coeff}
\begin{aligned}
\V{\mu}_{\Sop{C}} = \M{B}^{-1}\vectorize{\M{C}^{-1}}.
\end{aligned}
\end{equation}
Given the coefficients $\V{\mu}_{\Sop{C}}$, we can now express $\M{M}$ as
\begin{equation}
\M{M} = \sum_{i} [\V{\mu}_{\Sop{C}}]_i {\mathrm{vec}}^{-1}\{[\M{B}]_{:,i}\}.
\end{equation}
This can be realized by applying the candidate circuit corresponding to $[\M{B}]_{:,i}$ with probability $|[\V{\mu}_\Sop{C}]_i|\cdot \|\V{\mu}_{\Sop{C}}\|_1^{-1}$ and assigning a weight ${\mathrm{sgn}}([\V{\mu}_\Sop{C}]_i)\cdot \|\V{\mu}_{\Sop{C}}\|_1$ to the measurement outcome. In light of this, $\V{\mu}_{\Sop{C}}$ is referred to as the quasi-probability representation \cite{qem} of channel $\Sop{C}$.

\subsection{The Sampling Overhead of QEM}
\begin{figure}[t]
    \centering
    \includegraphics[width=.49\textwidth]{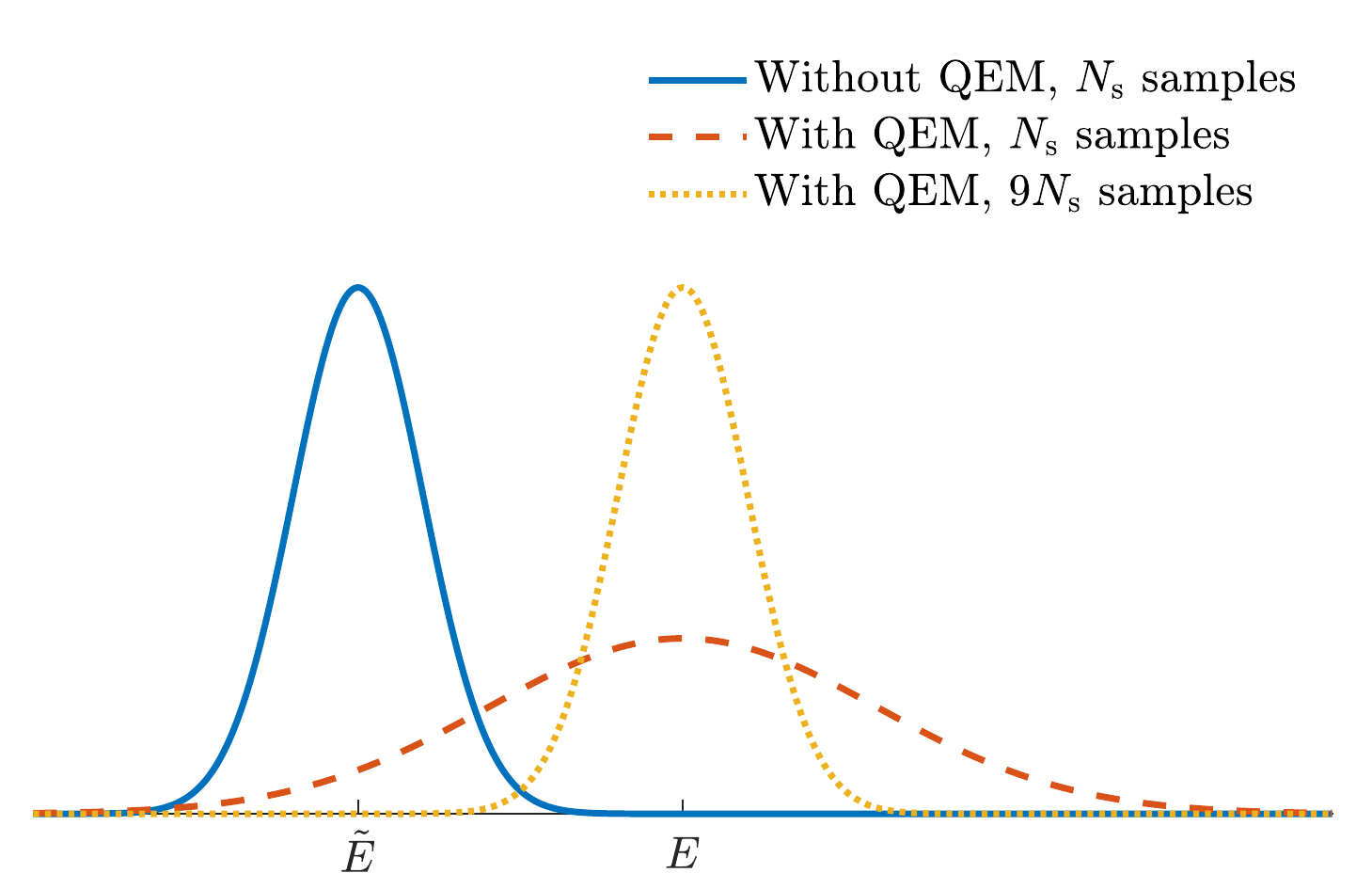}
    \caption{An illustration of the sampling overhead of \ac{qem}. The curves represent the distribution of the computational results with/without \ac{qem}.}
    \label{fig:sampling_overhead}
\end{figure}

In general, the probabilistic implementation of $\M{M}$ will incur a sampling overhead. To elaborate, if we wish to compute the expectation value $J(\V{\theta})$ to a certain accuracy, we have to operate the circuit a certain number of times (sampling from the output state vector). When the gates are perfect, the number of samples required is determined by the variance of the observable $\Sop{H}$ given by
\begin{equation}
{\mathrm{Var}}\{\Sop{H}\} = \bra{\psi(\V{\theta})}\Sop{H}^2\ket{\psi(\V{\theta})} - (\bra{\psi(\V{\theta})}\Sop{H}\ket{\psi(\V{\theta})})^2.
\end{equation}
In practice, the observable $\Sop{H}$ cannot be measured directly, but has to be estimated using the measurement outcome of several operators according to the decomposition \eqref{decompose_observable}. Therefore, if we have $\Sop{H}=\sum_i \Sop{H}_i$ where each $\Sop{H}_i$ can be measured directly, we have
$$
{\mathrm{Var}}\{\Sop{H}\}=\sum_i {\mathrm{Var}}\{\Sop{H}_i\}.
$$
and
$$
{\mathrm{Var}}\{\Sop{H}_i\} = \bra{\psi(\V{\theta})}\Sop{H}_i^2\ket{\psi(\V{\theta})} - (\bra{\psi(\V{\theta})}\Sop{H}_i\ket{\psi(\V{\theta})})^2.
$$
 Upon assuming that the required accuracy is quantified in terms of its variance $\sigma^2$, this may be achieved using $N_{\mathrm{s}}$ samples in the perfect gate scenario. After \ac{qem}, the expected value remains unchanged. However, if the number of samples is kept fixed, \ac{qem} will lead to a variance increase, since the entries in $\V{\mu}_{\Sop{C}}$ are not necessarily positive. Explicitly, the variance after \ac{qem} is given by
\begin{equation}
\sigma_{\mathrm{QEM}}^2 = \|\V{\mu}_{\Sop{C}}\|_1^2\sigma^2,
\end{equation}
where $\|\V{\mu}_{\Sop{C}}\|_1\ge 1$. Therefore, in order to achieve the same accuracy, we have to sample every quantum gate $N_{\mathrm{s}}(\|\V{\mu}_{\Sop{C}}\|_1^2-1)$ times additionally. To elaborate a little futher, we consider the 'toy' example portrayed in Fig. \ref{fig:sampling_overhead}. In this example, we assume that the error-free expectation value is $E$, and we assume that $\|\V{\mu}_{\Sop{C}}\|_1=3$. Observe from the figure that when the circuit is sampled $N_{\mathrm{s}}$ times, the computational result without \ac{qem} is randomly distributed around its mean value $\tilde{E}$, which deviates from the true value $E$. Having been corrected by \ac{qem}, the mean value of the computational result equals to $E$. However, the variance of the result is increased by a factor of $\|\V{\mu}_{\Sop{C}}\|_1^2=9$. To ensure that the accuracy meets our requirement, we have to sample the circuit $N_{\mathrm{s}}\|\V{\mu}_{\Sop{C}}\|_1^2=9N_{\mathrm{s}}$ times, as illustrated by the dotted curve of Fig. \ref{fig:sampling_overhead}. Empirical evidence has shown that applying quantum channel precoders is capable of reducing the sampling overhead for certain types of channels~\cite{practical_qem}.

In the previous example, we have considered the case of a single channel $\Sop{C}$. In general, a quantum circuit consists of $N_{\mathrm{g}}>1$ gates, which may be viewed as the cascade of $N_{\mathrm{g}}$ channels, denoted by $\Sop{C}_1,\dotsc,\Sop{C}_{N_{\mathrm{g}}}$. To achieve the required computational accuracy, we have to sample the circuit
\begin{equation}\label{sampling_overhead}
\widetilde{N_{\mathrm{s}}}=N_{\mathrm{s}}\prod_{i=1}^{N_{\mathrm{g}}} \|\V{\mu}_{\Sop{C}_i}\|_1^2
\end{equation}
times. To facilitate our analysis, we define the \textit{sampling overhead} of a circuit as the additional number of samples imposed by \ac{qem}, which equals to $\widetilde{N_{\mathrm{s}}}-N_{\mathrm{s}}$ in our previous example. We note that the sampling overhead of a circuit increases exponentially with the number of gates.

According to the previous discussions, we may use the following notion to characterize the sampling overhead incurred by a single channel $\Sop{C}$ when compensated by \ac{qem}.
\begin{definition}[Sampling Overhead Factor]
We define the sampling overhead factor (SOF) of a quantum channel $\Sop{C}$ as
\begin{equation}\label{sampling_overhead_def}
\gamma_{\Sop{C}} \triangleq \|\V{\mu}_{\Sop{C}}\|_1^2 - 1.
\end{equation}
\end{definition}
\begin{remark}
When there is only a single gate modelled by the associated channel $\Sop{C}$ in the circuit, we can see from \eqref{sampling_overhead} that the sampling overhead of the circuit may be represented in terms of the \ac{sof} $\gamma_{\Sop{C}}$ as $N_{\mathrm{s}}\gamma_{\Sop{C}}$. When there are several gates in the circuit, the sampling overhead of the circuit can be computed using the \acp{sof} of the channels as follows:
\begin{equation}\label{sof_exponential}
\widetilde{N_{\mathrm{s}}}-N_{\mathrm{s}}=N_{\mathrm{s}}\left[\prod_{i=1}^{N_{\mathrm{g}}} (1+\gamma_{\Sop{C}_i})-1\right].
\end{equation}
\end{remark}

\begin{figure}[t]
\begin{minipage}{\columnwidth}
\centering
\includegraphics[width=.85\textwidth]{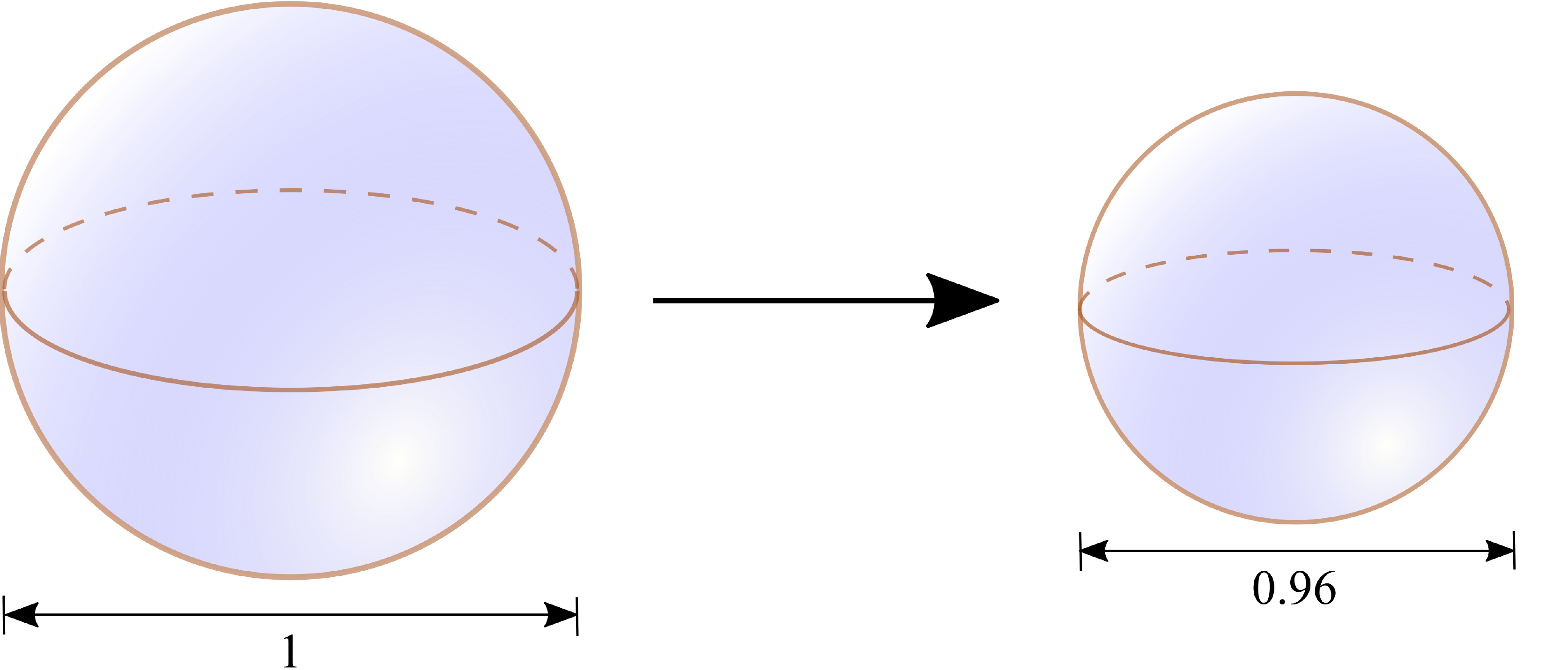}\hfill

{\footnotesize (a) The depolarizing channel shrinks the Bloch sphere.}
\vspace{3mm}
\end{minipage}
\begin{minipage}{\columnwidth}
\centering
\includegraphics[width=.85\textwidth]{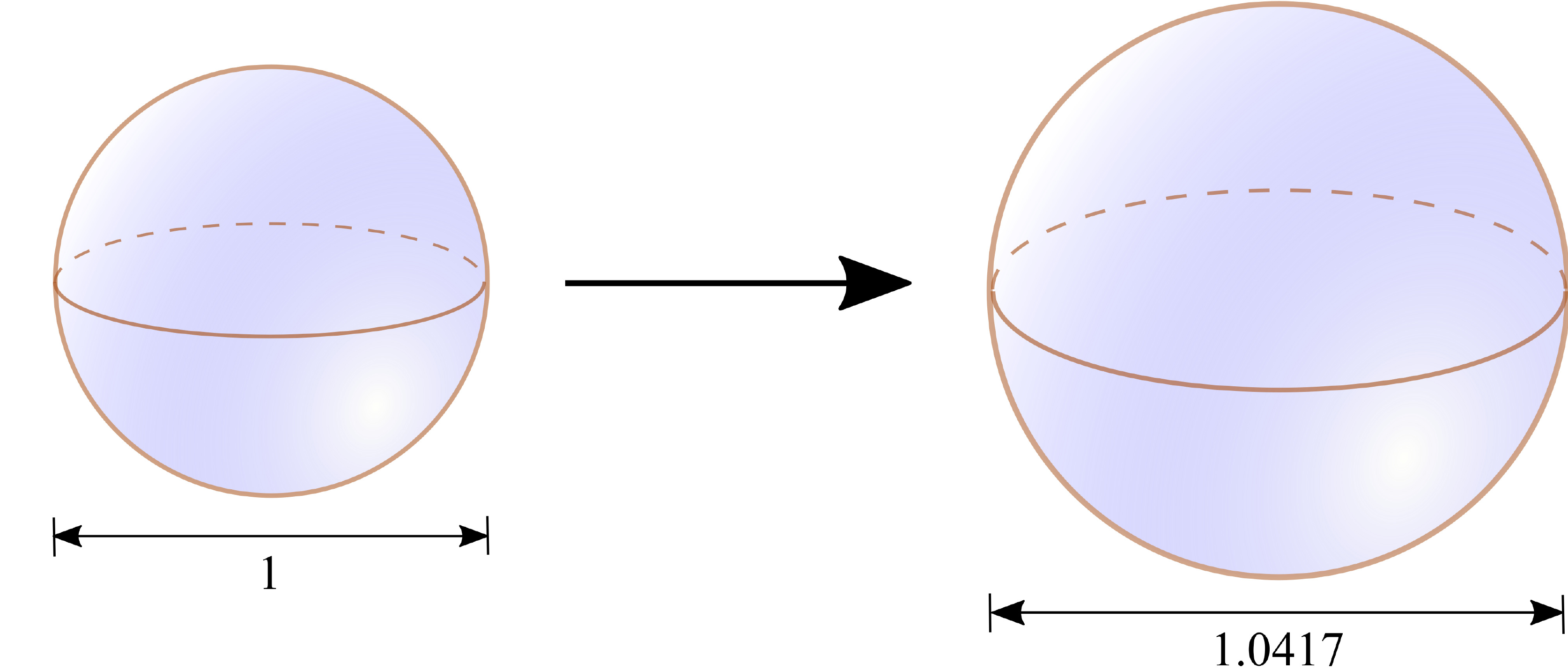}\hfill

{\footnotesize (b) The inverse of the depolarizing channel dilates the Bloch sphere.}
\vspace{3mm}
\end{minipage}
\caption{Geometric illustration of a single-qubit depolarizing channel as well as the corresponding inverse channel.}
\label{fig:dep_inv_dep}
\end{figure}

To provide further intuitions about the \ac{sof}, let us consider the 'toy example' of a single-qubit depolarizing channel $\Sop{C}$, having the following operator-sum representation
$$
\Sop{C}(\rho) = 0.97\rho + 0.01\M{S}_{\Sop{X}}\rho\M{S}_{\Sop{X}}+ 0.01\M{S}_{\Sop{Y}}\rho\M{S}_{\Sop{Y}}+ 0.01\M{S}_{\Sop{Z}}\rho\M{S}_{\Sop{Z}},
$$
which can be observed to have \ac{ggep} $\epsilon=0.03$. The corresponding Pauli transfer matrix takes the following form
$$
\M{C} = \diag{1,0.96,0.96,0.96},
$$
and the inverse Pauli transfer matrix is given by
$$
\M{C}^{-1} = \diag{1,1.0417,1.0417,1.0417}.
$$
Geometrically, the channel $\Sop{C}$ corresponds to a homogeneous shrinking of the Bloch sphere, making its radius $0.96$ times the original radius, while the inverse channel $\Sop{C}^{-1}$ corresponds to a homogeneous dilation extending the radius to $1/0.96=1.0417$ times the original radius, as portrayed in Fig.~\ref{fig:dep_inv_dep}.

To perform \ac{qem} on the channel $\Sop{C}$, we should first choose a basis. Note that the Pauli operators have the following Pauli transfer matrix representations
\begin{equation}\label{pauli_ptm}
\begin{aligned}
\M{P}_{\Sop{I}} &= \diag{1,1,1,1}, \\
\M{P}_{\Sop{X}} &= \diag{1,1,-1,-1}, \\
\M{P}_{\Sop{Y}} &= \diag{1,-1,1,-1}, \\
\M{P}_{\Sop{Z}} &= \diag{1,-1,-1,1}, \\
\end{aligned}
\end{equation}
which constitutes a complete basis of diagonal Pauli transfer matrices. In light of this, we may choose the Pauli operators as the basis. The corresponding quasi-probability representation can then be computed as
$$
\V{\mu}_{\Sop{C}} = [1.03125~-0.01042~-0.01042~-0.01042]^{\mathrm{T}},
$$
yielding $\|\V{\mu}_{\Sop{C}}\|_1=1.0625$. Therefore, given the basis we chose, the \ac{sof} of the channel $\Sop{C}$ is given by
$$
\gamma_{\Sop{C}} = \|\V{\mu}_{\Sop{C}}\|_1^2-1 = 0.1289.
$$

\section{Sampling Overhead Factor Analysis for Uncoded Quantum Gates}\label{sec:uncoded}
In this section, we investigate the \ac{sof} of quantum gates that are not protected by quantum codes. This will also lay the foundation for the analysis of coded quantum gates in Sec. \ref{sec:coded}.

The \ac{sof}, in essence, may also be viewed as a specific characteristic of the representation of a quantum channel under a specific basis, as implied by \eqref{sampling_coeff} and \eqref{sampling_overhead_def}. Hence, the particular choice of basis will certainly have an impact on it. Considering realistic restrictions and aiming for simplifying our analysis, we make the following assumptions concerning the choices of basis.
\begin{enumerate}
  \item The basis vectors should correspond to legitimate quantum operations in order to be implementable. Formally, we assume that the basis vectors are the vectorized Pauli transfer matrices of \ac{cptni} operators, for which the operation elements of \eqref{kraus_channel} satisfy
      \begin{equation}\label{kraus_cptni}
      \sum_{i}\M{K}_i^\dagger \M{K}_i \preceq \M{I}.
      \end{equation}
      This includes perfect gates (unitary operators), imperfect gates (\ac{cptp} operators), and measurements (trace-decreasing operators).
  \item We assume that the basis always includes all vectorized Pauli operators. This is not very restrictive for most existing quantum computers, since Pauli gates are one of their most fundamental building blocks.
\end{enumerate}

The other (potentially more important) factor influencing the \ac{sof} is the quantum channel itself. Various channel models have been proposed in the literature, such as depolarizing channels, phase damping channels, amplitude damping channels, etc. \cite{transversal,practical_qem,channel_model_win}. To maintain the generality of our treatment, we do not explicitly consider a specific channel model, but rather a general \ac{cptp} channel.

\subsection{Coherent--triangular decomposition of Memoryless CPTP channels}\label{ssec:cptp}
Without loss of generality, we assume that

\begin{assumption}\label{asu:identity}
The first row and the first column in a Pauli transfer matrix corresponds to the identity operator $\Sop{I}^{\otimes n}$ in the $n$-qubit Pauli group.
\end{assumption}

\begin{figure}[t]
    \centering
    \includegraphics[width=.45\textwidth]{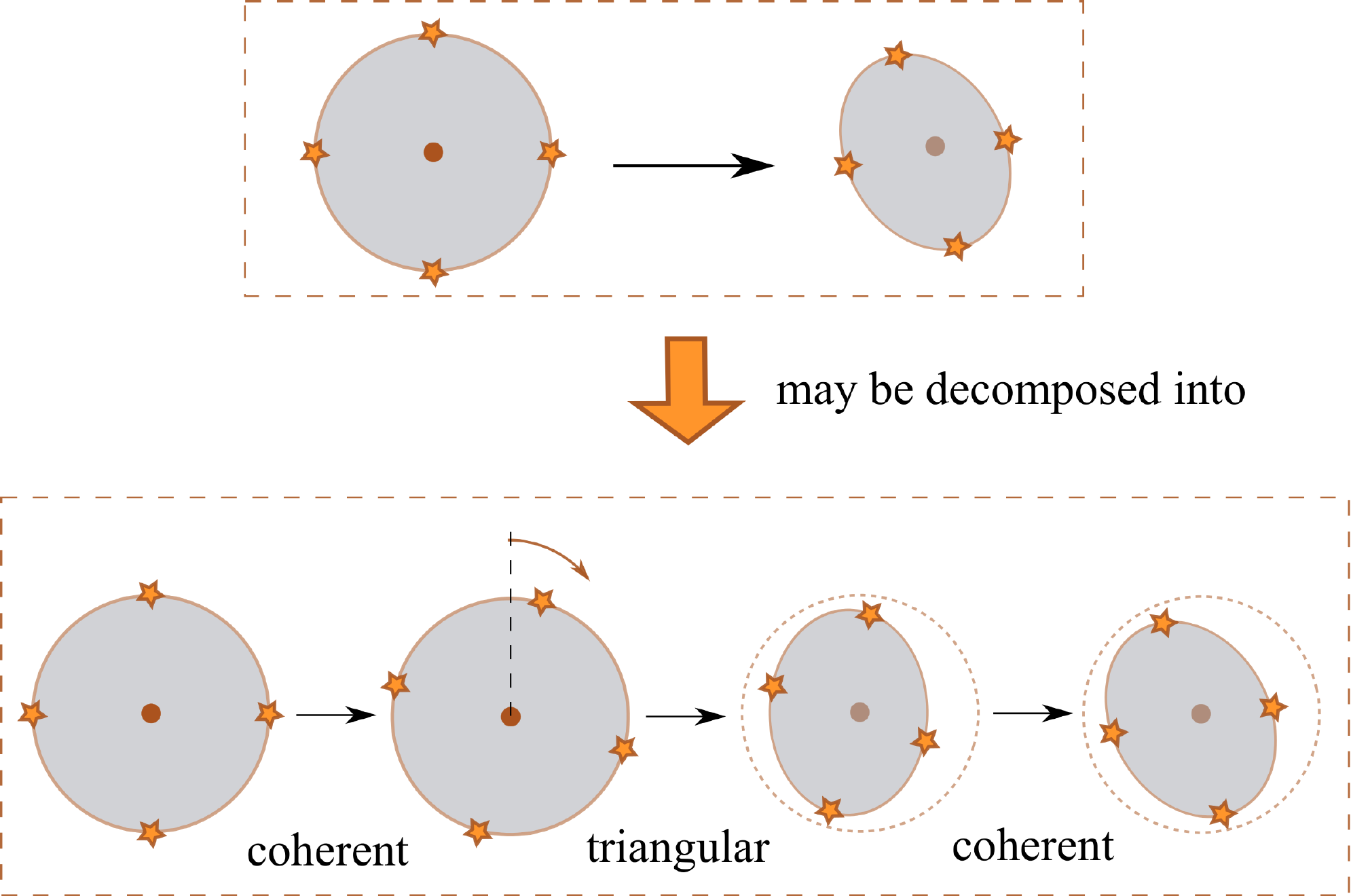}
    \caption{Geometric interpretation of the coherent--triangular decomposition of a signle-qubit \ac{cptp} channel on the Bloch sphere, detailed in Lemma \ref{lem:decomposition}. For better illustration, we only plot a cross-section of the Bloch sphere.}
    \label{fig:decomposition}
\end{figure}

Note that for a valid density matrix $\rho$, the condition ${\mathrm{Tr}}\{\rho\}=1$ is always satisfied. According to Assumption \ref{asu:identity}, this implies that for the corresponding vector representation $\V{x}$, we have
\begin{equation}\label{cptp_vec}
\V{x} = \frac{1}{\sqrt{2^n}}[1~\tilde{\V{x}}^{\mathrm{T}}]^{\mathrm{T}},
\end{equation}
where $\tilde{\V{x}}\in\mathbb{R}^{4^n-1}$, according to \eqref{ptm_vec}. The dimensionality of $\V{x}$ is $4^n$, because the number of Pauli operators over $n$ qubits is $4^n$. Thus for any trace-preserving channel, we have
\begin{equation}\label{cptp_mat}
\M{C} = \left[
          \begin{array}{cc}
            1 & \V{0}^{\mathrm{T}} \\
            \V{b} & \tilde{\M{C}} \\
          \end{array}
        \right],
\end{equation}
which amounts to the following result
\begin{equation}\label{cptp_result}
\M{C}\V{x} = \frac{1}{\sqrt{2^n}}[1~\tilde{\M{C}}\tilde{\V{x}}+\V{b}]^{\mathrm{T}}.
\end{equation}
It can now be seen from \eqref{cptp_vec}, \eqref{cptp_mat} and \eqref{cptp_result} that a \ac{cptp} channel can be viewed as an \emph{affine transformation} in the $(4^n-1)$-dimensional space spanned by the Pauli operators excluding the identity. In this regard, we have the following result for single-qubit channels, whose geometric interpretation is demonstrated in Fig. \ref{fig:decomposition}.

\begin{lemma}\label{lem:decomposition}
Any single-qubit \ac{cptp} channel can be expressed as the composition of (up to) two coherent channels\footnote{The term ``coherent channels'' refers to the channels having unitary matrix representations and themselves can be implemented using (error-free) unitary gates.} and a triangular channel, meaning that
\begin{equation}\label{cptp_decomposition}
\begin{aligned}
\M{C} &= \M{U}\M{D}\M{V}^{\mathrm{T}} \\
&=\left[
          \begin{array}{cc}
            1 & \V{0}^{\mathrm{T}} \\
            \V{0} & \tilde{\M{U}} \\
          \end{array}
        \right] \left[
          \begin{array}{cc}
            1 & \V{0}^{\mathrm{T}} \\
            \tilde{\M{U}}^{\mathrm{T}}\V{b} & \tilde{\M{D}} \\
          \end{array}
        \right]\left[
          \begin{array}{cc}
            1 & \V{0}^{\mathrm{T}} \\
            \V{0} & \tilde{\M{V}} \\
          \end{array}
        \right]^{\mathrm{T}},
\end{aligned}
\end{equation}
where $\tilde{\M{U}}$ and $\tilde{\M{V}}$ are a pair of unitary matrices, and $\tilde{\M{D}}$ is a diagonal matrix. The matrix $\M{D}$ corresponds to the triangular channel, while the matrices $\M{U}$ and $\M{V}$ represent the coherent channels.
\begin{IEEEproof}
Consider the singular value decomposition of $\tilde{\M{C}}$ given by
\begin{equation}
\tilde{\M{C}} = \tilde{\M{U}}\tilde{\M{D}}\tilde{\M{V}}^{\mathrm{T}}.
\end{equation}
From \eqref{cptp_mat} we obtain directly that $\M{D}$ is a triangular channel, and hence it now suffices to show that both $\M{U}$ and $\M{V}$ can be implemented by unitary gates. Since the entries of Pauli transfer matrices are all real numbers \cite{intro_gst}, the matrices $\tilde{\M{U}}$ and $\tilde{\M{V}}$ are both $3\times 3$ orthogonal matrices, corresponding to the three-dimensional rotations around the Bloch sphere belonging to the special orthogonal group $\mathrm{SO}(3)$. They can be implemented by single-qubit unitary gates belonging to $\mathrm{SU}(2)$ due to the $\mathrm{SO}(3)$-$\mathrm{SU}(2)$ homomorphism.

\end{IEEEproof}
\end{lemma}

Compared to the triangular component, the coherent component of a \ac{cptp} channel might be easier to deal with, since their effect may be compensated by using unitary gates. This implies that if the unitary gates designed for the compensation are error-free, the effect of coherent channels may be reversed without any sampling overhead. By contrast, the triangular component may have to be compensated by using probabilistic gates, hence imposes overhead.

It is known that Lemma \ref{lem:decomposition} does not hold for general multi-qubit channels \cite{polar_decomposition}. Nevertheless, it is applicable to the case where the channel $\M{C}$ is memoryless, hence it can be described by the tensor product of single-qubit channels. To see this, we may rewrite a memoryless channel as $\M{C} = \M{C}_1\otimes \M{C}_2 \otimes \dotsc \otimes \M{C}_n$, and for each $\M{C}_i$ we have $\M{C}_i = \M{U}_i\M{D}_i\M{V}_i^{\rm T}$. This further implies that
$$
\M{C} = \left(\bigotimes_{i=1}^n \M{U}_i\right)\left(\bigotimes_{i=1}^n \M{D}_i\right)\left(\bigotimes_{i=1}^n \M{V}_i\right)^{\rm T}.
$$
Observe that both $\bigotimes_{i=1}^n \M{U}_i$ and $\bigotimes_{i=1}^n \M{V}_i$ correspond to practically implementable single-qubit gates. Since the Kronecker product preserves the triangular structure, we see that $\bigotimes_{i=1}^n \M{D}_i$ also represents a triangular channel.

\subsection{Analysis on triangular channels}\label{ssec:pdc}
According to the discussions in Section \ref{ssec:cptp}, we are particularly interested in the quasi-probability representation of triangular channels, whose Pauli transfer matrices take the same form as the matrix $\M{D}$ in \eqref{cptp_decomposition}. More precisely, we define triangular channels as \ac{cptp} quantum channels whose Pauli transfer matrix can be written as follows
\begin{equation}
\M{D} = \left[
          \begin{array}{cc}
            1 & \V{0}^{\mathrm{T}} \\
            \V{b} & \M{L} \\
          \end{array}
        \right],
\end{equation}
where $\M{L}$ is a lower triangular matrix. This includes both amplitude damping channels and Pauli channels as representative examples. For example, a single-qubit amplitude damping channel having decay probability $p$ has the following Pauli transfer matrix
$$
\left[
  \begin{array}{cccc}
    1 & 0 & 0 & 0 \\
    0 & \sqrt{1-p} & 0 & 0 \\
    0 & 0 & \sqrt{1-p} & 0 \\
    p & 0 & 0 & 1-p \\
  \end{array}
\right],
$$
where the rows/columns are ordered for ensuring that they correspond to the Pauli-I, X, Y, and Z operators, respectively. Observe that the matrix does have a triangular structure, which is preserved under the permutation of the Pauli-X, Y, and Z operators. As a direct corollary, for a multi-qubit channel inflicting amplitude damping independently on each qubit, the Pauli transfer matrix is triangular, since the triangular structure is preserved under the Kronecker product.

For a fair comparison, we consider channels having the same \ac{ggep} $\epsilon$, meaning that $\M{D}\in\Set{D}_n(\epsilon)$, where we have:
\begin{equation}
\Set{D}_n(\epsilon) = \left\{\M{D}\in\Set{C}_n\right|\left. 4^{-n}{\mathrm{Tr}}\{\M{D}\}=1-\epsilon\right\},
\end{equation}
and $\Set{D}_n$ denotes the set of all \ac{cptp} triangular channels over $n$ qubits.

In the following proposition, we will show that regardless of the specific choice of the basis $\M{B}$, Pauli channels have the lowest \ac{sof} among all triangular \ac{cptp} channels. The Pauli channels are defined as channels that transform one Pauli matrix into another. Based on \eqref{ptm_rep}, this implies that the corresponding Pauli transfer matrices are diagonal matrices.
\begin{proposition}[Pauli channels have the lowest \ac{sof}]\label{prop:pauli}
Given a fixed \ac{ggep} $\epsilon$, for any full-rank basis matrix $\M{B}$ consisting of vectorized Pauli transfer matrix representation of \ac{cptni} operators, among all \ac{cptp} triangular channels over $n$ qubits, Pauli channels have the lowest \ac{sof}.
\begin{IEEEproof}
Please refer to Appendix \ref{sec:proof_pauli}.
\end{IEEEproof}
\end{proposition}

\begin{remark}
We note that there is some empirical evidence supporting that projecting a quantum channel onto the set of Pauli channels might help in reducing the sampling overhead \cite{practical_qem}. Here we formally show that this is indeed true, and it is true for the entire family of triangular channels.
\end{remark}

Next, we consider some notational simplifications for a further investigation in the context of Pauli channels. First of all, since the Pauli transfer matrices of Pauli channels are diagonal (as exemplified by \eqref{pauli_ptm}), we may rewrite the vector representation of a Pauli transfer matrix (or that of its inverse) in a reduced-dimensional manner. Specifically, we could represent the Pauli transfer matrix of a Pauli channel using merely the vector on its main diagonal, i.e. $\V{d} = {\mathrm{vec}}\{{\mathrm{mdiag}}\{\M{D}\}\}$. Hence we have $\M{D}\V{x} = \V{d}\circ\V{x}$.

Additionally, the $16^n\times 16^n$ basis matrix $\M{B}$ can be reduced to $\tilde{\M{B}}\in\mathbb{R}^{4^n\times 4^n}$ for Pauli channels. In light of this, we have
\begin{equation}
\V{\mu}_{\Sop{D}} = {\mathrm{vec}}\{{\mathrm{mdiag}}\{\tilde{\M{B}}^{-1}(1/\V{d})\}\}.
\end{equation}
For simplicity, we introduce the further notation of
\begin{equation}\label{reduce_qpr}
\tilde{\V{\mu}}_{\Sop{D}} = \tilde{\M{B}}^{-1}(1/\V{d}).
\end{equation}
Note that each column in $\tilde{\M{B}}$ represents the vectorized Pauli transfer matrix of a specific Pauli operator (as a quantum channel). According to the definitions of the Pauli operators and \eqref{ptm_rep}, under the computational basis, the vectorized Pauli transfer matrix representation of single-qubit Pauli operators can be expressed as
\begin{equation}
\begin{aligned}
\V{s}_{\Sop{I}} &= [1~1~1~1]^{\mathrm{T}},~\V{s}_{\Sop{X}} = [1~1~-1~-1]^{\mathrm{T}}, \\
\V{s}_{\Sop{Y}} &= [1~-1~1~-1]^{\mathrm{T}},~\V{s}_{\Sop{Z}} = [1~-1~-1~1]^{\mathrm{T}},
\end{aligned}
\end{equation}
respectively. In this case, it can be seen that the corresponding simplified basis matrix of
$$
\tilde{\M{B}}_1 = [\V{s}_{\Sop{I}}~\V{s}_{\Sop{X}}~\V{s}_{\Sop{Y}}~\V{s}_{\Sop{Z}}]
$$
has the form of the Hadamard transform matrix. In general, Pauli operators over $n$ qubits can be expressed as the tensor product of $n$ single-qubit Pauli operators, hence the corresponding $\tilde{\M{B}}$ takes the form of
$$
\tilde{\M{B}} = (\tilde{\M{B}}_1)^{\otimes n},
$$
which is simply a Hadamard transform matrix of higher dimensionality, where $\tilde{\M{B}}_1$ denotes the matrix $\tilde{\M{B}}$ for single-qubit systems.

By exploiting the properties of the Hadamard transform, we are now able to obtain the following result.
\begin{proposition}[Depolarizing channels have the lowest \ac{sof}]\label{prop:dep_channels}
Among all Pauli channels over $n$ qubits having the \ac{ggep} $\epsilon$, depolarizing channels have the lowest \ac{sof}.
\begin{IEEEproof}
Please refer to Appendix \ref{sec:proof_dep_channels}.
\end{IEEEproof}
\end{proposition}

According to Proposition \ref{prop:dep_channels}, depolarizing channels lend themselves most readily to be compensated by \ac{qem}. This means that the \ac{qem} method has a strikingly different nature compared to the family of quantum error correction schemes in terms of the overhead imposed, since depolarizing channels may be viewed as the channels most impervious for \acp{qecc}. To elaborate, depolarizing channels exhibit the lowest hashing bound among all Pauli channels, hence they would require the highest qubit overhead\footnote{Given a fixed \ac{gep}.} for \acp{qecc} \cite{ncbook}.

\subsection{Bounding the SOF of Pauli channels}\label{ssec:bound_overhead}
In Sec. \ref{ssec:pdc} we have shown that Pauli channels are preferable for \ac{qem} in the sense that they have the lowest \ac{sof}. In this subsection, we proceed by further investigating Pauli channels and bound their \ac{sof} for a given \ac{ggep} $\epsilon$. First of all, by explicitly calculating the \ac{sof} of depolarizing channels, we can readily obtain a lower bound on the \ac{sof} of triangular channels (hence also on Pauli channels), as stated below.
\begin{corollary}[\ac{sof} lower bound]\label{coro:qecc_precoder}
For an triangular channel $\Sop{C}$ having the \ac{ggep} $\epsilon$, the \ac{sof} incurred by \ac{qem} is bounded from below as
\begin{equation}\label{overhead_lb}
\gamma_{\Sop{C}} \ge 4\epsilon\cdot \frac{1}{(1-\epsilon)^2}.
\end{equation}
The lower bound is attained, when $\Sop{C}$ is a depolarizing channel.
\begin{IEEEproof}
According to Proposition \ref{prop:dep_channels}, it is clear that the channel having the lowest \ac{sof} among all $N$-qubit triangular channels is the $N$-qubit depolarizing channel. The \ac{sof} of this channel is given by
\begin{subequations}
\begin{align}
\gamma &= \left\|\M{H}^{-1}_{N}\left(1/\left(\M{H}_{N}\V{d}_{N}\right)\right)\right\|_1^2 - 1 \\
&= \left(\frac{(4^{N}-1)(1-2\epsilon)-\epsilon}{4^{N}(1-\epsilon)-1}\right)^2-1. \label{second_line_qecc}
\end{align}
\end{subequations}
where $\M{H}_{N}$ and $\M{H}^{-1}_{N}$ are the Hadamard transform matrix and the inverse Hadamard transform matrix having dimensionality of $16^N\times 16^N$, respectively, and we have $\V{d}_{N}=[1-\epsilon~\epsilon(4^{N}-1)^{-1}\V{1}^{\mathrm{T}}]$. It is straightforward to verify that the right hand side of \eqref{second_line_qecc} is a monotonically decreasing function of $N$, thus
\begin{subequations}
\begin{align}
\gamma &\le \lim_{N\rightarrow \infty}\left(\frac{(4^{N}-1)(1-2\epsilon)-\epsilon}{4^{N}(1-\epsilon)-1}\right)^2-1 \\
&= \frac{4\epsilon}{(1-\epsilon)^2}.
\end{align}
\end{subequations}
Hence the proof is completed.
\end{IEEEproof}
\end{corollary}

Let us now derive an upper bound on the \ac{sof} of Pauli channels. For this purpose, we consider a matrix representation specifically designed for Pauli channels, which will be referred to as the Pauli random walk (PRW) representation hereafter. More precisely, a Pauli channel $\Sop{C}$ over $n$ qubits can be represented by a matrix $\M{C}_{\mathrm{PRW}}$ having the following form:
\begin{equation}\label{ptm}
[\M{C}_{\mathrm{PRW}}]_{i,j} = \frac{1}{2^n} {\mathrm{Tr}}\{\Sop{P}_i\Sop{C}\Sop{P}_j\}.
\end{equation}
Conventionally, an $n$-qubit Pauli channel $\Sop{C}$ can be characterized using a vector $\V{\eta}_{\Sop{C}}$ satisfying
\begin{equation}\label{prob_vector}
\Sop{C}(\rho) = \sum_{i=1}^{4^n} [\V{\eta}_{\Sop{C}}]_i\M{S}_i^{(n)}\rho\left(\M{S}_i^{(n)}\right)^\dagger.
\end{equation}
We will refer to $\V{\eta}_{\Sop{C}}$ as the probability vector of $\Sop{C}$ in the rest of this paper. For examples of $\V{\eta}_{\Sop{C}}$ values corresponding to some simple channel models, please refer to Appendix \ref{sec:basic_channels}. In light of \eqref{prob_vector}, the PRW representation can be expressed as a function of $\V{\eta}_{\Sop{C}}$ as follows
$$
[\M{C}_{\mathrm{PRW}}(\V{\eta}_{\Sop{C}})]_{i,j} = [\V{\eta}_{\Sop{C}}]_l,~\Sop{P}_i\Sop{P}_j=\Sop{P}_l.
$$

\begin{figure}[t]
    \centering
    \includegraphics[width=.4\textwidth]{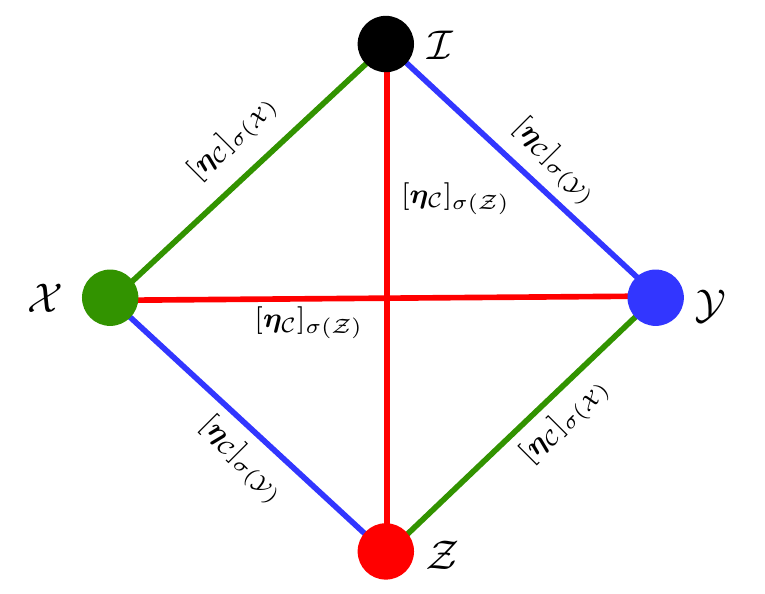}
    \caption{The weighted Cayley graph $\Set{G}$ of the single-qubit Pauli group $\Set{P}$ constructed based on channel $\Sop{C}$. The vertices represent the four operators in $\Set{P}$. The quantities next to the edges represent the corresponding edge weights.}
    \label{fig:pauli}
\end{figure}

To gain further insights into the PRW representation, we will rely on a weighted Cayley graph \cite{sgt} $\Set{G}$ of Pauli groups, in which the $i$-th vertex represents the $i$-th operator in $\Set{P}^n$. For a specific channel $\Sop{C}$, a pair of nodes $i$ and $j$ in the Cayley graph are connected with an edge having a weight of $[\V{\eta}_{\Sop{C}}]_l$, if we have $\Sop{P}_i\Sop{P}_j=\Sop{P}_l$. As a tangible example, the graph $\Set{G}$ corresponding to the single-qubit Pauli group is portrayed in Fig. \ref{fig:pauli}.
The function $\sigma(\Sop{O})$ denotes the index of the operator $\Sop{O}$ in $\Set{P}$, where we have $\sigma(\Sop{X}) = 2$, $\sigma(\Sop{Y}) = 3$, and $\sigma(\Sop{Z}) = 4$. For a fixed \ac{ggep} $\epsilon$, we can rewrite $\M{C}_{\mathrm{PRW}}(\V{\eta}_{\Sop{C}})$ as
\begin{equation}\label{decomposition_p}
\M{C}_{\mathrm{PRW}}(\V{\eta}_{\Sop{C}}) = (1-\epsilon)\M{I}+\M{A}(\Set{G},\V{\eta}_{\Sop{C}}),
\end{equation}
where $\M{A}(\Set{G},\V{\eta}_{\Sop{C}})$ is the weighted adjacency matrix of the graph $\Set{G}$ corresponding to the channel $\Sop{C}$, which satisfies
\begin{equation}\label{adjacency}
[\M{A}(\Set{G},\V{\eta}_{\Sop{C}})]_{i,j} = \left\{
                 \begin{array}{ll}
                   [\V{\eta}_{\Sop{C}}]_l, & \hbox{$i\neq j,~\Sop{P}_i\Sop{P}_j=\Sop{P}_l$;} \\
                   0, & \hbox{$i=j$.}
                 \end{array}
               \right.
\end{equation}
Whenever there is no confusion, we will simply denote $\M{C}_{\mathrm{PRW}}(\V{\eta}_{\Sop{C}})$ as $\M{C}$, and $\M{A}(\Set{G},\V{\eta}_{\Sop{C}})$ as $\M{A}$.

It can be observed from \eqref{adjacency} that the channel $\Sop{C}$ may be interpreted as a random walk over the graph $\Set{G}$, which maps an input state $|\psi\rangle\langle\psi|$ to $\Sop{P}_i|\psi\rangle\langle\psi|$ with probability $\eta_i$. The goal of the quasi-probability representation method is to find another operator that reverses the random walk process. Specifically, \eqref{sampling_coeff} can be simplified as follows
\begin{equation}\label{sampling_coeff_pauli}
\tilde{\V{\mu}}_{\Sop{C}} = \M{C}^{-1}\V{\alpha},
\end{equation}
where $\V{\alpha} = [1~\V{0}_{4^n-1}^{\mathrm{T}}]^{\mathrm{T}}$, and $\tilde{\V{\mu}}_{\Sop{C}}$ is obtained by extracting the $4^n$ entries corresponding to Pauli operators from $\V{\mu}_{\Sop{C}}$ in \eqref{sampling_coeff}.

With the aid of PRW representation, we are now ready to present the following \ac{sof} upper bound for Pauli channels.
\begin{proposition}[\ac{sof} upper bound]\label{prof:upper_pauli}
For an $n$-qubit Pauli channel $\Sop{C}$, given a \ac{ggep} $\epsilon$, the \ac{sof} can be upper bounded as
\begin{equation}\label{overhead_ub}
\gamma_{\Sop{C}} \le 4\epsilon \cdot \frac{1-\epsilon}{(1-2\epsilon)^2}.
\end{equation}
The equality is attained when there is only a single type of error, namely there is only one non-zero entry in $\V{\eta}_{\Sop{C}}$.
\begin{IEEEproof}
Please refer to Appendix \ref{sec:proof_upper_pauli}.
\end{IEEEproof}
\end{proposition}

Note that Pauli channels having only a single type of error correspond to the highest hashing bound, when mitigated by QECCs. Therefore, by considering both Corollary \ref{coro:qecc_precoder} and Proposition \ref{prof:upper_pauli}, one may intuitively conjecture that for Pauli channels having the same \ac{ggep}, the \ac{sof} increases as the hashing bound increases. The hashing bound of a Pauli channel $\Sop{C}$ can be expressed as \cite{hashing_bound1,hashing_bound2,hashing_bound3}
\begin{equation}\label{hashing_bound}
R_{\mathrm{hashing}} = 1 - H(\V{\eta}_{\Sop{C}}),
\end{equation}
where $R_{\mathrm{hashing}}$ is the highest affordable coding rate capable of satisfying the hashing bound, and $H(\V{\eta}_{\Sop{C}})$ denotes the entropy of $\V{\eta}_{\Sop{C}}$ viewed as a probability distribution. Mathematically, the entropy $H(\V{\eta}_{\Sop{C}})$ is a Schur-concave function \cite[Sec. 2.1]{majorization} with respect to the probability distribution $\V{\eta}_{\Sop{C}}$. To elaborate, a Schur-concave function $f(\V{x})$ is characterized by
$$
f(\V{x}) \le f(\M{Q}\V{x}),
$$
for any doubly stochastic matrix $\M{Q}$. This implies that doubly stochastic transformations on $\V{\eta}_{\Sop{C}}$ would lead to the increase of entropy \cite{doubly_stochastic}. The term $R_{\mathrm{hashing}}$ in \eqref{hashing_bound} can be seen to have the exactly opposite property termed as Schur-convex \cite[Sec. 2.1]{majorization}, hence the aforementioned conjecture can be formulated as follows: ``the \ac{sof} is a Schur-convex function with respect to the probability vector $\V{\eta}_{\Sop{C}}$''.

Next we show that the conjecture is correct, when the channels under consideration are memoryless channels.
\begin{proposition}\label{prop:memoryless}
For any $n$-qubit memoryless Pauli channel $\Sop{C}=\bigotimes_{i=1}^n \Sop{C}_i$, given a fixed \ac{ggep} $\epsilon$, the \ac{sof} is a Schur-convex function of $\V{\eta}_{\Sop{C}}$, meaning that
\begin{equation}
\|\V{\mu}(\V{\eta}_{\Sop{C}})\|_1 \ge \left\|\V{\mu}\left(\bigotimes_{i=1}^n\M{Q}_i\V{\eta}_{\Sop{C}_i}\right)\right\|_1
\end{equation}
holds for all doubly-stochastic matrices $\M{Q}_i$ preserving the \ac{ggep}, where $\V{\mu}(\V{x})$ denotes the quasi-probability representation vector of the Pauli channel having the probability vector of $\V{x}$.
\begin{IEEEproof}
Please refer to Appendix \ref{sec:proof_memoryless}.
\end{IEEEproof}
\end{proposition}

\begin{figure*}[t]
\subfloat[][Conceptual schematic of a \ac{qem}-protected uncoded Hadamard gate.]{
\begin{minipage}{.48\textwidth}
\centering
\includegraphics[width=.65\textwidth]{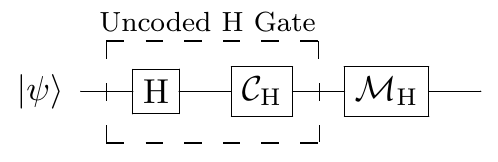}
\label{fig:uncoded_h}
\end{minipage}
}
\subfloat[][Conceptual schematic of a \ac{qem}-protected coded Hadamard gate.]{
\begin{minipage}{.48\textwidth}
\centering
\includegraphics[width=.95\textwidth]{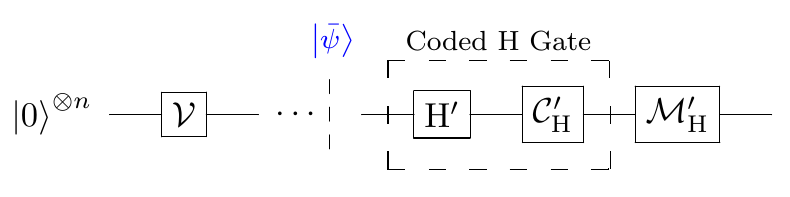}
\label{fig:coded_h}
\end{minipage}
}

\vspace{0.2cm}
\subfloat[][Detailed schematic of a \ac{qem}-protected coded Hadamard gate relying on the transversal gate configuration.]{
\begin{minipage}{1.0\textwidth}
\centering
\includegraphics[width=.8\textwidth]{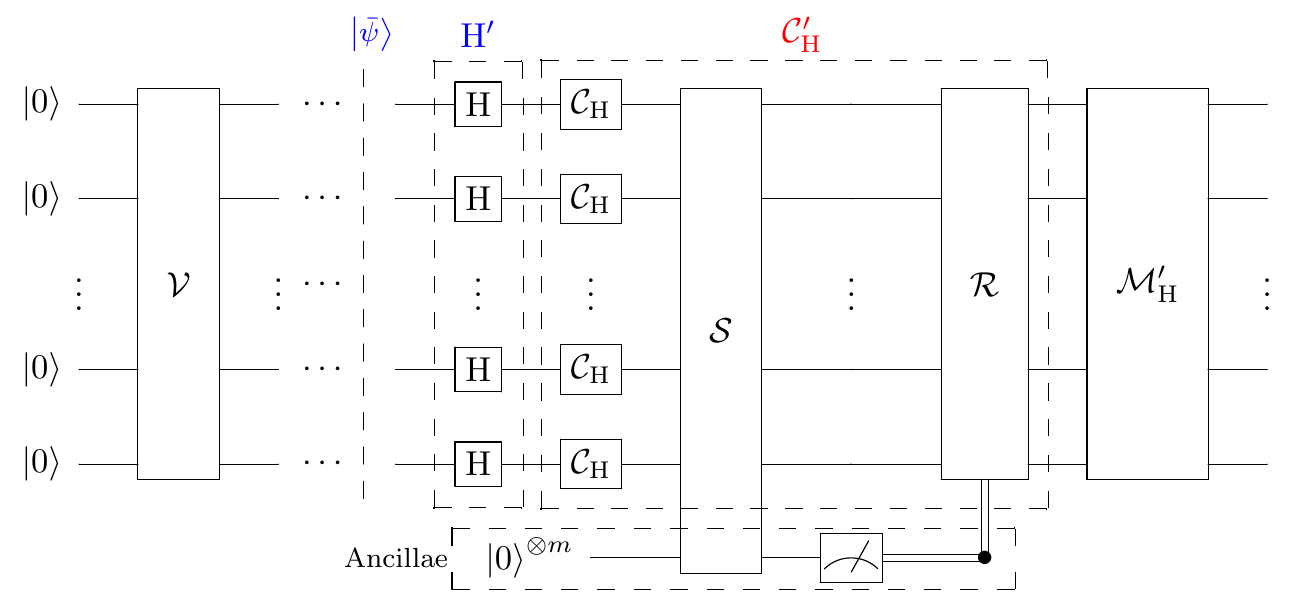}
\label{fig:coded_h_detail}
\end{minipage}
}
\caption{Comparison between an uncoded Hadamard gate and a coded Hadamard gate, both protected by \ac{qem}.}
\label{fig:test}
\end{figure*}

\subsection{SOF reduction using quantum channel precoders: Practical considerations}\label{ssec:qcp_practical}
\begin{figure}[t]
\centering
\subfloat[][Clifford twirling]{
\begin{minipage}{.5\columnwidth}
\centering
\includegraphics[width=.9\textwidth]{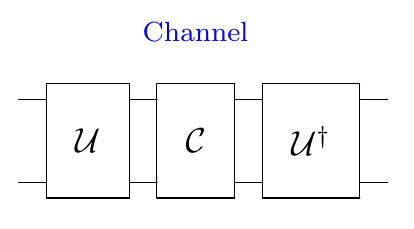}
\end{minipage}
\label{fig:c_twirling}
}
\subfloat[][Pauli twirling]{
\begin{minipage}{.4\columnwidth}
\centering
\includegraphics[width=.95\textwidth]{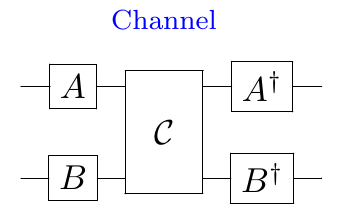}
\end{minipage}
\label{fig:p_twirling}
}
\caption{Schematics of Clifford twirling and Pauli twirling over two-qubit channels.}
\label{fig:twirling}
\end{figure}

Our previous analysis indicates that depolarizing channels are the most preferable channels in terms of having the lowest \ac{sof}. This implies that Clifford twirling \cite{ctwirl1,ctwirl2,ctwirl3}, a technique that turns an arbitrary channel into a depolarizing channel while preserving the original average fidelity, might be a quantum channel precoder enabling effective \ac{sof} reduction. Specifically, given a quantum channel $\Sop{C}$ over $n$ qubits, the Clifford twirling $\Sop{T}_{\Set{C}}$ transforms the channel such that the output state satisfies
\begin{equation}\label{c_twirl}
\Sop{T}_{\Set{C}}[\Sop{C}(\rho)] = \sum \M{U}^\dagger\Sop{C}(\M{U}\rho \M{U}^\dagger)\M{U} {\mathrm{d}}\M{U},
\end{equation}
where the summation is carried out over the Clifford group on $n$ qubits. Conceptually, the Clifford twirling over two-qubit channels can be implemented as demonstrated in Fig. \ref{fig:c_twirling}, where the gates comprising the circuits $\Sop{U}$ and $\Sop{U}^\dagger$ are chosen according to a uniform distribution over the set of Clifford gates.

In practice, however, the gates used for implementing Clifford twirling might be imperfect themselves. In light of this, a real-world Clifford twirling would in general impose an average fidelity reduction, and thus lead to additional \ac{sof}. For certain channels, the theoretical \ac{sof} reduction of Clifford twirling may be outweighed by this additional overhead. A representative example is constituted by the family of Pauli channels, whose \ac{sof} is rather close to that of depolarizing channels, according to Proposition \ref{prof:upper_pauli}.

The observation that the Pauli channels have similar \acp{sof} implies that Pauli twirling $\Sop{T}_{\Set{P}}$ might be a more practical quantum channel precoder, which turns an arbitrary channel into a Pauli channel in the following manner \cite{ptwirl}
\begin{equation}\label{pauli_twirling_definition}
\Sop{T}_{\Set{P}}[\Sop{C}(\rho)] = \frac{1}{4^n}\sum_{i=1}^{4^n} \M{S}_i^{\dagger}\Sop{C}(\M{S}_i\rho\M{S}_i^{\dagger})\M{S}_i.
\end{equation}

The implementation of Pauli twirling is portrayed in Fig. \ref{fig:p_twirling}, where the gates $A$ and $B$ are chosen according to a uniform distribution on the set of Pauli gates. In state-of-the-art quantum computers, two-qubit gates, as used in the Clifford twirling shown in Fig. \ref{fig:c_twirling}, would result in much more error than single-qubit gates (typically by a factor of $10$ or even higher\cite{fidelity_ratio}), hence Pauli twirling using single-qubit gates may introduce much lower additional \ac{sof} than Clifford twirling.

In practise, we cannot directly implement twirling at both sides of the channel. Instead, we have to twirl simultaneously both the perfect gate and the channel. Therefore, the techniques should be slightly modified in order to effectively apply the twirling to the channel. Specifically, if we wish to apply Pauli twirling to a channel $\Sop{C}$ associated with an imperfect gate $\tilde{\Sop{G}}=\Sop{C}\circ \Sop{G}$ where $\Sop{G}$ denotes the perfect gate, we may apply the following modified twirling to $\tilde{\Sop{G}}$
$$
\begin{aligned}
\widetilde{\Sop{T}}_{\Set{P}}[\tilde{\Sop{G}}(\rho)] &=\frac{1}{4^n}\sum_{i=1}^{4^n} \M{S}_i^{\dagger}\tilde{\Sop{G}}[(\M{G}^\dagger\M{S}_i\M{G})\rho(\M{G}^\dagger\M{S}_i\M{G})^{\dagger}]\M{S}_i \\
&= \frac{1}{4^n}\sum_{i=1}^{4^n} \M{S}_i^{\dagger}\Sop{C}[\M{S}_i\M{G}\rho\M{G}^\dagger\M{S}_i^{\dagger}]\M{S}_i \\
&=[\Sop{T}_{\Set{P}}(\Sop{C})\circ\Sop{G}](\rho).
\end{aligned}
$$
A similar procedure can also be applied to Clifford twirling. We note that the operation $(\M{G}^\dagger\M{S}_i\M{G})\rho(\M{G}^\dagger\M{S}_i\M{G})^{\dagger}$ can be simplified to $\M{S}_m\rho \M{S}_m^\dagger$ for some $m$, if the perfect gate $\Sop{G}$ is a Clifford gate, since the Clifford group $\Set{C}_n$ is the normalizer of the Pauli group $\Set{P}_n$ satisfying $\Set{C}_n = \{\M{U}\in \mathbb{U}(2^n)|\M{U}\Set{P}_n\M{U}^\dagger = \Set{P}_n\}.$

\section{Sampling Overhead Factor Analysis for Coded Quantum Gates}\label{sec:coded}
In this section, we investigate the \ac{sof} of gates protected by quantum channel codes, including \acp{qecc} and \acp{qedc}. These codes are designed to convert the original channel corresponding to the unprotected gate into an reduced-error-rate channel over more qubits, with the objective of having lower \ac{ggep}. Under the framework of \ac{qem}, they can also be viewed as channel precoders. Naturally, it is of great interest to us, whether an amalgam of quantum codes and \ac{qem} can benefit each other.

Specifically, we consider the scenario where every set of $k$ logical qubits is protected using $n$ physical qubits. Using the terminology of quantum coding, this means that we consider $[[n,k,d]]$ codes, where $d$ is the minimum distance of the code \cite{qecc_survey}. Furthermore, if not otherwise stated, we assume that Clifford gates are considered using the transversal gate scheme of \cite{transversal}, while non-Clifford gates are implemented via the magic state distillation process of \cite{magic_state}. These are conventional assumptions in the quantum fault-tolerant computing literature \cite[Sec. 10.6]{ncbook}. Furthermore, for the conciseness of discussion, we assume that the quantum channels encountered in this section are all Pauli channels, or had been turned into Pauli channels by means of Pauli twirling.

\subsection{Amalgamating quantum codes with QEM: A toy example}
To elaborate further on how \ac{qem} may be amalgamated with quantum codes, we consider the simple example of protecting a single Hadamard gate. As shown in Fig. \ref{fig:uncoded_h}, an uncoded imperfect Hadamard gate can be decomposed into a perfect Hadamard gate $\mathrm{H}$ and a quantum channel $\Sop{C}_{\mathrm{H}}$. Given that the channel $\Sop{C}_{\mathrm{H}}$ is known, we can apply the \ac{qem} circuit $\Sop{M}_{\mathrm{H}}$ to invert it. By contrast, in the coded scheme, the logical qubit is protected using an encoder $\Sop{V}$ exploiting $n$ physical qubits at the input of the circuit, as portrayed in Fig. \ref{fig:coded_h}. In the code space, the original input state $\ket{\psi}$ is expressed as the coded state $\ket{\bar{\psi}}$, while the coded Hadamard gate may be decomposed into an equivalent perfect Hadamard gate ${\mathrm{H}}^{\prime}$ and another quantum channel $\Sop{C}_{\mathrm{H}}^{\prime}$. Consequently, the \ac{qem} circuit $\Sop{M}_{\mathrm{H}}^{\prime}$ has to be designed for the transformed channel $\Sop{C}_{\mathrm{H}}^{\prime}$. More specifically, the Hadamard gate protected using the transversal gate configuration is depicted in more detail in Fig. \ref{fig:coded_h_detail}. The equivalent Hadamard gate is implemented simply by $n$ transversal Hadamard gates. As a result, each physical qubit experiences the same channel $\Sop{C}_{\mathrm{H}}$. Right after the transversal gates, with the help of $m$ ancillae, the integrity of the output state is examined by the stabilizer check $\Sop{S}$. The subsequent recovery circuit $\Sop{R}$ is capable of correcting a fixed number of Pauli errors, depending on the minimum distance of the code. For example, if Steane's codes is applied, $\Sop{R}$ can correct any single Pauli error that appeared within the circuit. The transversal gates along with the stabilizer check and the recovery circuit constitute the transformed channel $\Sop{C}_{\mathrm{H}}^{\prime}$.

Ideally, since $\Sop{S}$ and $\Sop{R}$ are able to correct errors, the transformed channel $\Sop{C}_{\mathrm{H}}^{\prime}$ might have a lower \ac{ggep} than the original channel $\Sop{C}_{\mathrm{H}}$. However, this might not be true in practice, because $\Sop{S}$ and $\Sop{R}$ themselves are also prone to errors. Intuitively, assuming that the \ac{ggep} of each gate in the circuit is at most $\epsilon$, as $\epsilon$ tends to zero, the \ac{ggep} of $\Sop{C}_{\mathrm{H}}^{\prime}$ is at most on the order of $O(\epsilon^2)$, since all single errors are corrected. Therefore, quantum codes are capable of reducing the channel \ac{ggep}, provided that $\epsilon$ is sufficiently small. Specifically, the value of physical gate \ac{ggep} $\epsilon_{\mathrm{th}}$, below which quantum codes become beneficial, is referred to as the fault-tolerance threshold \cite{pre_post_threshold}.

\begin{figure}[t]
    \centering
    \includegraphics[width=.385\textwidth]{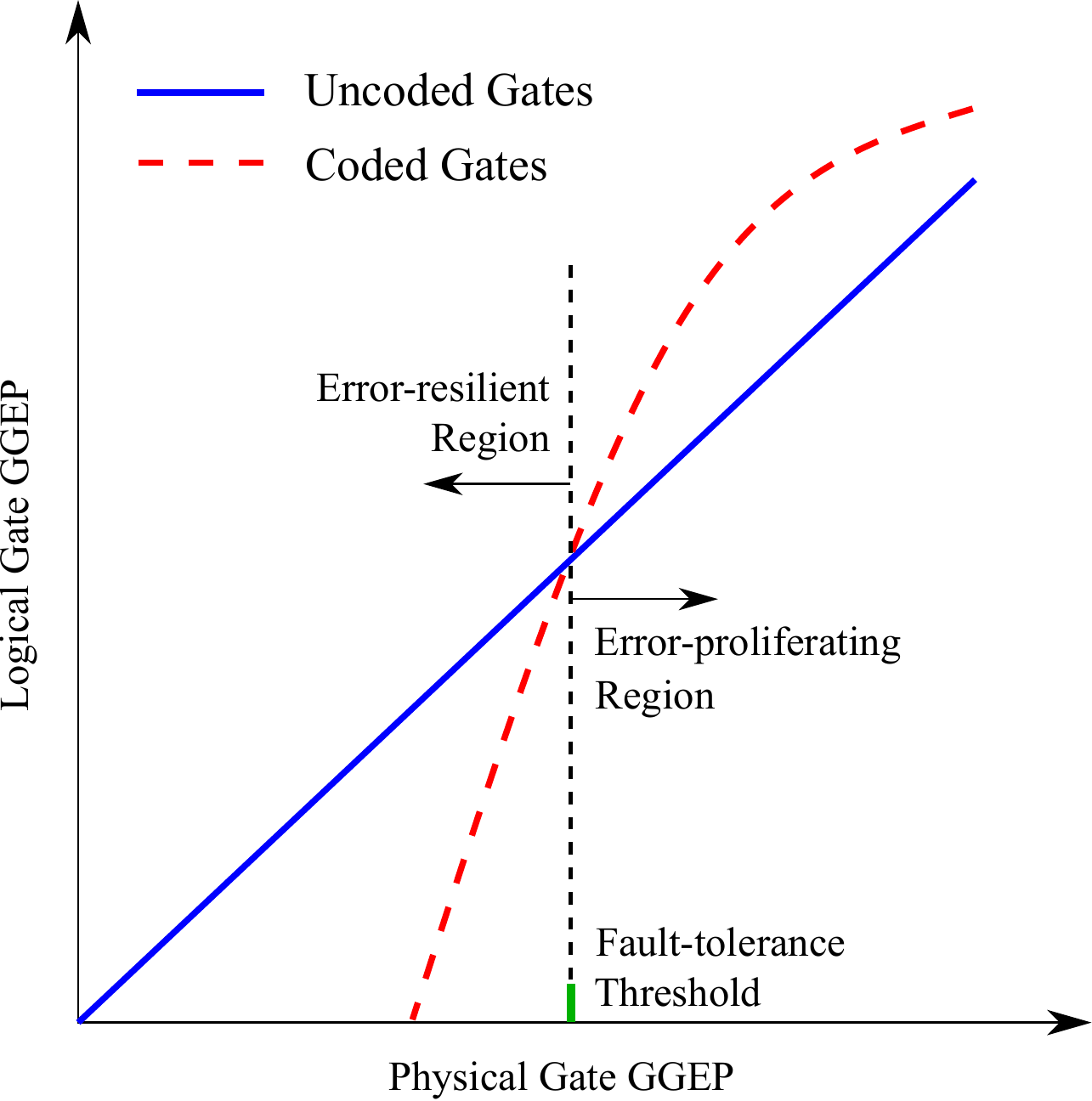}
    \caption{Illustration of the fault-tolerance threshold, the error-proliferating region and the error-resilient region of quantum codes.}
    \label{fig:illustration_threshold}
\end{figure}

In general, given a quantum code, the logical gate \ac{ggep} would be higher than the physical gate \ac{ggep}, when the physical gate \ac{ggep} is relatively high. As the physical gate \ac{ggep} decreases, it gradually becomes higher than the logical gate \ac{ggep}, as sketched in Fig.~\ref{fig:illustration_threshold} and detailed in~\cite[Sec. VI-B]{topological_threshold}. The physical gate \ac{ggep} corresponding to the cross-over point of the two curves is the fault-tolerance threshold of the quantum code. We will refer to the region where the quantum code is beneficial, namely where the logical gate \ac{ggep} is lower than the physical gate \ac{ggep}, as the error-resilient region. By contrast, the opposite region will be referred to as the 'error-proliferation' region.

In the following subsections, we will first analyse the \ac{sof} of coded gates when the code is operating in its error-proliferation region, followed by the opposite scenario.

\subsection{Quantum codes operating in their error-proliferating regions}
Using our previous results on uncoded gates in Section \ref{ssec:bound_overhead}, it may be readily shown that quantum codes operating in their error-proliferating regions may not lead to substantial \ac{sof} reduction. Formally, we have the following result.
\begin{corollary}\label{coro:pre_threshold}
For an uncoded gate having a \ac{ggep} of $\epsilon$ and \ac{sof} $\gamma$, the \ac{sof} is lower-bounded by $\gamma \cdot \frac{(1-2\epsilon)^2}{(1-\epsilon)^3}$, when the gate is protected by some quantum code operating in its error-proliferating region. Furthermore, provided that the channel corresponding to the uncoded gate is a depolarizing channel, the lower bound can be further refined to $\gamma \cdot \frac{(3-4\epsilon)^2}{3(1-\epsilon)^2(3-\epsilon)}$.
\begin{IEEEproof}
This is a direct corollary following from Corollary \ref{coro:qecc_precoder} and Proposition \ref{prof:upper_pauli}. To elaborate a little further, considering the extreme case where the threshold is met exactly so that the output \ac{ggep} of the quantum code is equal to $\epsilon$, the generic lower bound is obtained in the form of:
\begin{equation}
\gamma \cdot \frac{1}{(1-\epsilon)^2} \cdot \frac{(1-2\epsilon)^2}{1-\epsilon} = \gamma \cdot \frac{(1-2\epsilon)^2}{(1-\epsilon)^3}
\end{equation}
using \eqref{overhead_lb} and \eqref{overhead_ub}. The lower-bound valid for depolarizing channels is obtained as
\begin{equation}
\gamma \cdot \frac{4\epsilon\frac{1}{(1-\epsilon)^2}}{\left(\frac{(4-1)(1-2\epsilon)-\epsilon}{4(1-\epsilon)-1}\right)^2-1} =\gamma \cdot \frac{(3-4\epsilon)^2}{3(1-\epsilon)^2(3-\epsilon)}
\end{equation}
using \eqref{overhead_lb} and \eqref{second_line_qecc}, and by further exploiting the fact that single-qubit depolarizing channels have the highest \ac{sof} among all depolarizing channels sharing the same \ac{ggep}.
\end{IEEEproof}
\end{corollary}

\begin{figure}[t]
    \centering
    \includegraphics[width=.45\textwidth]{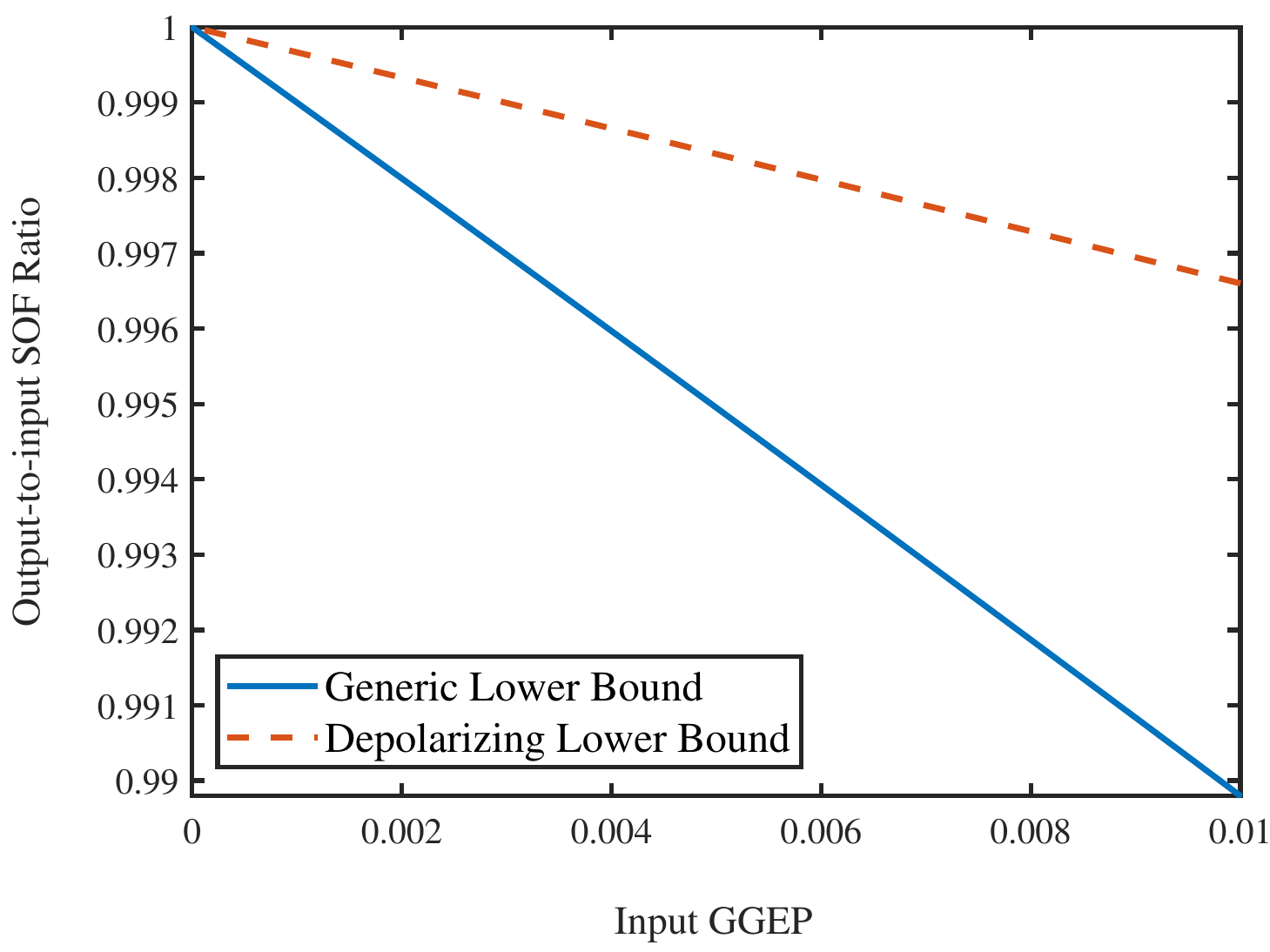}
    \caption{Lower bounds of the output-to-input \ac{sof} ratio for quantum codes operating in their error-proliferating regions.}
    \label{fig:lb_pre_threshold}
\end{figure}

To demonstrate the implications of Corollary \ref{coro:pre_threshold} more explicitly, we plot the lower bounds in Fig. \ref{fig:lb_pre_threshold}. Since the fault-tolerance thresholds of most \acp{qecc} are as low as $10^{-2}\sim 10^{-3}$, as it can be observed from Fig. \ref{fig:lb_pre_threshold}, even when the \ac{ggep} meets the threshold exactly, the quantum codes can only offer an overhead reduction of at most $1\%$. Therefore, amalgamating \ac{qem} with codes operating in their error-proliferating regions may not be mutually beneficial.

\subsection{QECCs operating in their error-resilient regions}\label{ssec:post_qecc}
In light of our previous discussions, it becomes plausible that \acp{qecc} operating in their error-resilient regions may contribute to \ac{sof} reduction by reducing the gate error probability. As stated in~\cite[Sec. 10.6.1]{ncbook} for example, their error-correcting capability can be further improved via concatenation. However, the price of concatenating codes is a drastic increase in the qubit overhead. It is thus interesting to investigate whether the amalgamation of \ac{qecc} and \ac{qem} would outperform pure concatenated \acp{qecc}, and if so, in what scenarios.

\begin{figure}[t]
\subfloat[][Concatenation configuration]{
\begin{minipage}{.462\columnwidth}
\centering
\includegraphics[width=.95\textwidth]{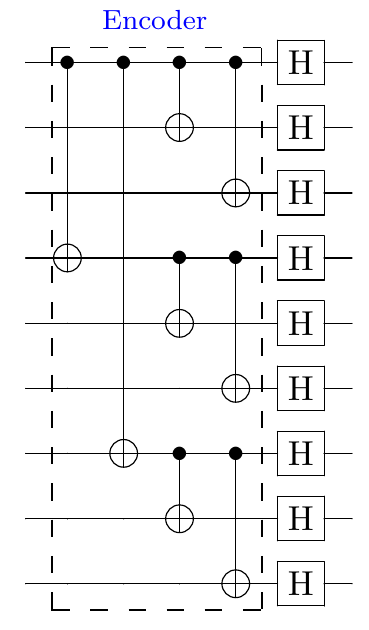}
\end{minipage}
}
\subfloat[][\ac{qecc}-\ac{qem} configuration]{
\begin{minipage}{.518\columnwidth}
\centering
\includegraphics[width=.95\textwidth]{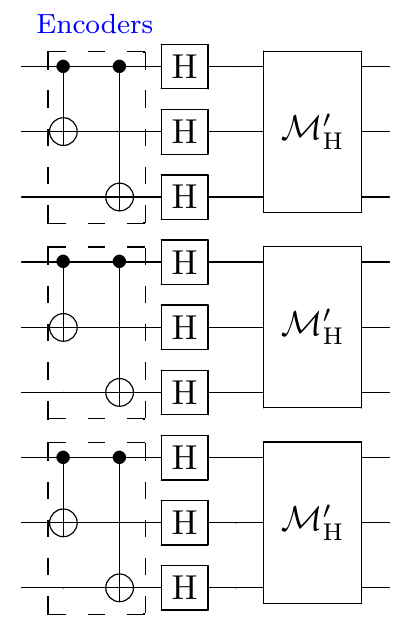}
\end{minipage}
}
\caption{Comparison between the concatenated \ac{qecc} scheme and the \ac{qecc}-\ac{qem} scheme protecting transversal Hadamard gates using the rate $1/3$ repetition code. In the example, the \ac{qecc}-\ac{qem} scheme runs three times as fast as the concatenation scheme.}
\label{fig:concatenation}
\end{figure}

To elaborate, we consider the simple example of transversal Hadamard gates protected by a rate $1/3$ repetition code, as portrayed in Fig. \ref{fig:concatenation} and detailed in \cite{topological_threshold}. By concatenating the repetition code twice, the number of physical qubits protecting a single logical qubit become three times that of the non-concatenated code. By contrast, if we amalgamate the rate $1/3$ code with \ac{qem}, the additional qubits can be used to parallelize the computation, leading to a computational acceleration by a factor of three. In this sense, the \ac{qecc}-\ac{qem} scheme outperform the concatenated scheme, when the \ac{sof} of \ac{qem} obeys $\gamma_{\Sop{C}}\le 2$.

As the number of gates in a circuit increases, the \ac{qem} sampling overhead grows exponentially (as indicated by \eqref{sof_exponential}), while the qubit overhead of the concatenated scheme remains constant. Therefore, there would be a critical point where the overall \ac{qem} sampling overhead escalation starts to outweigh the parallelization speedup benefits. This may be interpreted as a limitation imposed on the circuit size, beyond which full fault-tolerance becomes necessary. Next we provide an estimate of the critical point, given that the gate error probability is sufficiently low.

\begin{proposition}[Lower bound on the critical point]\label{prop:critical_point}
Consider a quantum circuit in which each gate has a \ac{ggep} at most $\epsilon$. If the gates are protected using $l$-stage concatenated (i.e., concatenated $l$ times) $[[n,k,d]]$ \ac{qecc} operating in its error-resilient region via the transversal gate configuration, amalgamating the code with \ac{qem} is more preferable than applying the $(l+1)$-stage concatenated code, when the number of gates $N_l$ satisfies
\begin{subequations}\label{breaking_lower_bound}
\begin{align}
N_l &\le \frac{\ln n}{\ln (1+4f^{(l)}(\epsilon))-\ln [1+4 f^{(l+1)}(\epsilon)]} \\
& \approx \frac{1}{f^{(l)}(\epsilon)-f^{(l+1)}(\epsilon)} \cdot \frac{\ln n}{4} \label{second_line_breaking}
\end{align}
\end{subequations}
when $\epsilon \ll 1$, and $f(\epsilon)$ is the output \ac{ggep} of the single-stage $[[n,k,d]]$ code given the input \ac{ggep} $\epsilon$, and $f^{(l)}(\epsilon)$ denotes the $l$-times self-composition of function $f(\epsilon)$, as exemplified by $f^{(2)}(\epsilon)=f[f(\epsilon)]$.
\begin{IEEEproof}
For a circuit in which every gate is protected using $l$-stage concatenated $[[n,k,d]]$ \acp{qecc}, the total computational overhead (including both the \ac{qem} overhead and qubit overhead) $\tilde{\gamma}_l$ can be bounded as
\begin{equation}
\tilde{\gamma}_l \le n^l (1+\gamma_l)^N,
\end{equation}
where $\gamma_l$ denotes the highest \ac{sof} of a single logical gate in the circuit, and $N$ is the total number of gates. Hence the critical point $N_l$ between $l$-stage concatenation and $(l+1)$-stage concatenation satisfies
\begin{equation}
(1+\gamma_l)^{N_l} = n(1+\gamma_{l+1})^{N_l}.
\end{equation}
According to Corollary \ref{coro:qecc_precoder} and Propostion \ref{prof:upper_pauli}, when $\epsilon \ll 1$, the upper and lower bounds for the \ac{sof} of a single gate tend to be equal. Thus the \ac{sof} of each uncoded gate can be upper bounded by $\gamma \le 4\epsilon$. For $l$-stage concatenated codes, we have $\gamma_l \le 4 f^{\circ l}(\epsilon)$. Therefore we obtain
\begin{equation}
N_l \le \frac{\ln n}{\ln (1+4f^{(l)}(\epsilon))-\ln [1+4 f^{(l+1)}(\epsilon)]}.
\end{equation}
Additionally, using the Maclaurin approximation of $\ln (1+x) \approx x$ when $x>0$ is sufficiently small, we obtain \eqref{second_line_breaking}. Hence the proof is completed.
\end{IEEEproof}
\end{proposition}

\begin{remark}
As a special case, the pure \ac{qem} (i.e., $l=0$) is more preferable than amalgamating a single-stage $[[n,k,d]]$ code with it, when the number of gates satisfies
\begin{subequations}
\begin{align}
N_0 &\le \frac{\ln n}{\ln (1+4\epsilon)-\ln [1+4 f(\epsilon)]} \\
& \approx \frac{1}{\epsilon-f(\epsilon)} \cdot \frac{\ln n}{4}.
\end{align}
\end{subequations}
Note that when $(l+1)$-stage \ac{qecc} concatenation cannot be implemented due to the associated physical limitations (e.g. total number of physical qubits), the amalgam of $l$-stage \ac{qecc} and \ac{qem} may be applied even beyond the critical point.
\end{remark}

\subsection{QEDCs operating in their error-resilient regions}\label{ssec:post_qedc}
Due to their smaller minimum distance than that of \acp{qecc}, \acp{qedc} are not capable of correcting any error. Nonetheless, they can be used as important building blocks in the scheme of \emph{post-selection fault-tolerance} \cite{post_selection1,post_selection2}. To expound a little further, post-selection fault-tolerance differs from its conventional counterpart in that it is implemented by detecting potential errors, and only accepting the results if no error is detected. Typically, \acp{qedc} have a shorter codeword length compared to \acp{qecc}, hence they often also possess a higher threshold. For instance, the $[[4,2,2]]$ QEDC can detect an error at the price of protecting a logical gate using four physical gates, while Steane's $[[7,1,3]]$ \ac{qecc} requires seven gates. This makes post-selection fault-tolerance a preferable scheme, when the gates are relatively noisy \cite{post_selection1,post_selection2,post_selection_exp}.

In the context of \ac{qem}, the high threshold of \acp{qedc} appears to make their amalgamation with \ac{qem} more beneficial. However, a subtle issue is that \acp{qedc} suffer from their own sampling overhead. To elaborate, if for every gate the probability of successful error detection is $p$, similar to that of \ac{qem}, the \ac{sof} of the \ac{qedc} may be defined as
\begin{equation}\label{so_qedc}
\gamma_{\mathrm{QEDC}} = \frac{1}{1-p}-1 = \frac{p}{1-p},
\end{equation}
which will be referred to as the \ac{qedc}-\ac{sof} in the following disucssions. Additionally, performing \ac{qem} on the post-selected channel (where single errors are eliminated) also incurs an \ac{sof}, which in this context will be referred to as the \ac{qem}-\ac{sof}. Thus, the total \ac{sof} can be calculated as follows:
\begin{equation}
\begin{aligned}
\gamma_{\mathrm{total}} &= (1+\gamma_{\mathrm{QEDC}})(1+\gamma_{\mathrm{QEM}})-1\\
&=\gamma_{\mathrm{QEDC}}+\gamma_{\mathrm{QEM}}+\gamma_{\mathrm{ QEDC}}\cdot\gamma_{\mathrm{QEM}},
\end{aligned}
\end{equation}
where $\gamma_{\mathrm{QEDC}}$ and $\gamma_{\mathrm{QEM}}$ represent the \ac{qedc}-\ac{sof} and the \ac{qem}-\ac{sof}, respectively. In this regard, the amalgamation of \ac{qedc} and \ac{qem} is only beneficial when the total \ac{sof} is lower than that of \ac{qem} applied directly to uncoded gates.

To compute the \ac{qedc}-\ac{sof} of each logical gate for a specific \ac{qedc}, it suffices to compute the probability that a single Pauli error occurs in the entire physical circuit corresponding to the logical gate. Let us assume that every single-qubit gate incurs a Pauli channel having a \ac{ggep} of $\epsilon_1$, and every two-qubit gate incurs a Pauli channel having \ac{ggep} $\epsilon_2$. For any logical gate implemented using the transversal gate configuration, it is clear that the occurrence probability of a single Pauli error in single-qubit and two-qubit gates, namely $p_1$ and $p_2$, can be expressed as
\begin{equation}
p_1 = n\epsilon_1 + O(\epsilon_1^2),~~p_2 = n\epsilon_2 + O(\epsilon_2^2),
\end{equation}
for any $[[n,k,d]]$ \ac{qedc}, since every logical gate is implemented using $n$ physical gates. The terms having the orders of $O(\epsilon_1^2)$ and $O(\epsilon_2^2)$ are negligible when the \acp{gep} are sufficiently small. According to \eqref{so_qedc}, we also have the following result for the corresponding \ac{qedc}-\ac{sof}
\begin{equation}
\gamma_1 = n\epsilon_1 + O(\epsilon_1^2),~~\gamma_2 = n\epsilon_2 + O(\epsilon_2^2).
\end{equation}
Among \acp{qedc} capable of detecting a single arbitrary Pauli error, the one having the lowest $n$ is the $[[4,2,2]]$ code. In this case, we have $\gamma_1 \approx 4\epsilon_1$ and $\gamma_2 \approx 4\epsilon_2$. The actual \ac{qedc}-\ac{sof} would be even higher due to the inevitable imperfections in the stabilizer measurements. On the other hand, for small $\epsilon_1$ and $\epsilon_2$, we can see from Corollary \ref{coro:qecc_precoder} and Proposition \ref{prof:upper_pauli} that when no \ac{qedc} is applied, the \ac{qem}-\ac{sof} is also approximately four times the \ac{ggep}. Therefore, we have the following remark.
\begin{remark}
Amalgamating \acp{qedc} and \ac{qem} might not be beneficial in the sense of \ac{sof} reduction, given that the logical gates are implemented using the transversal gate configuration.
\end{remark}

\begin{figure}[t]
\vspace{-0.4cm}
\subfloat[][Uncoded controlled-Z]{
\begin{minipage}{.45\columnwidth}
\centering
\vspace{0.74cm}
\includegraphics[width=.8\textwidth]{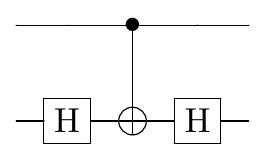}
\vspace{0.74cm}
\end{minipage}
}
\subfloat[][Coded controlled-Z]{
\begin{minipage}{.45\columnwidth}
\centering
\includegraphics[width=.6\textwidth]{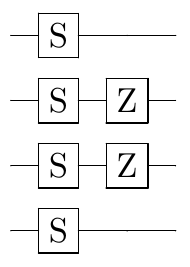}
\end{minipage}
}
\\
\subfloat[][Uncoded $\Sop{SWAP}\circ \Sop{H}^{\otimes 2}$]{
\begin{minipage}{.45\columnwidth}
\centering
\vspace{0.5cm}
\includegraphics[width=.9\textwidth]{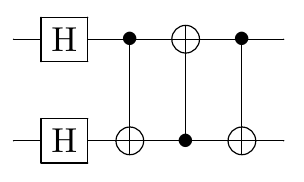}
\vspace{0.5cm}
\end{minipage}
}
\subfloat[][Coded $\Sop{SWAP}\circ \Sop{H}^{\otimes 2}$]{
\begin{minipage}{.45\columnwidth}
\centering
\includegraphics[width=.4\textwidth]{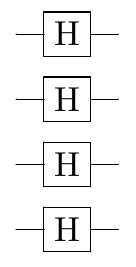}
\label{fig:swap_gate}
\end{minipage}
}
\caption{Illustration of logical gates on the two logical qubits within a $[[4,2,2]]$ \ac{qedc} code block. They are not implemented in a transversal manner.}
\label{fig:mapping_logical}
\end{figure}

This result may not be applicable to some logical gates, namely to those that are not implemented transversally. For example, some two-qubit gates processing the two logical qubits within a single block of the $[[4,2,2]]$ \ac{qedc} may be implemented using simple physical gates. As illustrated in Fig. \ref{fig:mapping_logical}, a logical controlled-Z gate can be implemented using six single-qubit gates $\Sop{S}\otimes (\Sop{Z}\circ\Sop{S})\otimes (\Sop{Z}\circ\Sop{S})\otimes \Sop{S}$. Since two-qubit gates typically have much higher \ac{ggep} compared to single-qubit gates, the \ac{qedc}-\ac{sof} may be even lower than the \ac{ggep} of a single controlled-Z gate. Similarly, the logical gate $\Sop{SWAP}\circ\Sop{H}^{\otimes 2}$ can be implemented by simply using four physical Hadamard gates, hence also has a low \ac{qedc}-\ac{sof}. Here, the operator $\Sop{SWAP}$ refers to the SWAP gate exchanging a pair of qubits \cite[Sec. 1.3.4]{ncbook}.

Unfortunately, non-transversal logical gates can only be designed in a case-by-case manner. Moreover, not all of them admit the nice and simple implementation as those shown in Fig. \ref{fig:mapping_logical}. For example, a CNOT gate between the two qubits in a $[[4,2,2]]$ code block has to be implemented via a SWAP gate, which has a high \ac{qedc}-\ac{sof}. By contrast, the transversal gate configuration is a general design paradigm that can be applied to all logical Clifford gates \cite{transversal}. In this regard, we may draw the conclusion that \ac{qedc}-\ac{qem} is only beneficial for certain specific non-transversal gate designs.

\section{Numerical Results}\label{sec:numerical}
\begin{figure*}[t]
    \centering
    \includegraphics[width=.75\textwidth]{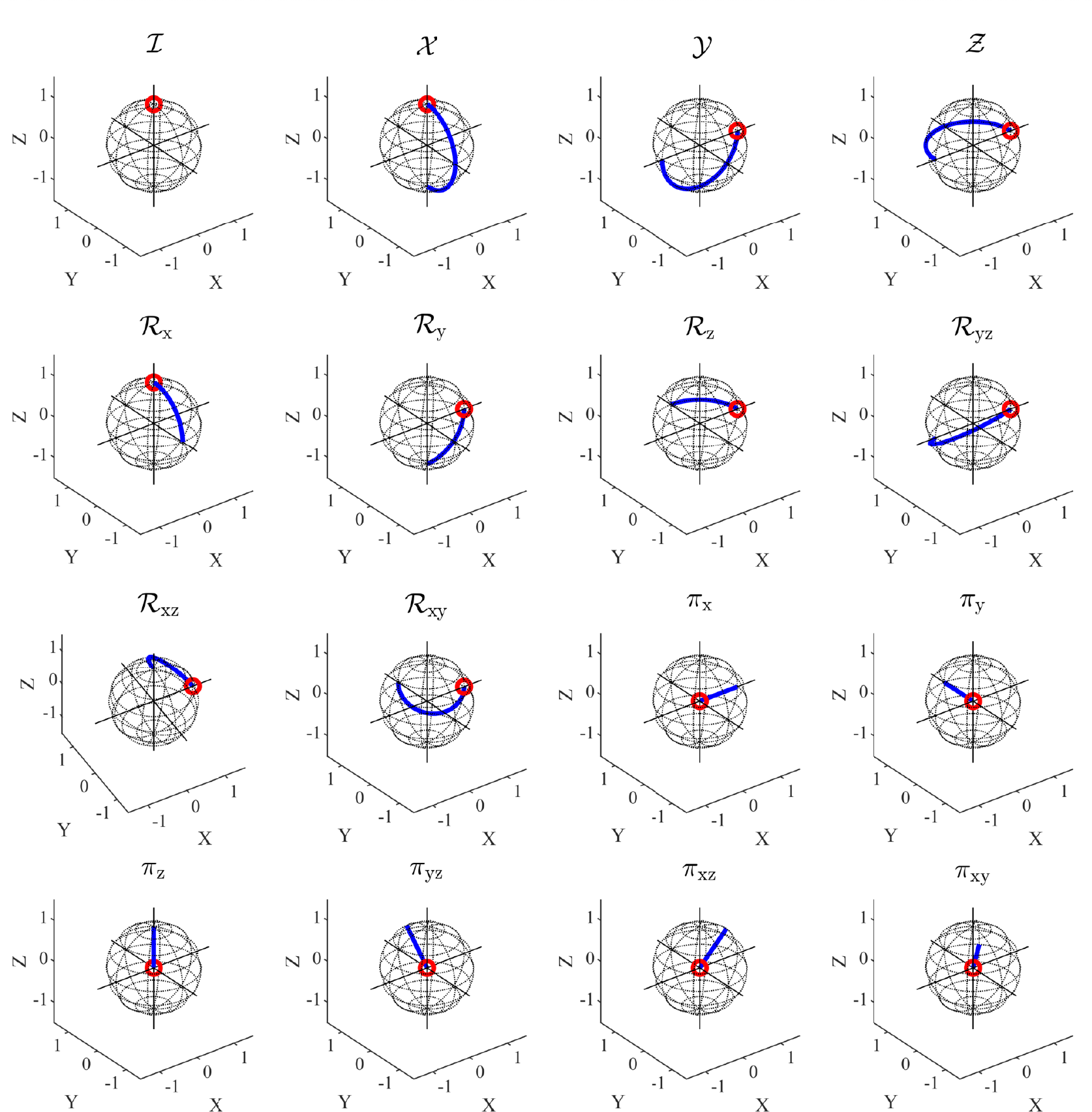}
    \caption{The trajectories of quantum states under the corresponding basis operations on the Bloch sphere. The circles represent the initial states, while the bold solid lines represent the trajectories.}
    \label{fig:illustration_basis_operations}
\end{figure*}

In this section, we augment our discussions throughout previous sections by numerical results. Throughout this section, for single-qubit channels, the basis matrix $\M{B}$ used for \ac{qem} is constituted by the conventional set of quantum operations listed in Table \ref{tbl:basis_operators} \cite{practical_qem}. The geometric interpretation of these operations is portrayed in Fig.~\ref{fig:illustration_basis_operations}. The basis operators of \ac{qem} for two-qubit channels are constituted by the tensor product of these operators. The operators $\Sop{R}_{\mathrm{x}}$, $\Sop{R}_{\mathrm{y}}$, and $\Sop{R}_{\mathrm{z}}$ represent $\pi/2$ rotations around the x-, y-, and z-axes of the Bloch sphere, respectively, while $\Sop{R}_{\mathrm{yz}}$, $\Sop{R}_{\mathrm{xz}}$, and $\Sop{R}_{\mathrm{xy}}$ represent $\pi$ rotations around the axes determined by the equations {\footnotesize$\left\{ \begin{array}{l}y=z \\ x=0 \end{array} \right.$}, {\footnotesize$\left\{ \begin{array}{l}x=z \\ y=0 \end{array} \right.$}, and {\footnotesize$\left\{ \begin{array}{l}x=y \\ z=0 \end{array} \right.$}, respectively. Similarly, the operators $\Sop{\pi}_{\mathrm{x}}$, $\Sop{\pi}_{\mathrm{y}}$, $\Sop{\pi}_{\mathrm{z}}$, $\Sop{\pi}_{\mathrm{yz}}$, $\Sop{\pi}_{\mathrm{xz}}$, and $\Sop{\pi}_{\mathrm{xy}}$ represent the measurement operations on the corresponding axes.

\subsection{Uncoded gates}
\begin{table}[t]
\captionsetup{justification=centering, labelsep=newline}
\centering
\caption{Sixteen basis operators used in \ac{qem} for single-qubit channels.}
\label{tbl:basis_operators}
\begin{tabular}{|c|c|c|}
  \hline
  \footnotesize
   & Operator & Output state \\ \hline
  1 & $\Sop{I}$ & $\Sop{I}(\rho) = \rho$ \\ \hline
  2 & $\Sop{X}$ & $\Sop{X}(\rho) = \uconj{\M{S}_{\Sop{X}}}{\rho}$ \\ \hline
  3 & $\Sop{Y}$ & $\Sop{Y}(\rho) = \uconj{\M{S}_{\Sop{Y}}}{\rho}$ \\ \hline
  4 & $\Sop{Z}$ & $\Sop{Z}(\rho) = \uconj{\M{S}_{\Sop{Z}}}{\rho}$ \\ \hline
  5 & $\Sop{R}_{\mathrm{x}}$ & $\Sop{R}_{\mathrm{x}}(\rho) = \frac{1}{2}\uconj{(\M{I}+\imath\M{S}_{\Sop{X}})}{\rho}$ \\ \hline
  6 & $\Sop{R}_{\mathrm{y}}$ & $\Sop{R}_{\mathrm{y}}(\rho) = \frac{1}{2}\uconj{(\M{I}+\imath\M{S}_{\Sop{Y}})}{\rho}$ \\ \hline
  7 & $\Sop{R}_{\mathrm{z}}$ & $\Sop{R}_{\mathrm{z}}(\rho) = \frac{1}{2}\uconj{(\M{I}+\imath\M{S}_{\Sop{Z}})}{\rho}$ \\ \hline
  8 & $\Sop{R}_{\mathrm{yz}}$ & $\Sop{R}_{\mathrm{yz}}(\rho) = \frac{1}{2}\uconj{(\M{S}_{\Sop{Y}}+\M{S}_{\Sop{Z}})}{\rho}$ \\ \hline
  9 & $\Sop{R}_{\mathrm{xz}}$ & $\Sop{R}_{\mathrm{xz}}(\rho) = \frac{1}{2}\uconj{(\M{S}_{\Sop{X}}+\M{S}_{\Sop{Z}})}{\rho}$ \\ \hline
  10 & $\Sop{R}_{\mathrm{xy}}$ & $\Sop{R}_{\mathrm{xy}}(\rho) = \frac{1}{2}\uconj{(\M{S}_{\Sop{X}}+\M{S}_{\Sop{Y}})}{\rho}$ \\ \hline
  11 & $\Sop{\pi}_{\mathrm{x}}$ & $\Sop{\pi}_{\mathrm{x}}(\rho) = \frac{1}{4}\uconj{(\M{I}+\M{S}_{\Sop{X}})}{\rho}$ \\ \hline
  12 & $\Sop{\pi}_{\mathrm{y}}$ & $\Sop{\pi}_{\mathrm{y}}(\rho) = \frac{1}{4}\uconj{(\M{I}+\M{S}_{\Sop{Y}})}{\rho}$  \\ \hline
  13 & $\Sop{\pi}_{\mathrm{z}}$ & $\Sop{\pi}_{\mathrm{z}}(\rho) = \frac{1}{4}\uconj{(\M{I}+\M{S}_{\Sop{Z}})}{\rho}$ \\ \hline
  14 & $\Sop{\pi}_{\mathrm{yz}}$ & $\Sop{\pi}_{\mathrm{yz}}(\rho) = \frac{1}{4}\uconj{(\M{S}_{\Sop{Y}}+\imath\M{S}_{\Sop{Z}})}{\rho}$ \\ \hline
  15 & $\Sop{\pi}_{\mathrm{xz}}$ & $\Sop{\pi}_{\mathrm{xz}}(\rho) = \frac{1}{4}\uconj{(\M{S}_{\Sop{X}}+\imath\M{S}_{\Sop{Z}})}{\rho}$ \\ \hline
  16 & $\Sop{\pi}_{\mathrm{xy}}$ & $\Sop{\pi}_{\mathrm{xy}}(\rho) = \frac{1}{4}\uconj{(\M{S}_{\Sop{X}}+\imath\M{S}_{\Sop{Y}})}{\rho}$ \\
  \hline
\end{tabular}
\end{table}

\begin{figure}[t]
    \centering
    \subfloat[][Single-qubit channels]{
    \includegraphics[width=.45\textwidth]{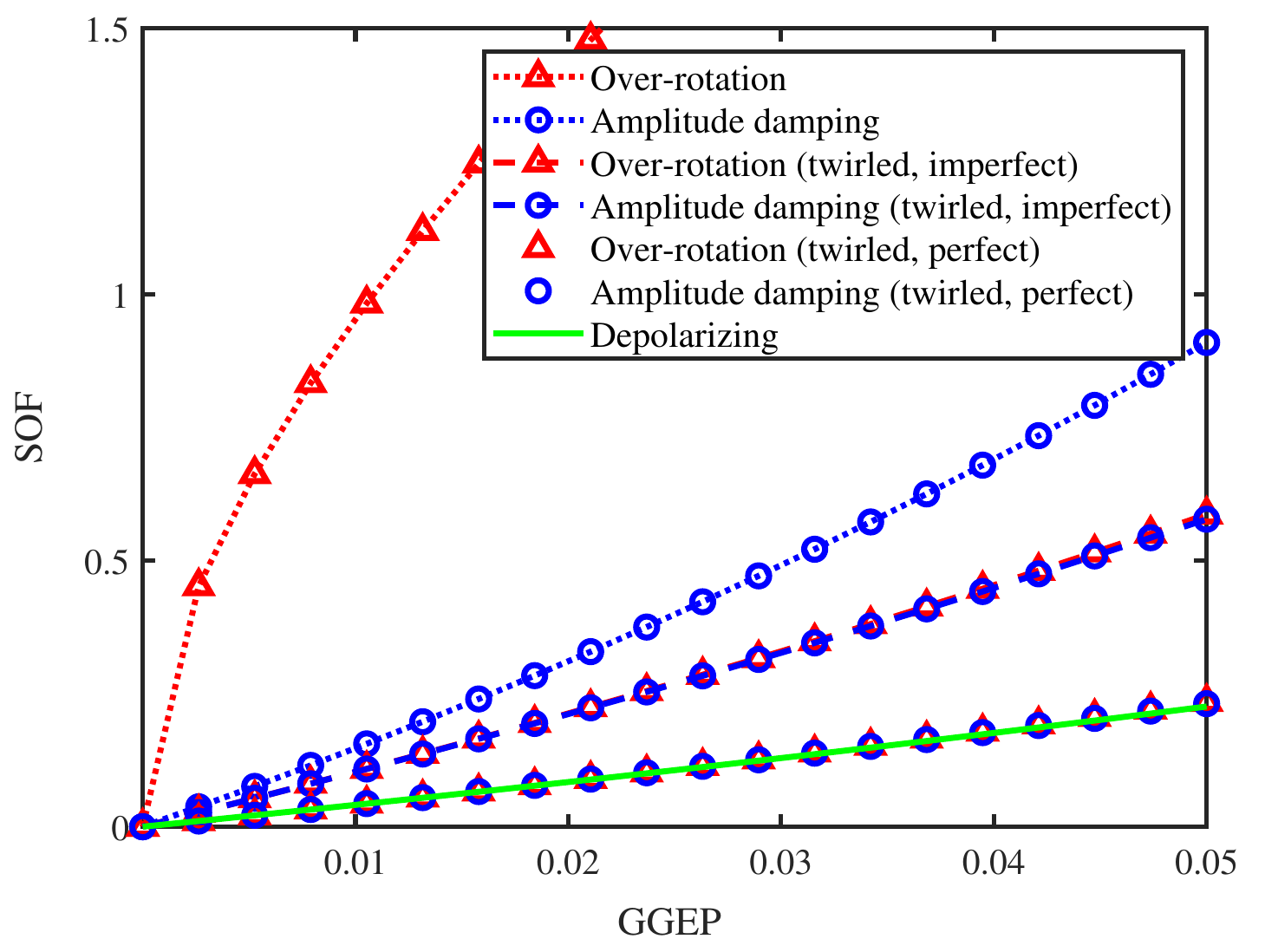}
    \label{fig:nonpauli_overhead_1}
    } \\
    \subfloat[][Two-qubit channels]{
    \includegraphics[width=.45\textwidth]{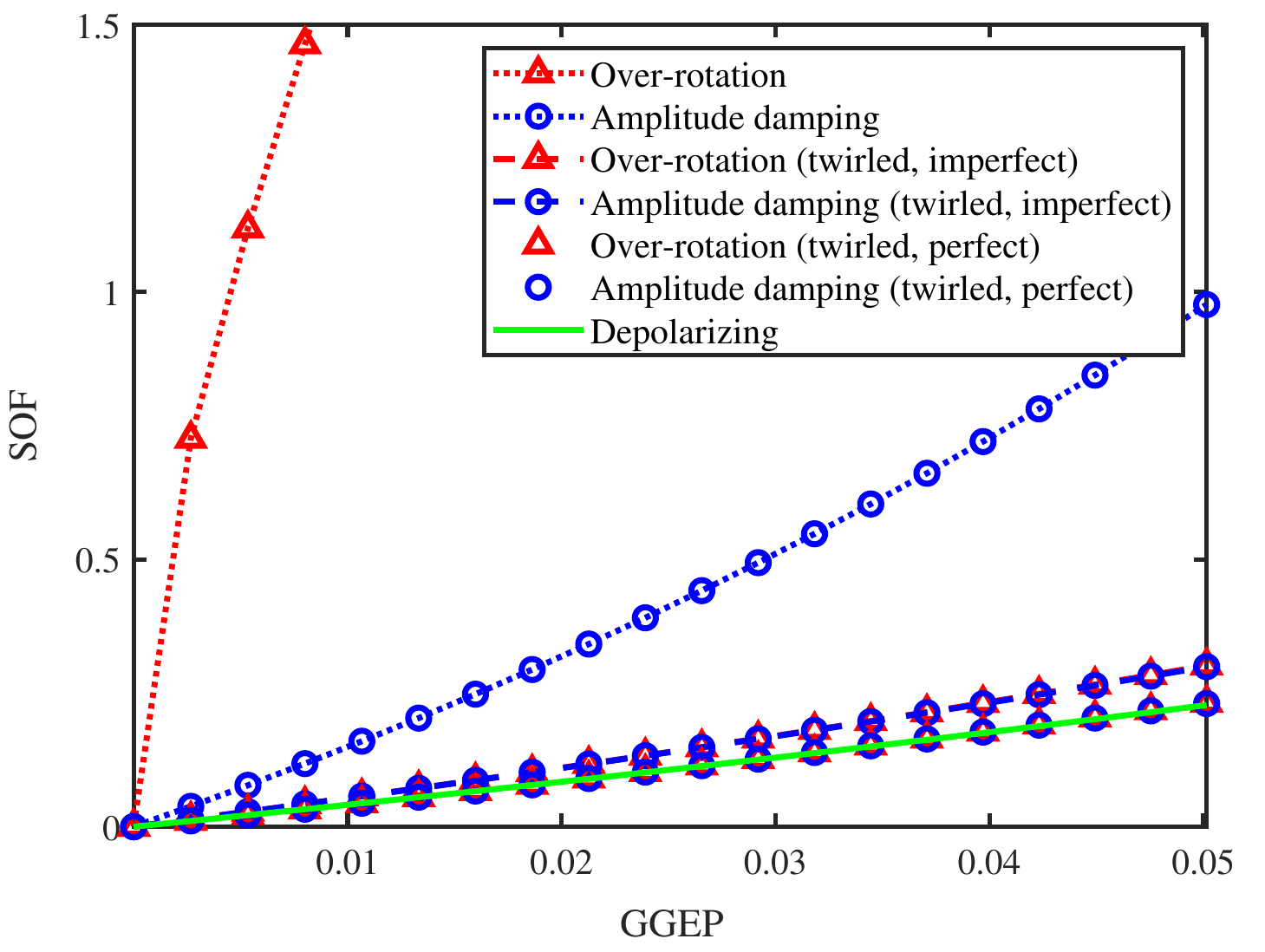}
    \label{fig:nonpauli_overhead_2}
    }
    \caption{\ac{sof} as a function of \ac{ggep} for uncoded gates inflicted by coherent error, amplitude damping and depolarizing channels having the same \ac{ggep}. The Pauli-twirled versions of coherent error and amplitude damping channels are also plotted for comparison.}
    \label{fig:nonpauli_overhead}
\end{figure}

We first characterize Proposition \ref{prop:pauli} and Proposition \ref{prop:dep_channels} via numerical examples. In Fig. \ref{fig:nonpauli_overhead}, the \ac{sof} vs. the \ac{ggep} is plotted for both single-qubit and two-qubit gates inflicted by coherent errors, amplitude damping and depolarizing channels, as detailed below. Here, the two-qubit channels are restricted to product channels, namely those constructed by the tensor product of two single-qubit channels. Specifically, a single-qubit amplitude damping channel $\Sop{C}_{\mathrm{d}amp}$ is characterized by \cite[Sec. 8.3.5]{ncbook}
\begin{equation}
\Sop{C}_{\mathrm{d}amp}(\rho) = \M{E}_0\rho\M{E}_0^\dagger + \M{E}_1\rho\M{E}_1^\dagger,
\end{equation}
where the operation elements are given by \cite[Sec. 8.3.5]{ncbook}
$$
\M{E}_0 = \left[
            \begin{array}{cc}
              1 & 0 \\
              0 & \sqrt{1-\delta} \\
            \end{array}
          \right],~\M{E}_1 = \left[
            \begin{array}{cc}
              0 & \sqrt{\delta} \\
              0 & 0 \\
            \end{array}
          \right],
$$
and the parameter $\delta$ is the amplitude damping probability of the channel, namely the probability that the channel turns an excited state $\ket{1}$ into the ground state $\ket{0}$. Notably, the amplitude damping channel is a non-Pauli triangular channel. The single-qubit coherent channel we consider here is the over-rotation channel, which takes the form of \cite{practical_qem}
\begin{equation}
\Sop{C}_{\mathrm{over}}(\rho) = \M{U}_{\mathrm{x}}\rho \M{U}_{\mathrm{x}}^\dagger,
\end{equation}
where
$$
\M{U}_{\mathrm{x}} = \left[
          \begin{array}{cc}
            \cos(\frac{4\phi}{\pi}) & \imath \sin(\frac{4\phi}{\pi}) \\
            \imath \sin(\frac{4\phi}{\pi}) & \cos(\frac{4\phi}{\pi}) \\
          \end{array}
        \right].
$$
The parameter $\phi$ controls the over-rotation angle of the channel, and $\imath=\sqrt{-1}$ denotes the imaginary unit.

Observe from Fig. \ref{fig:nonpauli_overhead} that the amplitude damping channels affecting both a single and two qubits have higher \acp{sof} than depolarizing channels. This corroborates Proposition \ref{prop:pauli} and Proposition \ref{prop:dep_channels}, which imply that depolarizing channels have the lowest \ac{sof} among all triangular channels. The over-rotation channels are not triangular channels, yet they exhibit the highest \ac{sof}. In general, their \acp{sof} would depend on the specific set of basis operators comprised by the matrix $\M{B}$. In fact, coherent channels are represented by unitary transformations. In light of this, in the ideal case that ``unitary rotation'' gates can be implemented without decoherence, they can be compensated in an overhead-free manner by simply applying its complex conjugate.

In Fig. \ref{fig:nonpauli_overhead}, we can also see the effect of quantum channel precoders, especially that of Pauli twirling. To elaborate, observe in both Fig. \ref{fig:nonpauli_overhead_1} and \ref{fig:nonpauli_overhead_2}, that the Pauli-twirled versions both of the coherent error and of the amplitude damping channels have almost the same \ac{sof} as depolarizing channels, provided that the gates used in the implementation of Pauli twirling are free from decoherence. By contrast, when Pauli twirling is implemented using realistic imperfect gates, the twirled channels have higher \ac{sof}, which is still lower than that of the amplitude damping channels. Another noteworthy fact is that imperfect Pauli twirling of two-qubit gates incurs relatively low overheads, compared to single-qubit gates. This is because two-qubit gates are more prone to decoherence than their single-qubit counterparts. Specifically, in this example we assume that every single-qubit gate (resp. two-qubit gate) has the same \ac{ggep}, and follow the convention that two-qubits gates have $10$ times higher \ac{ggep} than single-qubit gates \cite{fidelity_ratio}. In light of these results, Pauli twirling may be a preferable quantum channel precoder, especially for two-qubit gates.

\begin{figure}[t]
    \centering
   \begin{overpic}[width=.45\textwidth]{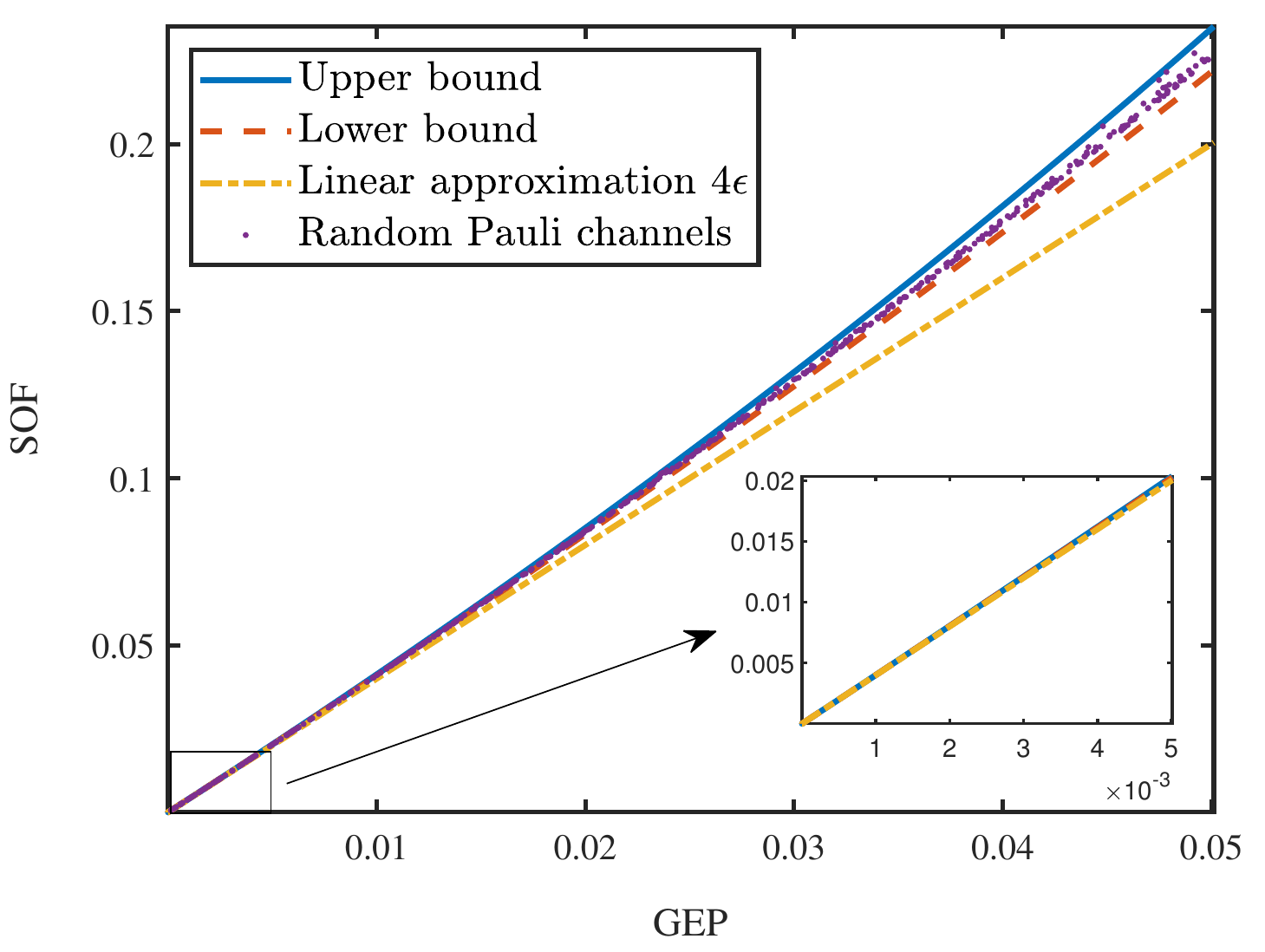}
        \put(43.75,67.85){\footnotesize  \eqref{overhead_ub}}
        \put(43.75,63.85){\footnotesize  \eqref{overhead_lb}}
    \end{overpic}
    \caption{Upper and lower bounds of the \ac{sof} as a function of \ac{gep} $\epsilon$ for Pauli channels over a single uncoded qubit.}
    \label{fig:pauli_overhead}
\end{figure}

Next we illustrate the bounds of the \ac{sof} of Pauli channels presented in Section \ref{ssec:bound_overhead}. As portrayed in Fig. \ref{fig:pauli_overhead}, given a fixed \ac{gep} $\epsilon$, all points representing the \acp{sof} of randomly produced Pauli channels fall between the upper bound \eqref{overhead_ub} and the lower bound \eqref{overhead_lb}. Moreover, it can be seen that when the \ac{gep} is less than $5\times 10^{-3}$, the upper and lower bounds are nearly identical, and a linear approximation of the \ac{sof} (i.e., $4\epsilon$) becomes rather accurate.

\subsection{Transversal gates protected by QECC}\label{ssec:numerical_qecc}
In this subsection, we investigate the \ac{sof} when \ac{qem} is applied to transversal logical gates protected by \acp{qecc} operating in their error-resilient regions. For the simplicity of presentation, we assume that all physical gates are subjected to the deleterious effect of depolarizing channels. Furthermore, we also assume that they can be decomposed into the tensor products of single-qubit depolarizing channels. Finally, we assume that all single-qubit gates have an identical \ac{gep}, and so do two-qubit gates.

As presented in Section \ref{ssec:post_qecc}, \ac{qem} and \acp{qecc} operating in their error-resilient regions can be beneficially amalgamated to reduce the \ac{sof}. Here we consider the amalgam of \ac{qem} and an $l$-stage concatenated Steane $[[7,1,3]]$ code. According to Proposition \ref{prop:critical_point}, for every \ac{qecc} operating in its error-resilient region, there would be several critical circuit sizes. To elaborate, if a quantum circuit contains gates that exceeds the $(l+1)$-th critical point, amalgamating \ac{qem} with the $l$-stage concatenated code will be more beneficial than relying on the $(l+1)$-stage concatenated code, and vice versa.

\begin{figure}[t]
    \centering
     \subfloat[][The critical points of circuit sizes (discussed in Proposition \ref{prop:critical_point})]{
    \includegraphics[width=.45\textwidth]{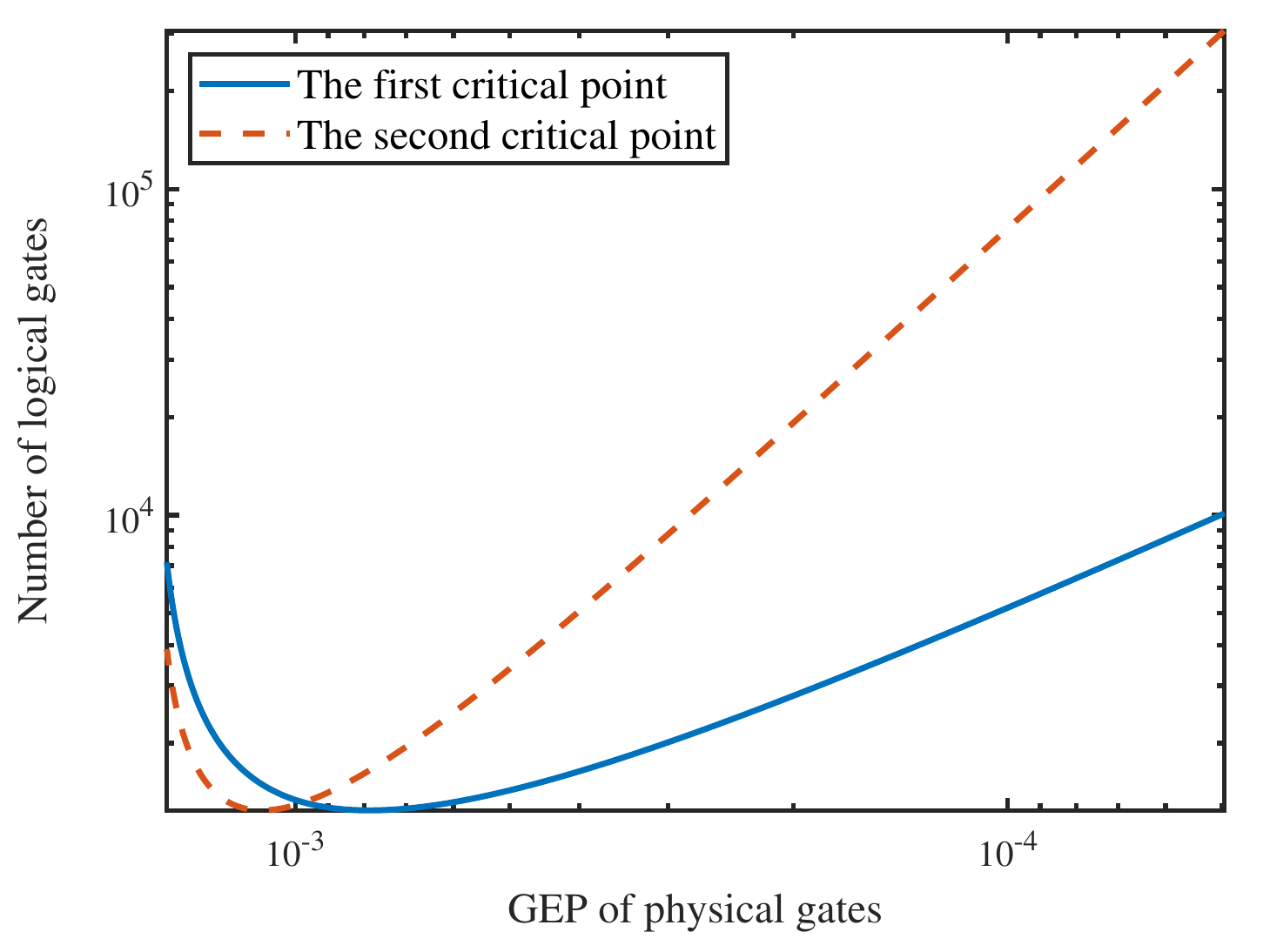}
    \label{fig:critical_lines}
    } \\
    \subfloat[][The most preferable \ac{qecc}-\ac{qem} scheme]{
    \includegraphics[width=.45\textwidth]{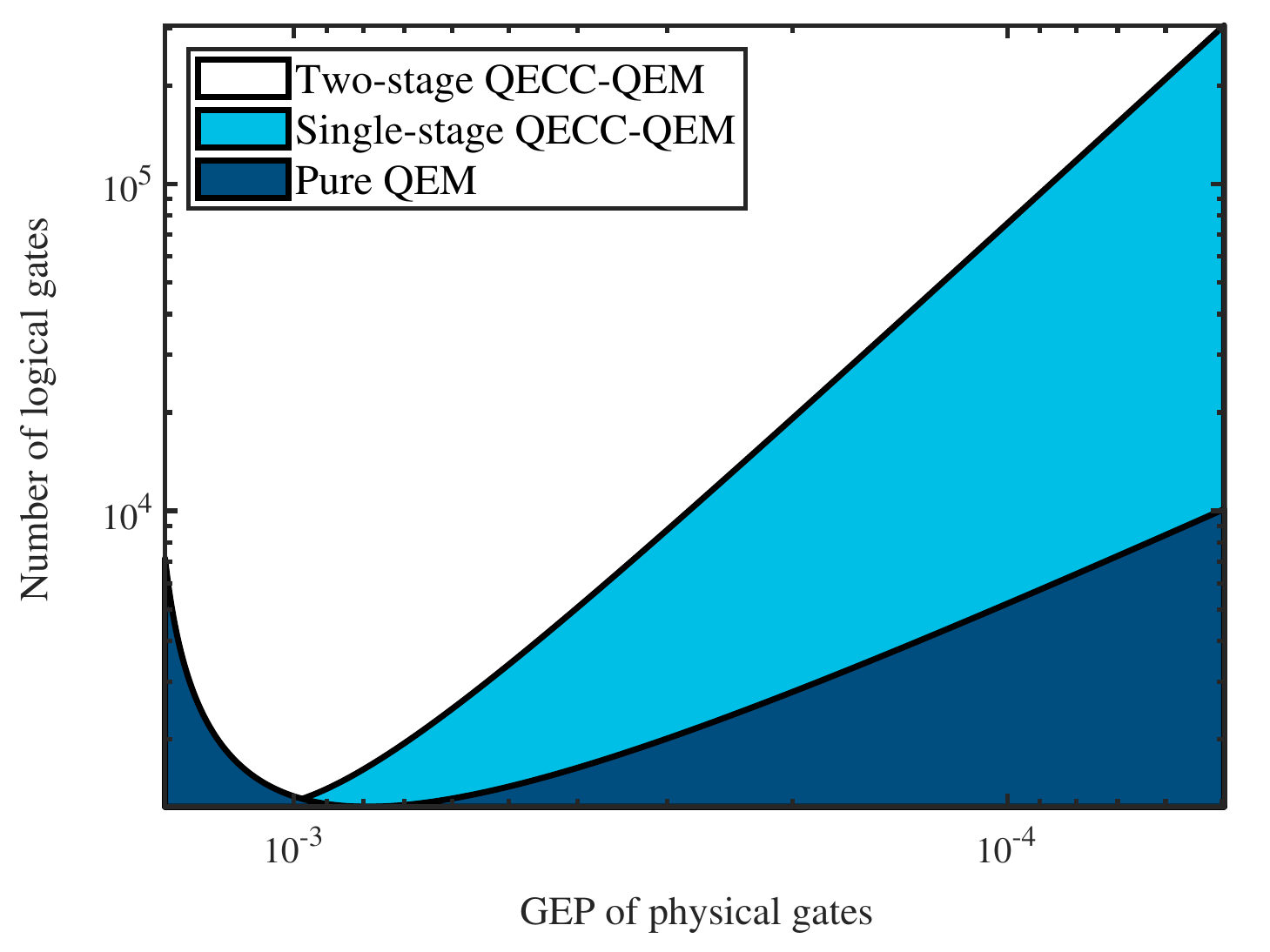}
    \label{fig:critical_areas}
    }
    \caption{Performance comparison between QECC-QEM schemes for quantum circuits containing various numbers of logical gates. Here the \ac{qecc} is chosen as Steane's $[[7,1,3]]$ code.}
    \label{fig:critical_pts}
\end{figure}

In Fig.~\ref{fig:critical_pts}, we compare the performance of three \ac{qecc}-\ac{qem} schemes for quantum circuits containing various number of logical gates. In Fig.~\ref{fig:critical_lines}, we demonstrate the aforementioned critical points of circuit size. In Fig.~\ref{fig:critical_areas} the areas having different shading represent the circuit configurations for which the corresponding \ac{qecc}-\ac{qem} scheme is the most preferable among the three candidates. As portrayed in the figures, for the case where the \ac{gep} of physical gates equals to $10^{-4}$, pure \ac{qem} is preferable when the circuit contains less than about $6\times 10^3$ gates. BY contrast, the single-stage \ac{qecc}-\ac{qem} combination may be a good choice for circuits containing between $6\times 10^3$ and $8\times 10^4$ gates. An interesting issue is that the pure \ac{qem} becomes the most preferable option when the \ac{gep} is higher than about $10^{-3}$, which is somewhat counter-intuitive. This may be attributed to the fact that the fault-tolerance threshold of the $[[7,1,3]]$ code under our assumptions used in this treatise is around $1.5 \times 10^{-3}$. When the \ac{gep} of physical gates is close to their threshold, the error-correction capability of \acp{qecc} is not fully exploited. To elaborate a little further, the term $[f^{(l)}(\epsilon)-f^{(l+1)}(\epsilon)]$ in \eqref{second_line_breaking} would typically be a non-monotonic function of $\epsilon$, with its maximum located close to the threshold. Hence, the critical circuit size increases as the \ac{gep} of physical gates decreases, provided that the \ac{gep} is sufficiently low.

\subsection{Gates protected by QEDC}
In this subsection, we consider a combined \ac{qedc}-\ac{qem} scheme, for which we make the same assumptions concerning the quantum gates as those stated in Section \ref{ssec:numerical_qecc}. Additionally, we assume that the \ac{gep} of two-qubit gates is $10$ times as high as that of single-qubit gates.

\begin{figure}[t]
    \centering
    \includegraphics[width=.45\textwidth]{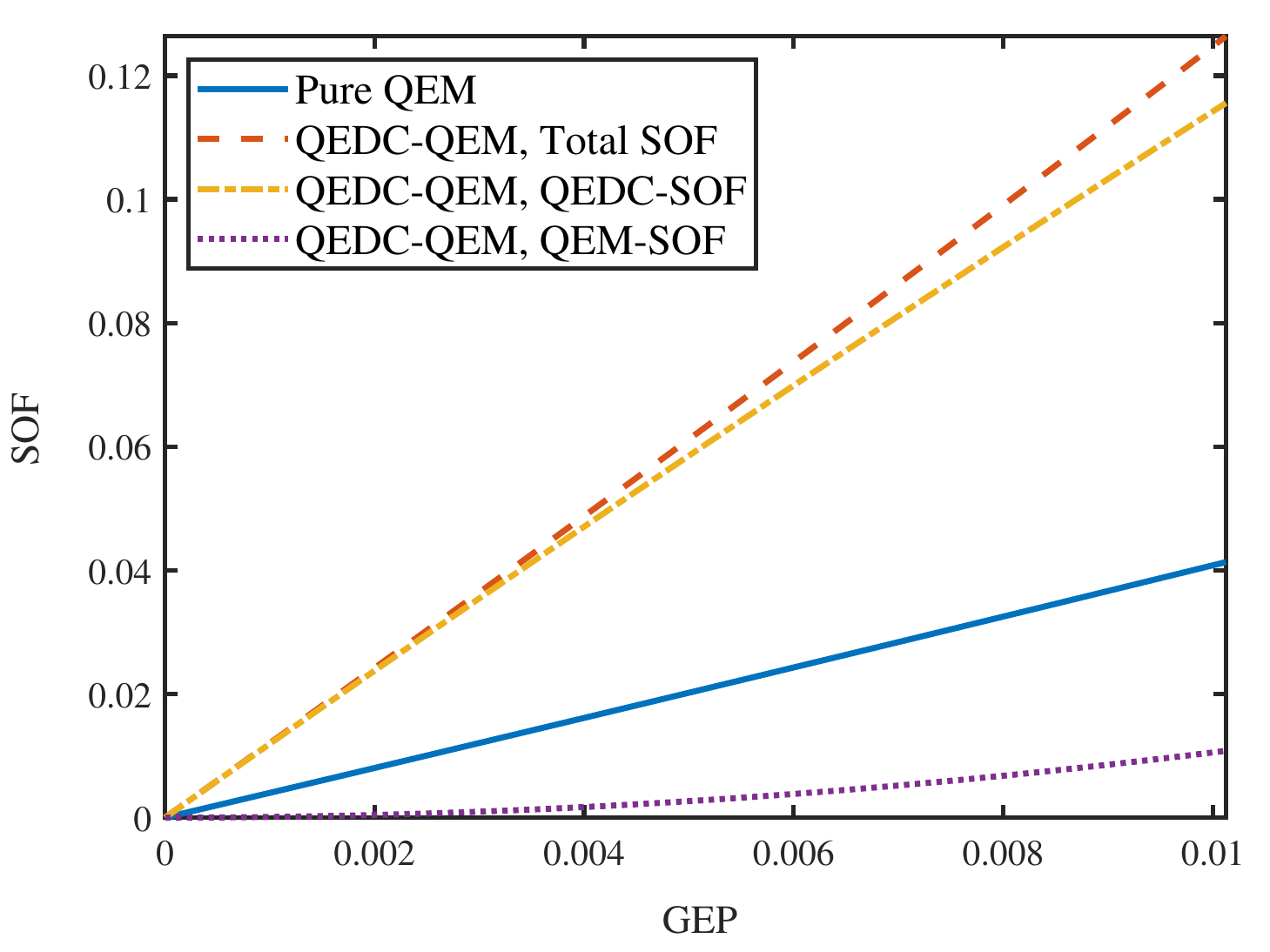}
    \caption{\ac{sof} comparison between pure \ac{qem} and amalgamated $[[4,2,2]]$ \ac{qedc}-\ac{qem} for a CNOT gate. The logical CNOT gates is implemented using the transversal gate configuration.}
    \label{fig:cnot_qedc}
\end{figure}

When logical gates are implemented transversally, according to the discussion in Section \ref{ssec:post_qedc}, the total \ac{sof} of the \ac{qedc}-\ac{qem} scheme would typically be even higher than that of pure \ac{qem}. This is demonstrated in Fig. \ref{fig:cnot_qedc}, where we consider the total \ac{sof} of a single transversal logical CNOT gate protected by the $[[4,2,2]]$ \ac{qedc}. It can be seen that most of the overhead is attributed to the \ac{qedc}-\ac{sof}, which is much higher than the overhead of pure \ac{qem}. By contrast, the \ac{qem} overhead in the \ac{qedc}-\ac{qem} scheme is significantly lower than that of pure \ac{qem}, implying that the post-selection fault-tolerance threshold of the $[[4,2,2]]$ code is higher than $0.01$.

\begin{figure}[t]
    \centering
    \includegraphics[width=.45\textwidth]{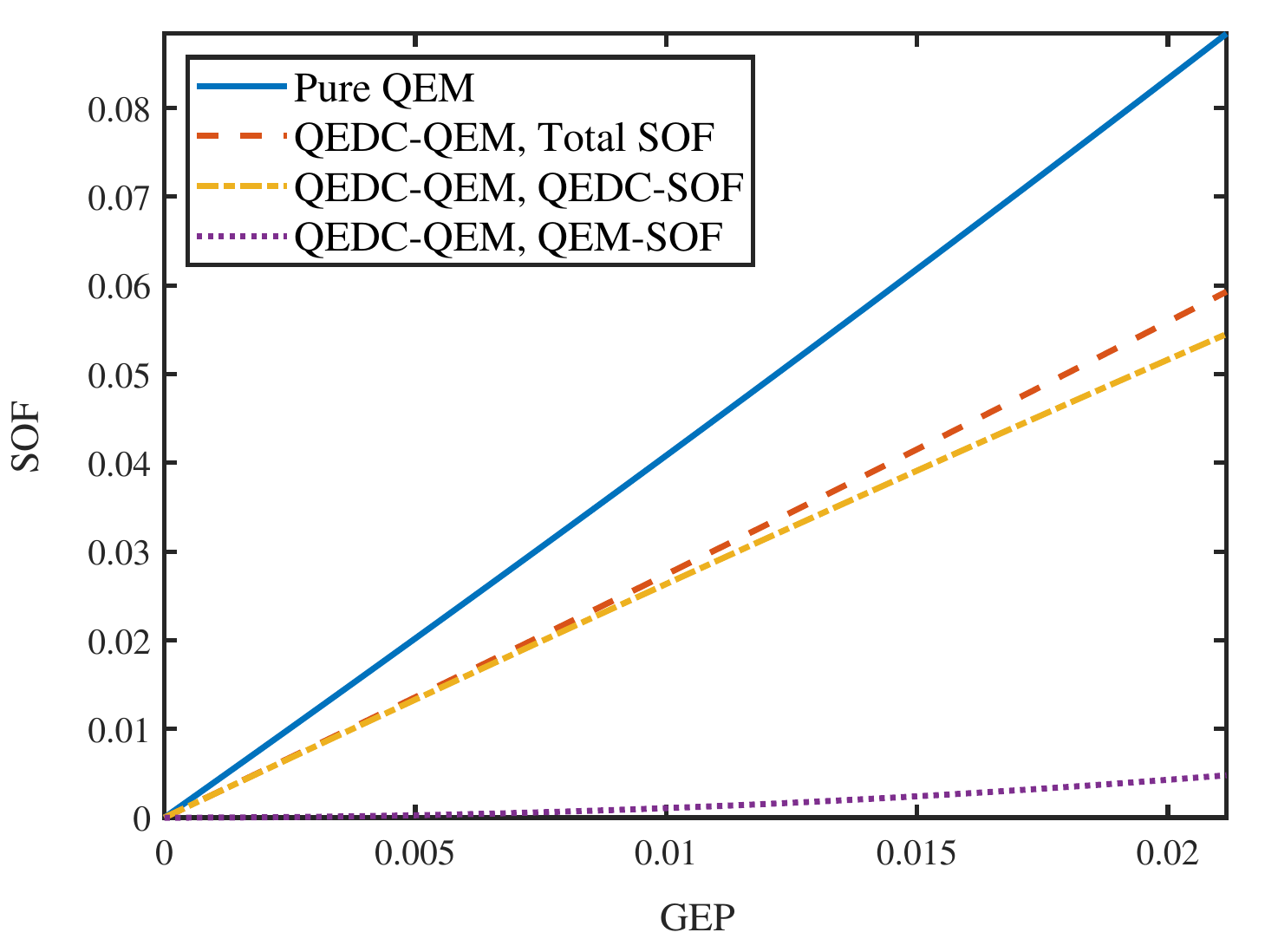}
    \caption{\ac{sof} comparison between pure \ac{qem} and amalgamated $[[4,2,2]]$ \ac{qedc}-\ac{qem} for a $\Sop{SWAP}\circ \Sop{H}^{\otimes 2}$ gate. The logical $\Sop{SWAP}\circ \Sop{H}^{\otimes 2}$ gate is implemented as illustrated in Fig. \ref{fig:swap_gate} instead of transversally.}
    \label{fig:swap_qedc}
\end{figure}

As suggested by Fig. \ref{fig:mapping_logical}, some non-transversal logical gates may even outperform transversal gates in terms of requiring lower \ac{qedc}-\ac{qem} \ac{sof}. In particular, we consider the specific logical gate of $\Sop{SWAP}\circ \Sop{H}^{\otimes 2}$ implemented in the manner illustrated in Fig. \ref{fig:swap_gate}. Since this implementation only involves single-qubit gates, the \ac{qedc}-\ac{sof} is significantly lower compared to the transversal implementation. Consequently, as portrayed in Fig. \ref{fig:swap_qedc}, the total \ac{qedc}-\ac{qem} overhead is lower than that of pure \ac{qem}. However, it is still not clear, whether this fact can justify the practical value of the \ac{qedc}-\ac{qem} scheme, since designing a low-sampling-overhead non-transversal implementation of all Clifford gates would require substantial effort. This may be an interesting topic deserving further investigation.

\section{Conclusions}\label{sec:conclusions}
We have presented a comprehensive analysis on the \ac{sof} of \ac{qem} under various channel conditions. For uncoded gates affected by errors modelled by general \ac{cptp} channels, we have shown that Pauli channels have the lowest \ac{sof} among all triangular channels (which includes the amplitude damping channels) having the same \ac{ggep}. Following this line of reasoning, we have shown furthermore that depolarizing channels have the lowest \ac{sof} in the family of all Pauli channels.

We have also conceived the \ac{qecc}-\ac{qem} as well as the \ac{qedc}-\ac{qem} schemes, and have shown that there exist several critical quantum circuits sizes, beyond which sophisticated codes having more concatenation stages is more preferable, and vice versa. Specifically, for \ac{qedc}-\ac{qem}, we have demonstrated that it may not be compatible with the popular transversal gate configuration, but they may still have beneficial applications, when the logical gates are appropriately designed.

\appendices
\section{Proof of Proposition 1}\label{sec:proof_pauli}
\begin{IEEEproof}
To prove our claim, it suffices to show that Pauli twirling does not increase the \ac{sof}. Without loss of generality, we assume that the specific columns corresponding to Pauli operators in $\M{B}$ are the first $4^n$ columns. First we note that the quasi-probability representation corresponding to any \ac{cptp} channel having Pauli transfer matrix representation $\M{C}$ may be expressed as:
\begin{equation}\label{ptm_original}
\V{\mu}_{\M{C}} = \M{B}^{-1}\vectorize{\M{C}^{-1}},
\end{equation}
while the quasi-probability representation corresponding to the Pauli-twirled channel is given by
\begin{equation}\label{ptwirled}
\V{\mu}_{\Sop{T}_{\Set{P}}\M{C}} = \M{B}^{-1} \vectorize{[\Sop{T}_{\Set{P}}\M{C}]^{-1}},
\end{equation}
where the super-operator $\Sop{T}_{\Set{P}}$ represents the Pauli twirling operation. We can rewrite \eqref{ptwirled} in a matrix form as
\begin{equation}\label{matrix_ptwirled}
\V{\mu}_{\Sop{T}_{\Set{P}}\M{C}} = \M{B}^{-1}\vectorize{[\ivectorize{\M{T}_{\Set{P}}\vectorize{\M{C}}}]^{-1}},
\end{equation}
where $\M{T}_{\Set{P}}$ denotes the matrix representation of the Pauli twirling operator. Recall that the Pauli-twirled channel is given by
\begin{equation}\label{ptwirler}
\Sop{T}_{\Set{P}}\M{C} = \frac{1}{4^n} \sum_{i=1}^{4^n} \Sop{P}_i \M{C}.
\end{equation}
Upon introducing $\M{B} = [\V{b}_1~\V{b}_2~\dotsc~\V{b}_{16^n}]$, the Pauli operator $\Sop{P}_i$ can be expressed in a matrix form as $\M{P}_i = \V{b}_i\V{b}_i^{\mathrm{T}}$. Thus the Pauli twirling operator can be represented as
\begin{equation}\label{ptwirler_ortho}
\M{T}_{\Set{P}}=\frac{1}{4^n}(\M{B}\M{I}_{\Set{T}})(\M{B}\M{I}_{\Set{T}})^{\mathrm{T}},
\end{equation}
with $\M{I}_{\Set{P}}$ being the following matrix
$$
\M{I}_{\Set{P}} = \left[
                    \begin{array}{cc}
                      \M{I}_{4^n} & \M{0}_{4^n\times (16^n-4^n)} \\
                      \M{0}_{(16^n-4^n) \times 4^n} & \M{0}_{(16^n-4^n)\times(16^n-4^n)} \\
                    \end{array}
                  \right].
$$
Thus we have
$$
\M{T}_{\Set{P}}\vectorize{\M{C}} = \vectorize{\mdiag{\M{C}}}.
$$
Since the Pauli transfer matrix $\M{C}$ is triangular, we have
$$
(\mdiag{\M{C}})^{-1} = (\mdiag{\M{C}^{-1}}),
$$
hence \eqref{matrix_ptwirled} can be simplified as
\begin{equation}\label{triangular_matrix}
\V{\mu}_{\Sop{T}_{\Set{P}}\M{C}} = \M{B}^{-1}\M{T}_{\Set{P}}\vectorize{\M{C}^{-1}}.
\end{equation}

Using \eqref{ptm_original} and \eqref{triangular_matrix}, we can show that if the statement of
$$
\V{\mu}_{\Sop{T}_{\Set{P}}\M{C}} = \M{T}_{\mathrm{L}} \M{B}^{-1} \vectorize{\M{C}^{-1}} = \M{T}_{\mathrm{L}} \V{\mu}_{\M{C}}
$$
holds for a certain matrix $\M{T}_{\mathrm{L}}$, the proof can be completed by showing that $\|\M{T}_{\mathrm{L}}\|_1\le 1$, since we have:
$$
\|\M{T}_{\mathrm{L}}\|_1 = \sup_{\V{x}} \frac{\|\M{T}_{\mathrm{L}}\V{x}\|_1}{\|\V{x}\|_1} \ge \frac{\|\V{\mu}_{\Sop{T}_{\Set{P}}\M{C}}\|_1}{\|\V{\mu}_{\M{C}}\|_1}.
$$
Next we construct the matrix $\M{T}_{\mathrm{L}}$ explicitly. Let us consider the QR decomposition \cite{GoluVanl96} of the matrix $\M{B}$
\begin{equation}\label{qrd}
\M{B} = \M{Q} \M{R},
\end{equation}
where $\M{Q}$ is an orthogonal matrix and $\M{R}$ is an upper triangular matrix. Substituting \eqref{ptwirler_ortho} and \eqref{qrd} into \eqref{triangular_matrix}, we have
\begin{equation}\label{twirled_cost}
\V{\mu}_{\Sop{T}_{\Set{P}}\M{C}} = \frac{1}{4^n}\M{I}_{\Set{P}}\M{R}^{\mathrm{T}} \M{Q}^{\mathrm{T}}\vectorize{\M{C}^{-1}}.
\end{equation}
Similarly, we can obtain
\begin{equation}\label{original_cost}
\V{\mu}_{\M{C}} = \M{R}^{-1} \M{Q}^{\mathrm{T}} \vectorize{\M{C}^{-1}}.
\end{equation}
Having compared \eqref{twirled_cost} and \eqref{original_cost}, we may observe that
$$
\M{T}_{\mathrm{L}} = \frac{1}{4^n} \M{I}_{\Set{P}} \M{R}^{\mathrm{T}}\M{R}.
$$
Upon introducing $\M{Q}=[\V{q}_1~\V{q}_2~\dotsc~\V{q}_{16^n}]$, the matrix $\M{R}$ can be represented by
\begin{equation}\label{qr_r}
\M{R} = \left[
          \begin{array}{cccc}
            \V{q}_1^{\mathrm{T}}\V{b}_1 & \V{q}_1^{\mathrm{T}}\V{b}_2 &\V{q}_1^{\mathrm{T}}\V{b}_3 & \dotsc \\
            0 & \V{q}_2^{\mathrm{T}}\V{b}_2 & \V{q}_2^{\mathrm{T}}\V{b}_3 & \dotsc \\
            0 & 0 & \V{q}_3^{\mathrm{T}}\V{b}_3 & \dotsc \\
            \vdots & \vdots & \vdots &\ddots
          \end{array}
        \right].
\end{equation}
Since Pauli operators are orthogonal to each other, the first $4^n$ columns in $\M{B}$  (i.e., $\V{b}_1$ through $\V{b}_{4^n}$) are also mutually orthogonal, meaning that
$$
\V{q}_i^{\mathrm{T}}\V{b}_j = \left\{
                           \begin{array}{ll}
                             \|\V{b}_i\|_2, & \hbox{$i=j$;} \\
                             0, & \hbox{$i\neq j$,}
                           \end{array}
                         \right.
$$
when $i=1,2,\dotsc,4^n$ and $j=1,2,\dotsc,4^n$.
Therefore we have
$$
\M{I}_{\Set{P}} \M{R}^{\mathrm{T}} = [\diag{\|\V{b}_1\|_2,\|\V{b}_2\|_2,\dotsc,\|\V{b}_{4^n}\|_2}~\V{0}_{4^n\times 16^n}].
$$

Upon introducing $m_{\M{B}} = \max_{i=1,\dotsc,16^n} \|\V{b}_i\|_2$, from \eqref{qr_r} we have
\begin{subequations}\label{norm1_1}
\begin{align}
\|\M{T}_{\mathrm{L}}\|_1&\le \frac{m_{\M{B}}}{4^n}\max_{i=1,\dotsc,16^n} \sum_{j=1}^{\min\{4^n,i\}} |\V{q}_j^{\mathrm{T}}\V{b}_i|\\
&\le \frac{m_{\M{B}}}{4^n}  \max_{i=1,\dotsc,16^n}\|\V{w}_i\|_1\max_{j=1,\dotsc,4^n}\|\V{b}_j\|_2 \\
&\le \frac{m_{\M{B}}^2}{4^n}  \max_{i=1,\dotsc,16^n}\|\V{w}_i\|_1,
\end{align}
\end{subequations}
where $\V{w}_i\in \mathbb{R}^{\min\{4^n,i\}}$, and
\begin{equation}\label{tilde_b_i}
[\V{w}_i]_j = \frac{\V{b}_j^{\mathrm{T}}\V{b}_i}{\|\V{b}_j\|_2^2}.
\end{equation}
Using \eqref{tilde_b_i}, the projection of $\V{b}_i$ onto the space spanned by Pauli operators can be expressed as
$$
\begin{aligned}
\M{T}_{\Set{P}}\V{b}_i &= [\M{Q}]_{:,1:4^n}[\M{Q}]_{:,1:4^n}^{\mathrm{T}} \V{b}_i \\
&= \sum_{j=1}^{\min\{4^n,i\}} (\V{q}_j^{\mathrm{T}}\V{b}_i) \V{q}_j\\
&= \sum_{j=1}^{\min\{4^n,i\}}[\V{w}_i]_j \V{b}_j.
\end{aligned}
$$
Since $\M{B}_i=\ivectorize{\V{b}_i}$ corresponds to a \ac{cptni} operator, the complete positiveness\footnote{We say that an operator is complete positive if it maps positive semidefinite matrices to positive semidefinite matrices.} implies $[\V{w}_i]_j\ge 0$ for all $i$ and $j$, and the ``trace--non-increasing'' property implies $\sum_{j}[\V{w}_i]_j\le 1$. Therefore we have $\|\V{w}_i\|_1\le 1$ for all $i$, and hence
\begin{equation}\label{norm1_3}
\|\M{T}_{\mathrm{L}}\|_1\le 4^{-n} m_{\M{B}}^2.
\end{equation}
Note that $m_{\M{B}}^2 = \max_i \|\M{B}_i\|_{\mathrm{F}}^2$. By consider the operator-sum decomposition of $\M{B}_i$, we see that
\begin{subequations}\label{fro}
\begin{align}
\|\M{B}_i\|_{\mathrm{F}}^2 &=\|\sum_i (\M{K}_i^*\otimes \M{K}_i) \|_{\mathrm{F}}^2 \\
&\le \left(\sum_i \|\M{K}_i^*\otimes \M{K}_i\|_{\mathrm{F}}\right)^2 \\
&=\left(\sum_i {\mathrm{Tr}}\{\M{K}_i^\dagger \M{K}_i\}\right)^2 \\
&\le 4^n, \label{last_line_fro}
\end{align}
\end{subequations}
where \eqref{last_line_fro} follows from \eqref{kraus_cptni}.

Combining \eqref{norm1_3} and \eqref{fro}, we can see that $\|\M{T}_{\mathrm{L}}\|_1 \le 1$, hence the proof is completed.
\end{IEEEproof}

\section{Proof of Proposition 2}\label{sec:proof_dep_channels}
To facilitate our further analysis, we denote the Hadamard transform matrix on the space of Pauli transfer matrix representation of an $n$-qubit system as $\M{H}_n\in\mathbb{R}^{4^n\times 4^n}$. The corresponding inverse Hadamard transform is denoted by $\M{H}^{-1}_{\mathrm{n}} = \frac{1}{4^n}\M{H}$. We omit the subscript $n$, whenever there is no confusion.  Given these notations, according to \eqref{reduce_qpr}, the simplified quasi-probability representation vector for a Pauli channel $\Sop{C}$ can be expressed as
\begin{equation}
\tilde{\V{\mu}}_{\Sop{C}}= \M{H}^{-1}(1/\V{c}).
\end{equation}
Since the channel is \ac{cptp}, the vector $1/\V{c}$ satisfies $[1/\V{c}]_1 = 1$. Hence we define $\V{\zeta}\in\mathbb{R}^{4^n-1}$ so that
$$
1/\V{c} = \V{1}+[0~\V{\zeta}^{\mathrm{T}}]^{\mathrm{T}}.
$$
For comparison, we consider a depolarizing channel $\Sop{L}$ having quasi-probability representation corresponding to $\tilde{\V{\mu}}_{\Sop{L}}$ given by
\begin{equation}\label{dep}
\tilde{\V{\mu}}_{\Sop{L}} = \M{H}^{-1}(\V{1}+[0~\bar{\zeta}\V{1}^{\mathrm{T}}]^{\mathrm{T}}),
\end{equation}
where $\bar{\zeta} \triangleq \frac{1}{4^n-1}\sum_{i=1}^{4^n-1}\zeta_i$. One may verify that the channel $\Sop{L}$ characterized by \eqref{dep} is a depolarizing channel with its \ac{ggep} satisfying
\begin{equation}\label{dep_fidelity}
1-\epsilon(\Sop{L}) = \frac{1}{4^n}\Big(1+\frac{4^n-1}{1+\bar{\zeta}}\Big).
\end{equation}
By contrast, the \ac{ggep} of the channel $\Sop{C}$ satisfies
\begin{equation}\label{original_fidelity}
\begin{aligned}
1-\epsilon(\Sop{C}) &= \frac{1}{4^n} \V{1}^{\mathrm{T}}\V{c} \\
&=\frac{1}{4^n}\Big(1+\sum_{i=1}^{4^n-1}\frac{1}{1+\zeta_i}\Big).
\end{aligned}
\end{equation}
Since $\zeta_i>-1,~\forall i$, from \eqref{dep_fidelity} and \eqref{original_fidelity} we have
\begin{equation}
\epsilon(\Sop{L}) \ge \epsilon(\Sop{C})
\end{equation}
due to the convexity of $f(x) = (1+x)^{-1}$, when $x>-1$.

Next we show that $\|\tilde{\V{\mu}}_{\Sop{C}}\|_1\ge \|\tilde{\V{\mu}}_{\Sop{L}}\|_1$. Note that the vector $\tilde{\V{\mu}}_{\mathrm{C}}$ can be decomposed as
\begin{equation}
\tilde{\V{\mu}}_{\Sop{C}} = \tilde{\V{\mu}}_{\Sop{L}} + \M{H}^{-1}\V{r},
\end{equation}
where
$$
\begin{aligned}
\V{r} &= 1/\V{c}-1/\V{l}\\
&= [0~(\V{\zeta} - \bar{\zeta}\V{1})^{\mathrm{T}}]^{\mathrm{T}}.
\end{aligned}
$$
From the definition of $\bar{\zeta}$ we see that $\V{1}^{\mathrm{T}}\V{r}=0$, hence $[\tilde{\V{\mu}}_{\Sop{C}}]_1 = [\tilde{\V{\mu}}_{\Sop L}]_1\triangleq \mu_1$. In addition, we have
\begin{equation}
\V{1}^{\mathrm{T}}\tilde{\V{\mu}}_{\Sop{C}} = \V{1}^{\mathrm{T}}\tilde{\V{\mu}}_{\Sop{L}} = [1/\V{c}]_1 = 1,
\end{equation}
since the channels are \ac{cptp}. Therefore we obtain
\begin{equation}\label{equal_sum}
\V{1}^{\mathrm{T}}[\tilde{\V{\mu}}_{\Sop{C}}]_{2:4^n} = \V{1}^{\mathrm{T}}[\tilde{\V{\mu}}_{\Sop{L}}]_{2:4^n}.
\end{equation}
The $1$-norm of $\tilde{\V{\mu}}_{\Sop{L}}$ can be calculated explicitly as
\begin{equation}
\|\tilde{\V{\mu}}_{\Sop{L}}\|_1 = \mu_1 - \sum_{i=2}^{4^n} [\tilde{\V{\mu}}_{\Sop{L}}]_i.
\end{equation}
For $\tilde{\V{\mu}}_{\Sop{C}}$, we denote the sign of its $i$-th entry as $s_i$, thus
\begin{equation}
\begin{aligned}
\|\tilde{\V{\mu}}_{\Sop{C}}\|_1 &= \mu_1 + \sum_{i=2}^{4^n} s_i[\tilde{\V{\mu}}_{\Sop{C}}]_i \\
&=\mu_1 +\V{1}_{+}^{\mathrm{T}}[\tilde{\V{\mu}}_{\Sop{C}}]_{2:4^n} - \V{1}_{-}^{\mathrm{T}}[\tilde{\V{\mu}}_{\Sop{C}}]_{2:4^n},
\end{aligned}
\end{equation}
where
$$
[\V{1}_{+}]_i = \left\{
                  \begin{array}{ll}
                    1, & \hbox{$s_{i-1}>0$;} \\
                    0, & \hbox{$s_{i-1}<0$.}
                  \end{array}
                \right. {\mathrm{and}}~[\V{1}_{-}]_i = \left\{
                  \begin{array}{ll}
                    1, & \hbox{$s_{i-1}<0$;} \\
                    0, & \hbox{$s_{i-1}>0$.}
                  \end{array}
                \right.
$$
From \eqref{equal_sum} we have
$$
\V{1}_{+}^{\mathrm{T}}[\tilde{\V{\mu}}_{\Sop{C}}]_{2:4^n} + \V{1}_{-}^{\mathrm{T}}[\tilde{\V{\mu}}_{\Sop{C}}]_{2:4^n} = \sum_{i=2}^{4^n}[\tilde{\V{\mu}}_{\Sop{L}}]_i.
$$
Hence
\begin{equation}
\begin{aligned}
\|\tilde{\V{\mu}}_{\Sop{C}}\|_1 &= \|\tilde{\V{\mu}}_{\Sop{L}}\|_1 + 2\V{1}_{+}^{\mathrm{T}}[\tilde{\V{\mu}}_{\Sop{C}}]_{2:4^n} \\
& \ge \|\tilde{\V{\mu}}_{\Sop{L}}\|_1.
\end{aligned}
\end{equation}

Finally, since $\epsilon(\Sop{L})\ge \epsilon(\Sop{C})$, we may construct a depolarizing channel $\Sop{L}^\prime$ characterized by $\epsilon(\Sop{L}^{\prime})=\epsilon(\Sop{C})$, while satisfying
$$
\|\tilde{\V{\mu}}_{\Sop{L}}\|_1 \ge \|\tilde{\V{\mu}}_{\Sop{L}^\prime}\|_1.
$$
Hence the proof is completed.

\section{The Values of $\V{\eta}_{\Sop{C}}$ For Basic Pauli Channels}\label{sec:basic_channels}
To elaborate further on the intuition about $\V{\eta}_{\Sop{C}}$, the values of $\V{\eta}_{\Sop{C}}$ corresponding to some single-qubit Pauli channels are listed in Table \ref{tbl:basic_channels}.

\begin{table}[h]
\captionsetup{justification=centering, labelsep=newline}
\centering
\caption{ $\V{\eta}_{\Sop{C}}$ values of single-qubit Pauli channels.}
\label{tbl:basic_channels}
\begin{tabular}{|c|c|c|}
  \hline
  \footnotesize
  Channel & Parameters & $\V{\eta}_{\Sop{C}}$ \\ \hline
  Depolarizing & Depolarizing prob. $p$ & $[1-p,~\frac{p}{3},~\frac{p}{3},~\frac{p}{3}]^{\mathrm{T}}$ \\ \hline
  Bit-flip & Bit-flip prob. $p_{\mathrm{x}}$ & $[1-p_{\mathrm{x}},~p_{\mathrm{x}},~0,~0]^{\mathrm{T}}$  \\ \hline
  Phase-flip & Phase-flip prob. $p_{\mathrm{z}}$ & $[1-p_{\mathrm{z}},~0,~0,~p_{\mathrm{z}}]^{\mathrm{T}}$  \\ \hline
\end{tabular}
\end{table}

\section{Proof of Proposition 3}\label{sec:proof_upper_pauli}
\begin{IEEEproof}
From \eqref{sampling_coeff_pauli}, we have
\begin{equation}
\gamma_{\Sop{C}} = \|\M{C}^{-1}\V{\alpha}\|_1^2-1,
\end{equation}
where
\begin{equation}\label{decomp_prw}
\M{C} = (1-\epsilon)\M{I}+\M{A}.
\end{equation}
 Since the graph $\Set{G}$ is symmetric, each column (resp. row) of $\M{C}$ can be obtained by permuting the first column (resp. row) of $\M{C}$. Thus we have
\begin{equation}\label{simplified_overhead_pauli}
\gamma_{\Sop{C}} = \|\M{C}^{-1}\|_1^2-1.
\end{equation}
From \eqref{decomp_prw} we can obtain
\begin{subequations}\label{inv_prw}
\begin{align}
\M{C}^{-1} &= ((1-\epsilon)\M{I} + \epsilon\tilde{\M{A}})^{-1} \\
&=\frac{1}{1-\epsilon} \left(\M{I}+\frac{\epsilon}{1-\epsilon} \tilde{\M{A}}\right)^{-1} \\
&=\frac{1}{1-\epsilon} \sum_{n=0}^{\infty} (-1)^n\left(\frac{\epsilon}{1-\epsilon}\right)^n \tilde{\M{A}}^n, \label{lastline_prw}
\end{align}
\end{subequations}
where $\tilde{\M{A}}\triangleq \epsilon^{-1} \M{A}$, and \eqref{lastline_prw} is obtained using the matrix inversion lemma. Exploiting the sub-multiplicativity of matrix $p$-norms \cite[Chap. 5]{matrix_analysis}, we have
\begin{equation}\label{norm_prw}
\|\tilde{\M{A}}^n\|_1 \le \|\tilde{\M{A}}\|_1^n =1,
\end{equation}
where the equality follows from the fact that $\V{1}^{\mathrm{T}} \tilde{\M{A}}=1$ and that all entries in $\tilde{\M{A}}$ are non-negative. Substituting \eqref{norm_prw} into \eqref{inv_prw}, we have
\begin{subequations}
\begin{align}
\|\M{C}^{-1}\|_1 &\le \frac{1}{1-\epsilon}\sum_{n=0}^{\infty} \left\|(-1)^n \left(\frac{\epsilon}{1-\epsilon}\right)^n \tilde{\M{A}}^n\right\|_1 \\
&\le \frac{1}{1-\epsilon} \sum_{n=0}^{\infty} \left(\frac{\epsilon}{1-\epsilon}\right)^n \label{secondline_norm2}= \frac{1}{1-2\epsilon}.
\end{align}
\end{subequations}
Therefore, from \eqref{simplified_overhead_pauli} we obtain
\begin{equation}
\gamma_{\Sop{C}} \le \left(\frac{1}{1-2\epsilon}\right)^2-1 = 4\epsilon\cdot\frac{1-\epsilon}{(1-2\epsilon)^2}.
\end{equation}

To show that channels having a single type of error achieve the equality, we note that
\begin{equation}
\Sop{P}_i^{2n} = \Sop{I},~\Sop{P}_i^{2n+1} = \Sop{P}_i
\end{equation}
holds for any Pauli operator $\Sop{P}_i$. In light of this, the inverse of their PRW matrix can be shown to satisfy
$$
\M{C}^{-1} = \frac{1}{1-\epsilon}\sum_{n=0}^{\infty}\left\{\left(\frac{\epsilon}{1-\epsilon}\right)^{2n} \M{I}-\left(\frac{\epsilon}{1-\epsilon}\right)^{2n+1}\tilde{\M{A}}\right\}.
$$
Therefore we have
$$
\begin{aligned}
\|\M{C}^{-1}\|_1 &=\frac{1}{1-\epsilon}\sum_{n=0}^{\infty}\left\{\left(\frac{\epsilon}{1-\epsilon}\right)^{2n}+\left(\frac{\epsilon}{1-\epsilon}\right)^{2n+1}\left\|\tilde{\M{A}}\right\|_1\right\} \\
&=\frac{1}{1-\epsilon} \sum_{n=0}^{\infty} \left(\frac{\epsilon}{1-\epsilon}\right)^n,
\end{aligned}
$$
which is identical to \eqref{secondline_norm2}. Hence the proof is completed.
\end{IEEEproof}

\section{Proof of Proposition 4}\label{sec:proof_memoryless}
\begin{IEEEproof}
Let $\V{\eta}$ be the probability vector of a Pauli channel. We first show that the function
\begin{equation}
f(\V{\eta}) = \left\|\M{H}^{-1}_1\left(1/(\M{H}_1[1-\epsilon~\V{\eta}^{\mathrm{T}}]^{\mathrm{T}})\right)\right\|_1
\end{equation}
is Schur-convex with respect to $\V{\eta}$. We proceed by first decomposing $f(\V{\eta})$ as
\begin{equation}
f(\V{\eta}) = g\{h_1[h_2(\V{\eta})]\},
\end{equation}
where
\begin{subequations}
\begin{align}
g(\V{x}) &=\left\|\M{H}^{-1}_1[1~\V{x}^{\mathrm{T}}]^{\mathrm{T}}\right\|_1~~(\V{x}\succeq \V{1}),\\
h_1(\V{x}) &= 1/\V{x}~~(\V{0}\preceq \V{x}\preceq \V{1}), \\
h_2(\V{\eta}) &= \M{H}_1[1-\epsilon~\V{\eta}^{\mathrm{T}}]^{\mathrm{T}}~~\left(\V{0}\preceq \V{\eta}\preceq \frac{\epsilon}{3}\V{1}\right).
\end{align}
\end{subequations}
Since $h_2(\V{\eta})$ is an affine function of $\V{\eta}$, we see that
\begin{equation}
\begin{aligned}
h(\V{\eta}) &= h_1[h_2(\V{\eta})] \\
&=1/\left(\M{H}_1[1-\epsilon~\V{\eta}^{\mathrm{T}}]^{\mathrm{T}}\right)
\end{aligned}
\end{equation}
is element-wise convex with respect to $\V{\eta}$. Therefore, to show that $f(\V{\eta})$ is Schur-convex, it suffices to show that $g(\V{x})$ is Schur-convex and increasing for $\V{x}$ satisfying $\V{x}\succeq \V{1}$.

Next we show the Schur-convexity of $g(\V{x})$. Note that
\begin{subequations}
\begin{align}
g(\V{x}) &= \left\|\M{H}^{-1}_1[1~\V{x}^{\mathrm{T}}]^{\mathrm{T}}\right\|_1 \\
&= 1+\V{1}^{\mathrm{T}}(\V{x}-1) + \frac{1}{4}\left\|\tilde{\M{H}}_1(\V{x}-1)\right\|_1,
\end{align}
\end{subequations}
where
\begin{equation}\label{tildeh}
\tilde{\M{H}}_1 = \left[
  \begin{array}{ccc}
    -1 & 1 & -1 \\
    1 & -1 & -1 \\
    -1 & -1 & 1 \\
  \end{array}
\right]
\end{equation}
is obtained by removing the first row and the first column from $\M{H}_1$. Since doubly stochastic transformations do not affect the term $\V{1}^{\mathrm{T}}(\V{x}-1)$, the problem is reduced to showing the Schur-convexity of $\|\tilde{\M{H}}_1\V{x}\|_1$ for $\V{x}\succeq \V{0}$. To facilitate the analysis, we utilize $\epsilon = \V{1}^{\mathrm{T}}\V{x}$ and define $\V{x} = [x_1~x_2~x_3]^{\mathrm{T}}$. Now we see that
\begin{equation}
\tilde{\M{H}}_1\V{x} = [2x_2-\epsilon~2x_1-\epsilon~2x_3-\epsilon]^{\mathrm{T}},
\end{equation}
hence
\begin{equation}
\|\tilde{\M{H}}_1\V{x}\|_1 = \|2\V{x}-\epsilon\|_1.
\end{equation}
For fixed $\epsilon$, $\|2\V{x}-\epsilon\|_1$ is convex with respect to $\V{x}$. In addition, it is also a symmetric function of $\V{x}$, meaning that its value is unchanged upon permutation of $\V{x}$. Therefore, $g(\V{x})$ is Schur-convex.

To show that $g(\V{x})$ is increasing, we calculate the gradient of $g(\V{x})$ as
\begin{equation}
\begin{aligned}
\nabla_{\V{x}}g(\V{x}) &= \frac{\partial}{\partial \V{x}}\left(\V{1}^{\mathrm{T}}(\V{x}-1) + \|\tilde{\M{H}}_1(\V{x}-1)\|_1\right) \\
&= \V{1} + \tilde{\M{H}}_1{\mathrm{sgn}}(\tilde{\M{H}}_1(\V{x}-1))
\end{aligned}
\end{equation}
where ${\mathrm{sgn}}(\cdot)$ is the sign function satisfying
$$
\begin{aligned}
\left[{\mathrm{sgn}}(\V{x})\right]_i = \left\{
                         \begin{array}{ll}
                           -1, & \hbox{$x_i<0$;} \\
                           1, & \hbox{$x_i>0$;} \\
                           0, & \hbox{$x_i=0$.}
                         \end{array}
                       \right.
\end{aligned}
$$
After some manipulation, one can verify that $ \tilde{\M{H}}_1{\mathrm{sgn}}(\tilde{\M{H}}_1(\V{x}-1))\succeq -\V{1}$ according to \eqref{tildeh}, hence $g(\V{x})$ is increasing.

Given that the \ac{sof} of single-qubit Pauli channels is Schur-convex, we may generalize the result to $n$-qubit memoryless Pauli channels. To elaborate, the PRW matrix of an $n$-qubit memoryless Pauli channel can be expressed as
\begin{equation}
\M{C}(\V{\eta}_{\Sop{C}}) = \bigotimes_{i=1}^n \M{C}_i(\V{\eta}_{\Sop{C}_i}),
\end{equation}
where $\M{C}_i(\V{\eta}_{\Sop{C}_i})$ corresponds to the partial channel of the $i$-th qubit. Thus
\begin{equation}
\left\|\M{C}^{-1}(\V{\eta}_{\Sop{C}})\right\|_1= \prod_{i=1}^n\|\M{C}_i^{-1}(\V{\eta}_{\Sop{C}_i})\|_1.
\end{equation}
Since $\|\M{C}_i^{-1}(\V{\eta}_{\Sop{C}_i})\|_1$ is Schur-convex with respect to the corresponding probability vector $\V{\eta}_{\Sop{C}_i}$, we have
\begin{equation}
\|\M{C}_i^{-1}(\V{\eta}_{\Sop{C}_i})\|_1 \ge \|\M{C}_i^{-1}(\M{Q}_i\V{\eta}_{\Sop{C}_i})\|_1
\end{equation}
for any doubly stochastic matrix $\M{Q}_i$. Therefore
\begin{equation}
\left\|\M{C}^{-1}(\V{\eta}_{\Sop{C}})\right\|_1 \ge \prod_{i=1}^n\left\|\M{C}_i^{-1}(\M{Q}_i\V{\eta}_{\Sop{C}_i})\right\|_1.
\end{equation}
Hence the proof is completed.
\end{IEEEproof}

\bibliographystyle{ieeetr}
\bibliography{IEEEabrv,QEM_analysis}

\begin{IEEEbiography}{\bf Yifeng Xiong} received his B.S. degree in information engineering, and the M.S. degree in information and communication engineering from Beijing Institute of Technology (BIT), Beijing, China, in 2015 and 2018, respectively. He is currently pursuing the PhD degree with Next-Generation Wireless within the School of Electronics and Computer Science, University of Southampton. His research interests include quantum computation, quantum information theory, graph signal processing, and statistical inference over networks.
\end{IEEEbiography}

\begin{IEEEbiography} {\bf Daryus Chandra} received the B.Eng. and M.Eng. degree in electrical engineering from Universitas Gadjah Mada, Indonesia, in 2013 and 2014, respectively. He obtained his PhD degree with the Next-Generation Wireless Research Group, School of Electronics and Computer Science, University of Southampton, UK, in 2020. Currently, he is a postdoctoral research fellow with the Quantum Internet Research Group, University of Naples Federico II, Italy. His research interests include classical and quantum error-correction codes, quantum information, and quantum communications.
\end{IEEEbiography}

\begin{IEEEbiography}{\bf Soon Xin Ng} (S'99-M'03-SM'08) received the B.Eng. degree (First class) in electronic engineering and the Ph.D. degree in telecommunications from the University of Southampton, Southampton, U.K., in 1999 and 2002, respectively. From 2003 to 2006, he was a postdoctoral research fellow working on collaborative European research projects known as SCOUT, NEWCOM and PHOENIX. Since August 2006, he has been a member of academic staff in the School of Electronics and Computer Science, University of Southampton. He is involved in the OPTIMIX and CONCERTO European projects as well as the IU-ATC and UC4G projects. He is currently an Associate Professor in telecommunications at the University of Southampton.

His research interests include adaptive coded modulation, coded modulation, channel coding, space-time coding, joint source and channel coding, iterative detection, OFDM, MIMO, cooperative communications, distributed coding, quantum error correction codes and joint wireless-and-optical-fibre communications. He has published over 200 papers and co-authored two John Wiley/IEEE Press books in this field. He is a Senior Member of the IEEE, a Chartered Engineer and a Fellow of the Higher Education Academy in the UK.
\end{IEEEbiography}

\begin{IEEEbiography}
{\bf Lajos Hanzo} (M'91-SM'92-F'04) (\url{http://www-mobile.ecs.soton.ac.uk}, \url{https://en.wikipedia.org/wiki/Lajos_Hanzo}) (FIEEE'04, Fellow of the Royal Academy of Engineering F(REng), of the IET and of EURASIP), received his Master degree and Doctorate in 1976 and 1983, respectively from the Technical University (TU) of Budapest. He was also awarded the Doctor of Sciences (DSc) degree by the University of Southampton (2004) and Honorary Doctorates by the TU of Budapest (2009) and by the University of Edinburgh (2015).  He is a Foreign Member of the Hungarian Academy of Sciences and a former Editor-in-Chief of the IEEE Press.  He has served several terms as Governor of both IEEE ComSoc and of VTS.  He has published 1900+ contributions at IEEE Xplore, 19 Wiley-IEEE Press books and has helped the fast-track career of 123 PhD students. Over 40 of them are Professors at various stages of their careers in academia and many of them are leading scientists in the wireless industry.
\end{IEEEbiography}
\end{document}

%% file: SupportDocuments/YFPreamble.tex
\usepackage{amssymb}
\usepackage{amsmath}
\usepackage{mathrsfs}
\usepackage{rotating}
\usepackage{overpic}
\usepackage{algorithm}
\usepackage{algorithmic}
\usepackage{multirow}
\usepackage{array}
\usepackage{graphicx}
\usepackage[caption=false]{subfig}
\usepackage{bm}
\usepackage{tikz}
\usepackage{epstopdf}
\usepackage{stfloats}
\usepackage{cite}
\usepackage{psfrag}
\usepackage{acronym}
\usepackage{enumerate}
\usepackage[braket]{qcircuit}
\xyoption{dvips}
\xyoption{ps}

\usepackage{amssymb}
\usepackage{amsmath}

\definecolor{BLUE}{rgb}{0,0,1}

\newtheorem{corollary}{Corollary}
\newtheorem{proposition}{Proposition}
\newtheorem{remark}{Remark}
\newtheorem{lemma}{Lemma}
\newtheorem{definition}{Definition}

\newtheorem{assumption}{Assumption}

\newcommand{\diag}[1]{{\mathrm{diag}}\left\{#1\right\}}

\acrodef{cptni}[CPTnI]{completely positive trace-nonincreasing}
\acrodef{cptp}[CPTP]{completely positive trace-preserving}
\acrodef{ggep}[GGEP]{generalized gate error probability}
\acrodef{nisq}[NISQ]{noisy intermediate-scale quantum}
\acrodef{gep}[GEP]{gate error probability}
\acrodef{qaoa}[QAOA]{quantum approximate optimization algorithm}
\acrodef{qcp}[QCP]{quantum channel precoder}
\acrodef{qem}[QEM]{quantum error mitigation}
\acrodef{qpr}[QPR]{quasi-probability representation}
\acrodef{qecc}[QECC]{quantum error correction code}
\acrodef{qedc}[QEDC]{quantum error detection code}
\acrodef{ptm}[PTM]{Pauli transfer matrix}
\acrodef{sof}[SOF]{sampling overhead factor}

\acrodefplural{gep}[GEPs]{gate error probabilities}
\acrodefplural{qcp}[QCPs]{quantum channel precoders}
\acrodefplural{qecc}[QECCs]{quantum error correction codes}
\acrodefplural{qedc}[QEDCs]{quantum error detection codes}
\acrodefplural{ptm}[PTMs]{Pauli transfer matrices}
\acrodefplural{sof}[SOFs]{sampling overhead factor}

\renewcommand{\IEEEQED}{\IEEEQEDopen}